\documentclass[twocolumn,trackchanges]{aastex7}

\usepackage{natbib}
\usepackage{graphicx}	
\usepackage{rotating}
\usepackage{amsmath}	
\usepackage{amssymb}	
\usepackage{xspace}	
\usepackage{color}
\usepackage{lineno}
\usepackage{tablefootnote}
\usepackage{fleqn}
\usepackage{threeparttable}
\usepackage{xcolor}
\usepackage{dcolumn}
\usepackage{float}
\usepackage{multirow,longtable,supertabular}
\usepackage{hyperref}

\newcommand{\thisgrb}{GRB~240825A\xspace}
\newcommand{\AstroSat}{{\em AstroSat}\xspace}
\newcommand{\fermi}{{\em Fermi}\xspace}
\newcommand{\kw}{{\em Konus}-Wind\xspace}

\newcommand{\swiftT}{{T$_{0}$}\xspace}
\newcommand{\fermiT}{{T$_{0}$}\xspace}
\newcommand{\keV}{{\rm keV}\xspace}

\newcommand{\s}{{\rm ~s}\xspace}
\newcommand{\swift}{{\em Swift}\xspace}
\newcommand{\tninty}{{$T_{\rm 90}$}\xspace}

\newcommand{\mvts}{{$t_{\rm mvts}$}\xspace}
\newcommand{\Ep}{$E_{\rm p}$\xspace}
\newcommand{\sw}[1]{\texttt{#1}}

\begin{document}

\title{Deciphering the Physical Origin of GRB 240825A: A Long GRB Lacking a Bright Supernova}

\correspondingauthor{Rahul Gupta}
\email[show]{rahul.gupta@nasa.gov, rahulbhu.c157@gmail.com}

\author[0000-0003-4905-7801]{Rahul Gupta}
\affiliation{Astrophysics Science Division, NASA Goddard Space Flight Center, Mail Code 661, Greenbelt, MD 20771, USA}
\affiliation{NASA Postdoctoral Program Fellow}
\email{rahulbhu.c157@gmail.com, rahul.gupta@nasa.gov}

\author[0000-0002-4744-9898]{Judith Racusin}
\affiliation{Astrophysics Science Division, NASA Goddard Space Flight Center, Mail Code 661, Greenbelt, MD 20771, USA}
\email{judith.racusin@nasa.gov}

\author[0000-0002-7158-5099]{R. S{\'a}nchez-Ram{\'i}rez}
\affiliation{Instituto de Astrof\'isica de Andaluc\'ia (IAA-CSIC), Glorieta de la Astronom\'ia s/n, E-18008, Granada, Spain}
\email{ruben@iaa.es}

\author[0000-0002-7400-4608]{Y.-D. Hu}
\affiliation{Guangxi Key Laboratory for Relativistic Astrophysics, School of Physical Science and Technology, Guangxi University, Nanning 530004, China}
\affiliation{INAF-Osservatorio Astronomico di Brera, Via E. Bianchi 46, 23807 Merate, LC, Italy}
\email{huyoudong072@hotmail.com}

\author[0000-0002-8860-6538]{A. Rossi}
\affiliation{INAF – Osservatorio di Astrofisica e Scienza dello Spazio, Via Piero Gobetti 93/3, 40129 Bologna, Italy}
\email{andrea.rossi@inaf.it}

\author[0000-0001-7920-4564]{M. D. Caballero-Garc{\'i}a}
\affiliation{Instituto de Astrof\'isica de Andaluc\'ia (IAA-CSIC), Glorieta de la Astronom\'ia s/n, E-18008, Granada, Spain}
\email{mcaballero@iaa.es}

\author[0000-0001-5939-644X]{Pi Nuessle}
\affiliation{Department of Physics, The George Washington University, 725 21st St. NW, Washington, DC 20052, USA}
\affiliation{Astrophysics Science Division, NASA Goddard Space Flight Center, Mail Code 661, Greenbelt, MD 20771, USA}
\email{nuessle167@gwu.edu}

\author[0000-0003-2999-3563]{A. J. Castro-Tirado}
\affiliation{Instituto de Astrof\'isica de Andaluc\'ia (IAA-CSIC), Glorieta de la Astronom\'ia s/n, E-18008, Granada, Spain}
\affiliation{Ingenier\'ia de Sistemas y Autom\'atica, Universidad de M\'alaga, Unidad Asociada al CSIC por
el IAA, Escuela de Ingenier\'ias Industriales, Arquitecto Francisco
Pe\~nalosa, 6, Campanillas, 29071 M\'alaga, Spain}
\email{ajct@iaa.es}

\author[0000-0001-9309-7873]{S. R. Oates}
\affiliation{Department of Physics, Lancaster University, Lancaster, LA1 4YB, UK}
\email{s.r.oates@lancaster.ac.uk}

\author[0009-0005-6858-6356]{P. P. Bordoloi}
\affiliation{Indian Institute of Science Education and Research, Thiruvananthapuram, Kerala, India, 695551}
\email{pragyan.bordoloi21@iisertvm.ac.in}

\author[0000-0002-9928-0369]{A. Aryan}
\affiliation{Graduate Institute of Astronomy, National Central University, 300 Jhongda Road, 32001 Jhongli, Taiwan}
\email{amararyan941@gmail.com}

\author[0000-0001-6849-1270]{S. Dichiara}
\affiliation{Penn State Univ, Dept Astron \& Astrophys, 525 Davey Lab, University Pk, PA 16802 USA}
\email{sbd5667@psu.edu}

\author[0000-0002-2149-9846]{P. Veres}
\affiliation{Department of Space Science, University of Alabama in Huntsville, Huntsville, AL 35899, USA}
\affiliation{Center for Space Plasma and Aeronomic Research, University of Alabama in Huntsville, Huntsville, AL 35805, USA}
\email{peter.veres@uah.edu}

\author[0000-0002-7465-0941]{N. Klingler}
\affiliation{Astrophysics Science Division, NASA Goddard Space Flight Center, Mail Code 661, Greenbelt, MD 20771, USA}
\email{noelklin@umbc.edu}

\author[0000-0002-5448-7577]{N. Omodei}
\affiliation{SLAC National Accelerator Laboratory, Stanford University, Stanford, CA 94305, USA}
\email{nicola.omodei@gmail.com}

\author{E. Maiorano}
\affiliation{INAF – Osservatorio di Astrofisica e Scienza dello Spazio, Via Piero Gobetti 93/3, 40129 Bologna, Italy}
\email{elisabetta.maiorano@inaf.it}

\author[0000-0002-9852-2469]{D. Tak}
\affiliation{Department of Physics, Kyung Hee University, Seoul 02447, Republic of Korea}
\email{donggeun.tak@gmail.com}

\author[0009-0009-4622-7749]{S. Shilling}
\affiliation{Department of Physics, Lancaster University, Lancaster, LA1 4YB, UK}
\email{s.shilling@lancaster.ac.uk}

\author[0000-0002-3842-965X]{J. E. Adsuara}
\affiliation{Image Processing Laboratory, University of Valencia, Paterna, Valencia, Spain}
\email{Jose.Adsuara@uv.es}

\author{P. H. Connell}
\affiliation{Image Processing Laboratory, University of Valencia, Paterna, Valencia, Spain}
\email{paul.connell@uv.es}

\author[0009-0009-4604-9639]{E. Fern{\'a}ndez-Garc{\'\i}a}
\affiliation{Instituto de Astrof\'isica de Andaluc\'ia (IAA-CSIC), Glorieta de la Astronom\'ia s/n, E-18008, Granada, Spain}
\email{emifdez@iaa.es}

\author{G. Garc{\'i}a-Segura}
\affiliation{Instituto de Astronom\'{\i}a, Universidad Nacional Aut\'onoma de M\'exico,  Carr. Tijuana-Ensenada km.107, 22860 Ensenada, B.C., Mexico}
\affiliation{Instituto de Astrof\'isica de Andaluc\'ia (IAA-CSIC), Glorieta de la Astronom\'ia s/n, E-18008, Granada, Spain}
\email{ggs@astro.unam.mx}

\author[0000-0003-2265-0381]{A. Ghosh}
\affiliation{Centre for Astro-Particle Physics (CAPP) and Department of Physics, University of Johannesburg, PO Box 524, Auckland Park 2006, South Africa}
\email{ghosh.ankur1994@gmail.com}

\author[0000-0002-5274-6790]{E. G{\"o}{\v{g}}{\"u}{\c{s}}}
\affiliation{Sabanc\i~University, Faculty of Engineering and Natural Sciences, \.Istanbul 34956 T\"urkiye}
\email{ersin.gogus@sabanciuniv.edu}

\author[0000-0002-0719-8203]{F. J. Gordillo-V\'azquez}
\affiliation{Instituto de Astrof\'isica de Andaluc\'ia (IAA-CSIC), Glorieta de la 
Astronom\'ia s/n, E-18008, Granada, Spain}
\email{vazquez@iaa.es}

\author[0000-0003-4268-6277]{M. Gritsevich}
\affiliation{Swedish Institute of Space Physics (IRF), Bengt Hultqvists v{\"{a}}g 1, 981 92 Kiruna, Sweden}
\affiliation{University of Helsinki, Faculty of Science, Gustav Hällströmin katu 2, FI-00014, Finland}
\affiliation{Institute of Physics and Technology, Ural Federal University, Mira str. 19, 620002 Ekaterinburg}
\email{maria.gritsevich@helsinki.fi}

\author{A. N. Guelbenzu} 
\affiliation{Thüringer Landessternwarte Tautenburg, Sternwarte 5, 07778 Tautenburg, Germany} 
\email{ananicuesa@gmail.com}

\author[0000-0003-2628-6468]{S. Guziy}
\affiliation{Instituto de Astrof\'isica de Andaluc\'ia (IAA-CSIC), Glorieta de la Astronom\'ia s/n, E-18008, Granada, Spain}
\affiliation{Petro Mohyla Black Sea National University, Mykolaiv 54000, Ukraine}\affiliation{Mykolaiv Astronomical Observatory, Mykolaiv 54030, Ukraine}
\email{gss@iaa.es}

\author[0000-0003-2931-3732]{L. Hanlon}
\affiliation{School of Physics and Centre for Space Research, University College 
Dublin, Belfield, Dublin 4, Ireland}
\email{lorraine.hanlon@ucd.ie}

\author[0009-0004-9747-7215]{H. J. van Heerden}
\affiliation{Department of Physics, University of the Free State, 205 Nelson 
Mandela Drive, Bloemfontein, 9300, South Africa}
\email{VanHeerdenHJ@ufs.ac.za}

\author[0000-0003-3220-7543]{S. Iyyani}
\affiliation{Indian Institute of Science Education and Research, Thiruvananthapuram, Kerala, India, 695551}
\affiliation{High Performance Computing Centre, IISER Thiruvananthapuram, Kerala, India, 695551}
\email{shabnam@iisertvm.ac.in}

\author[0000-0001-5108-0627]{A. Martin-Carrillo}
\affiliation{School of Physics and Centre for Space Research, University College 
Dublin, Belfield, Dublin 4, Ireland}
\email{antonio.martin-carrillo@ucd.ie}

\author[0000-0001-8890-5418]{P. J. Meintjes}
\affiliation{Department of Physics, University of the Free State, 205 Nelson 
Mandela Drive, Bloemfontein, 9300, South Africa}
\email{MeintjPJ@ufs.ac.za}

\author[0000-0002-3799-1237]{J. Navarro-Gonz\'alez}
\affiliation{Image Processing Laboratory, University of Valencia, Paterna, Valencia, Spain}
\email{Javier.Navarro-Gonzalez@uv.es}

\author[0000-0001-7851-7788]{T. Neubert}
\affiliation{National Space Institute, Technical University of Denmark, Kgs. 
Lyngby, Denmark}
\email{neubert@space.dtu.dk}

\author[0000-0002-2572-7033]{N. {\O}stgaard}
\affiliation{Birkeland Centre for Space Science, Department of Physics and 
Technology, University of Bergen, Norway}
\email{Nikolai.Ostgaard@uib.no}

\author{S. B. Pandey}
\affiliation{Aryabhatta Research Institute of Observational Sciences (ARIES), Manora Peak, Nainital-263002, India} 
\email{shashiaries0@gmail.com}

\author[0000-0002-7273-3671]{I. P{\'e}rez-Garc{\'i}a}
\affiliation{Instituto de Astrof\'isica de Andaluc\'ia (IAA-CSIC), Glorieta de la Astronom\'ia s/n, E-18008, Granada, Spain}
\email{ipg@iaa.es}

\author[0000-0002-0130-2460]{S.~Razzaque}
\affiliation{Centre for Astro-Particle Physics (CAPP) and Department of Physics, University of Johannesburg, PO Box 524, Auckland Park 2006, South Africa}
\affiliation{Department of Physics, The George Washington University, 725 21st St. NW, Washington, DC 20052, USA}
\affiliation{National Institute for Theoretical and Computational Sciences (NITheCS), Private Bag X1, Matieland, South Africa}
\email{soebur.razzaque@gmail.com}

\author[0000-0002-6909-192X]{E. Sonbas}
\affiliation{Department of Physics, The George Washington University, 725 21st St. NW, Washington, DC 20052, USA}
\email{edasonbas@gwu.edu}

\author[0009-0004-7113-8258]{Si-Yu Wu}
\affiliation{Instituto de Astrof\'isica de Andaluc\'ia (IAA-CSIC), Glorieta de la Astronom\'ia s/n, E-18008, Granada, Spain}
\email{wusiyu.11@outlook.com}

\author[0000-0001-9435-1327]{A. Pozanenko}
\affiliation{Faculty of Physics, Higher School of Economics, Moscow 101000, Russia}
\email{grb123@mail.ru}

\author[0000-0003-3554-1037]{A. Volnova}
\affiliation{Sternberg Astronomical Institute, Moscow State University, Russia}
\email{alinusss@gmail.com}

\author[0000-0003-3244-6616]{A. Moskvitin}
\affiliation{Special Astrophysical Observatory, Russian Academy of Sciences, Nizhny Arkhyz 369167, Russia}
\email{moskvitin.alexander@gmail.com}

\author[0000-0002-9798-029X]{S. Belkin}
\affiliation{School of Physics \& Astronomy, Monash University, Clayton VIC 3800, Australia}
\email{astrolltime@gmail.com}

\author{O. Spiridonova}
\affiliation{Special Astrophysical Observatory, Russian Academy of Sciences, Nizhny Arkhyz 369167, Russia}
\email{ospir@sao.ru}

\author{O. Burkhonov}
\affiliation{Ulugh Beg Astronomical Institute, Uzbek Academy of Sciences, Tashkent 100052, Uzbekistan}
\email{boa@astrin.uz}

\author{Sh. Egamberdiyev}
\affiliation{Ulugh Beg Astronomical Institute, Uzbek Academy of Sciences, Tashkent 100052, Uzbekistan}
\affiliation{National University of Uzbekistan, Tashkent, Uzbekistan}
\email{shuhrat@astrin.uz}

\author{E. Klunko}
\affiliation{Institute of Solar-Terrestrial Physics, Russian Academy of Sciences (Siberian Branch), Irkutsk 664033, Russia}
\email{eklunko@gmail.com}

\author{V. Rumyantsev}
\affiliation{Crimean Astrophysical Observatory, Russian Academy of Sciences,  Nauchny 298409, Crimea}
\email{rum@craocrimea.ru}

\author{I. Sokolov}
\affiliation{Institute of Astronomy of the Russian Academy of Science, Moscow, Russia}
\email{ilia.v.sokolov@gmail.com}

\author{A. Novichonok}
\affiliation{Petrozavodsk State University, 185910, Petrozavodsk, Karelia, Russia}
\email{cometsnov01@gmail.com}

\author{I. Reva}
\affiliation{Fesenkov Astrophysical Institute, Observatory street 23, Almaty 050020, Kazakhstan}
\email{alfekka@gmail.com}

\author{A. Volvach}
\affiliation{Crimean Astrophysical Observatory, Russian Academy of Sciences, Nauchny 298409, Crimea}
\email{volvach@bk.ru}

\author{L. Volvach}
\affiliation{Crimean Astrophysical Observatory, Russian Academy of Sciences, Nauchny 298409, Crimea}
\email{voe@inbox.ru}

\author[0000-0003-4044-5357]{Jan Snigula}
\affiliation{Max-Planck-Institut f\"ur Extraterrestrische Physik, Giessenbachstra\ss{}e 1, 85748 Garching, Germany}
\email{snigula@mpe.mpg.de}

\author[0009-0009-9751-9215]{Chiara Ventura}
\affiliation{INAF-Osservatorio Astronomico di Roma, via Frascati 33, I-00040 Monte Porzio Catone, Italy}
\email{chiara.ventura@inaf.it}


\begin{abstract}

We present a comprehensive multiwavelength analysis of GRB 240825A, a bright gamma-ray burst (GRB) detected by \textit{Fermi} and \textit{Swift}, with a prompt duration ($T_{\rm 90}$ $\sim$ 4~sec in 50-300 \keV in GBM) near the boundary separating short and long GRBs, prompting a detailed investigation into its classification and progenitor. We use classical prompt metrics (duration, minimum variability timescale (MVT), lag, and spectral hardness) and modern classification techniques (machine-learning (ML) based t-SNE, support vector machine, energy-hardness-duration, and $\varepsilon \equiv E_{\gamma,\mathrm{iso},52} / E_{p,z,2}^{5/3}$) and find most properties (prompt energetics, placement on the Amati relation, and spectral lag) of GRB 240825A consistent with a collapsar origin. However, extensive late-time optical and NIR follow-up with the 10.4\,m GTC and 8.4\,m binocular LBT telescopes reveals no bright supernova (like SN 1998bw) is detected down to stringent limits (e.g., $m_r > 25.0$~mag at 17.59 days), despite a redshift of $z = 0.659$ measured from GTC spectroscopy. Host galaxy SED modeling with \texttt{Prospector} indicates a massive, and star-forming galaxy—typical of collapsar GRB hosts, though with a large offset. We compare these findings with hybrid events like GRB 211211A, GRB 230307A, GRB 200826A, including SNe-GRBs, and conclude that GRB~240825A most likely originated from a massive star collapse, possibly with the associated SN heavily obscured or intrinsically faint. This study emphasizes the need for multiwavelength follow-up and a multi-layered classification to determine GRB progenitors.

\end{abstract}


\section{Introduction} 

Gamma-ray bursts (GRBs) are broadly divided into two main classes—short and long—based on their observed durations (\tninty, the time interval containing 90\% of the burst's fluence, \citealt{1993ApJ...413L.101K}), and spectral hardness \citep{2016ApJ...829....7L}. This dichotomy, first revealed by the Burst and Transient Source Experiment (BATSE) onboard the \textit{Compton Gamma Ray Observatory}, has served as a foundation for understanding GRB progenitors. Long-duration GRBs (LGRBs, \tninty $>$ 2 sec) typically have softer photon spectra than short-duration GRBs (SGRBs, \tninty $\leq$ 2 sec). 

The physical association of LGRBs with the core-collapse of massive, rapidly rotating stars \citep{1993ApJ...405..273W, 1999ApJ...524..262M}, sometimes accompanied by broad-lined Type Ibc supernovae (SNe) \citep{2003Natur.423..847H, 2006ARA&A..44..507W}, and SGRBs with compact binary mergers such as binary neutron star or neutron star–black hole systems \citep{2007PhR...442..166N, 2014ARA&A..52...43B}, provided an initial physical basis for the \tninty-based dichotomy. However, the \tninty-based classification scheme has significant limitations. The duration of a GRB is not a purely intrinsic property but is instead dependent on the energy band of the detector, the sensitivity threshold, the signal-to-noise (S/N) ratio of the event, and the redshift of the source \citep{2013ApJ...763...15Q, 2013ApJ...764..179B, 2022ApJ...927..157M}. For instance, \tninty tends to be longer at lower energies, and bursts with extended soft tails may be misclassified as long GRBs when observed with sensitive instruments like \swift-Burst Alert Telescope (BAT). Such instrumental dependencies introduce ambiguities, particularly for GRBs near the empirical boundary of $T_{90} \sim 2$ seconds. Moreover, the observed duration distributions of short and long GRBs significantly overlap, further complicating the classification of events near this boundary.

The complexity has only increased with the discovery of more such ambiguous or ``hybrid" events. For example, GRB~060614, a long-duration GRB ($T_{90} \sim 100$ sec), showed no accompanying SN emission down to deep optical limits \citep{2006Natur.444.1044G, 2006Natur.444.1050D}, challenging the long-GRB/SN association. \cite{2006Natur.444.1047F} reported another nearby long-duration GRB 060505, with \tninty values of 4 sec, that exhibited no SN emission down to limits hundreds of times fainter than SN 1998bw (a prototypical GRB-SN). These nearby ``SN-less" GRBs, located in star-forming regions, suggest that not all long GRBs conform to the collapsar model. More recently, long-duration GRBs like GRB 191019A, GRB 211211A, GRB 211227A, and GRB 230307A exhibit signatures (e.g., kilonova (KN)-like emission) consistent with compact binary mergers \citep{2022Natur.612..223R, 2022Natur.612..228T, 2024Natur.626..737L, 2022ApJ...931L..23L, 2022Natur.612..232Y, 2024Natur.626..742Y, 2025ApJ...979..159S}. Conversely, a short-duration burst GRB 200826A, despite $T_{90} \lesssim 2$ sec, is associated with young, star-forming environments and a SN component, suggesting a collapsar origin \citep{2021NatAs...5..917A, 2022ApJ...932....1R}. These events highlight that duration alone is not a reliable proxy for progenitor type \citep{2013ApJ...764..179B, 2025JHEAp..45..325Z}.

Such hybrid events inspired the development of physically motivated classification systems such as the ``Type I (associated with the mergers of compact objects)/ Type II (associated with the collapsars)" scheme \citep{2009ApJ...703.1696Z}, which incorporates prompt emission, afterglow, host galaxy properties, local environments, and the presence (or absence) of SNe. To overcome the limitations of duration-based classification, the community increasingly employs multiwavelength and multidimensional analyses. Key discriminants include prompt emission properties such as spectral lag \citep{2000ApJ...534..248N}, and minimum variability timescale \citep{2013MNRAS.432..857M, 2023A&A...671A.112C}, afterglow properties \citep{2025JHEAp..4700384L, 2020ApJ...904...97D}, as well as host galaxy properties such as morphology, star formation rate, and metallicity \citep{2022JApA...43...82G, 2006Natur.441..463F, 2022ApJ...940...56F}. Furthermore, machine learning (ML) techniques have been leveraged to integrate these diverse observational parameters into more robust, probabilistic classification frameworks \citep{2020ApJ...896L..20J, 2024MNRAS.532.1434Z, 2025ApJ...981...14N, 2024ApJ...974..120N, 2023ApJ...959...44L}.

In this study, we present a comprehensive multiwavelength investigation (classical and ML-based) of GRB 240825A, a long-duration event ($T_{90} \sim 4$ sec) detected by \fermi. We examine the prompt emission characteristics (including ML-based classification), afterglow temporal and spectral evolution, and host galaxy properties to establish its classification and physical origin. We used 10.4\,m \textit{Gran Telescopio Canarias} (GTC) and the Large Binocular Telescope (LBT) telescopes to search for associated SN and decipher the physical origin of GRB 240825A. Despite extensive monitoring, no SN emission is detected, reminiscent of GRBs 060505, 060614, and GRB 111005A \citep{2018A&A...616A.169M, 2018A&A...615A.136T}. The available limits provide only mild constraints. They mainly disfavor a luminous SN 1998bw-like component at the epochs covered, while fainter, or more dust-obscured GRB-SNe remain possible. Section \ref{subsec:Gamma-ray Observations} presents the observations and data analysis, Section \ref{results} details the results, and Section \ref{sec:Discussion} discusses the physical interpretation of the results. Conclusions are drawn in Section~\ref{sec:Conclusion}.

\section{Observations and Data Analysis} 
\label{sec:Observations}

\subsection{Prompt emission detection} 
\label{subsec:Gamma-ray Observations}

\begin{figure}[ht]
    \centering
        \includegraphics[width = 0.45\textwidth]{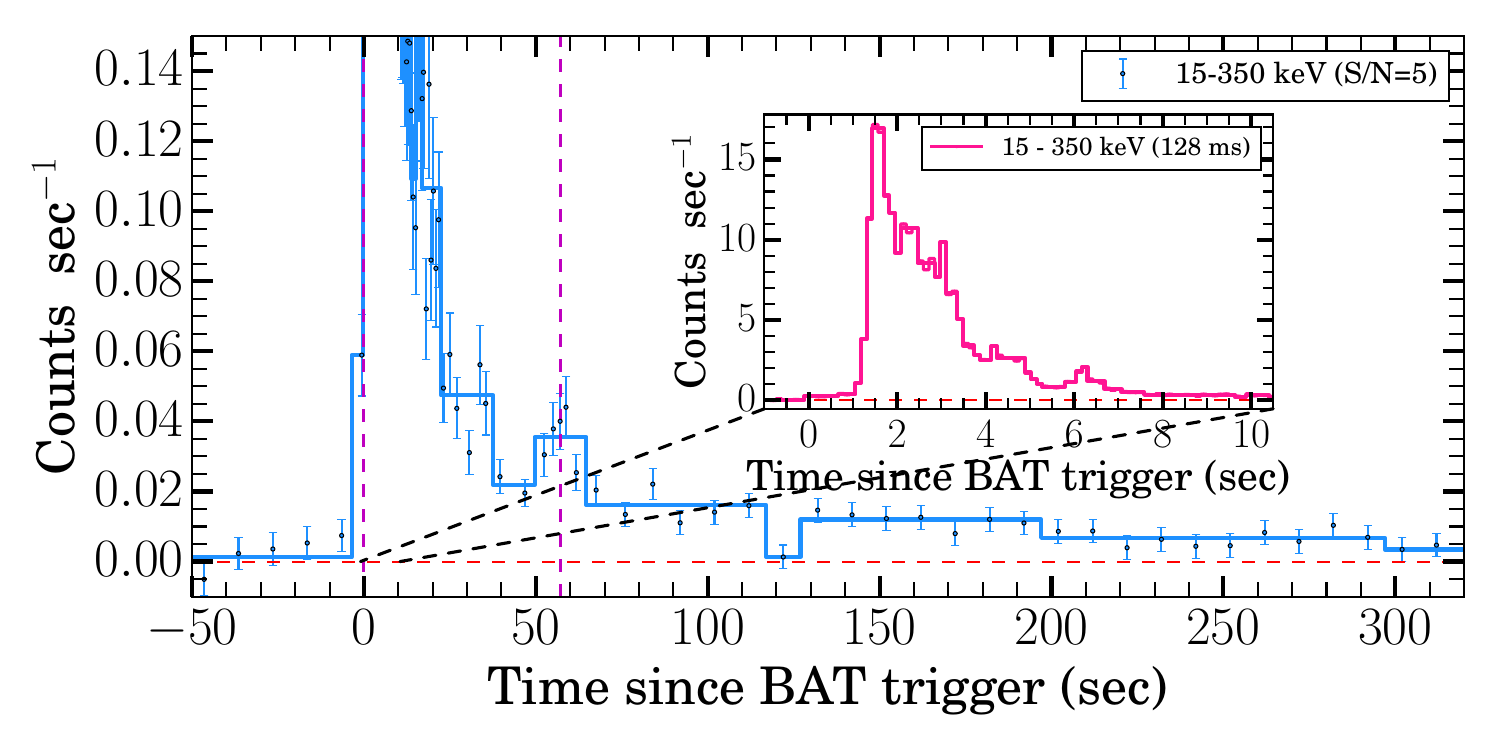}
    \includegraphics[width = 0.47\textwidth]{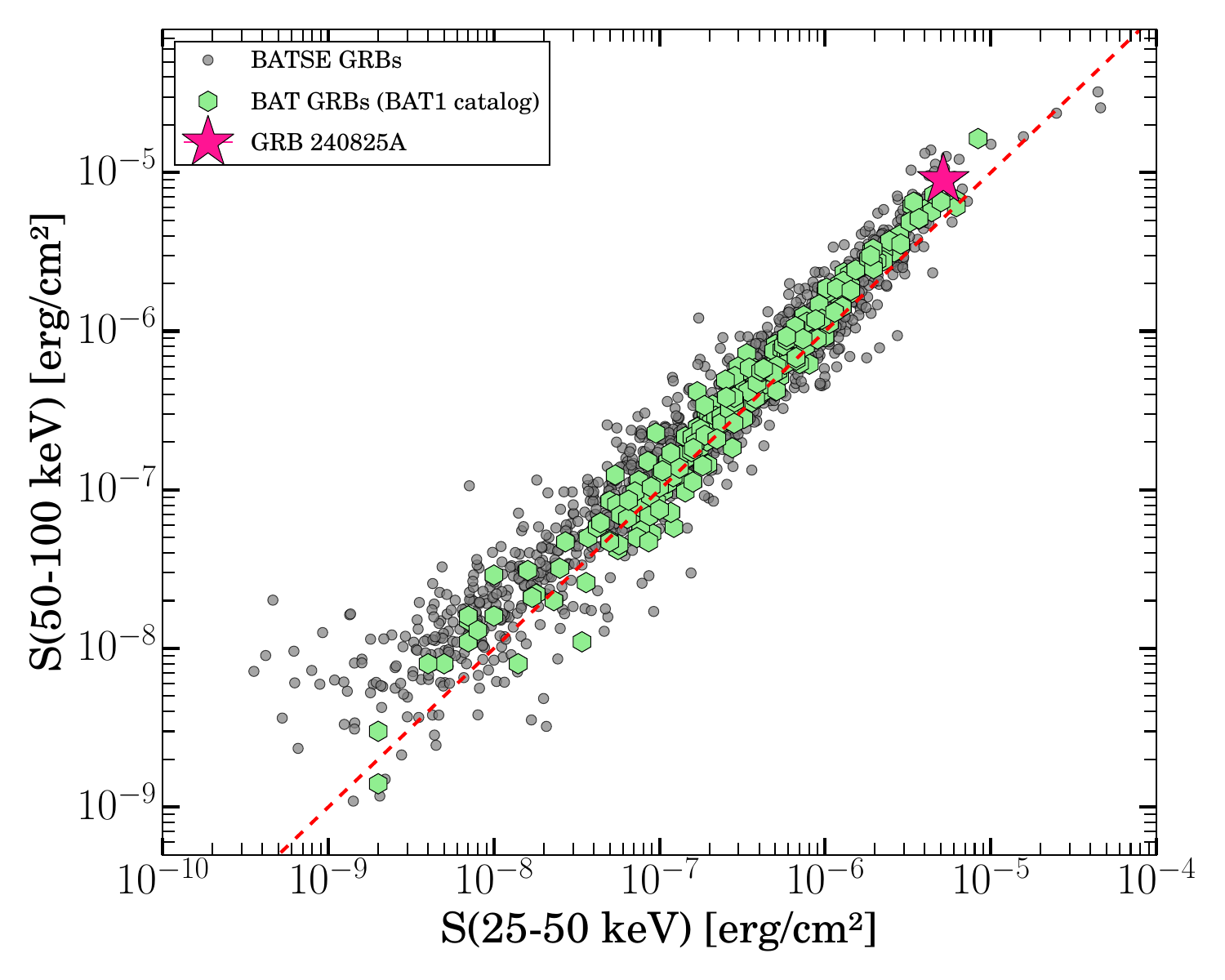}
    \caption{Top: BAT light curve in the 15-350 keV energy range reveals the presence of a faint extended emission post the main burst, observed after employing S/N criteria with 5 or 10-second binning. The vertical dashed lines represent a BAT \tninty duration of 57.20 sec. The inset shows the prompt zoomed light curve of GRB 240825A in the 15-350 keV energy range, observed using the \swift-BAT and presented in 128ms time bins, which highlights the burst's immediate temporal behavior. Bottom: The comparative analysis of GRB 240825A's fluence (pink star) with other GRBs detected by the BAT and BATSE instruments\footnote{\url{https://swift.gsfc.nasa.gov/results/batgrbcat/}}, showing its high fluence ($\sim$ 10$^{-5}$ erg cm$^{-2}$), typically consistent with collapsar GRBs. The red dashed line marks that fluence in the S(50-100 keV) band is equal to fluence in the S(25-50 keV) band.}
    \label{FIG1}
\end{figure}

GRB 240825A was detected by the \fermi Gamma-ray Burst Monitor (GBM) at 15:53:00 UTC on August 25, 2024 \citep{2024GCN.37273....1F}. The burst was also detected by the \fermi Large Area Telescope (LAT), which detected significant high-energy emission above 100 MeV \citep{2024GCN.37288....1D}. The \swift-BAT independently triggered on GRB 240825A at 15:52:59 UTC, localizing the burst to coordinates RA (J2000) = 22h 58m 12s, Dec (J2000) = +01d 02' 10" with an uncertainty of 3 arcminutes \citep{2024GCN.37274....1G}. We carried out the temporal analysis of the \swift-BAT data of GRB 240825A (see method in \citealt{2024arXiv240904871G}). The mask-weighted light curve (binned at 128 ms in the 15–350 keV band), displayed in Figure \ref{FIG1}, shows a bright, fast-rising, exponential decay (FRED)-like main prompt emission pulse, starting at \swiftT sec (\swift-BAT trigger time), peaking at \swiftT+1.8 sec, and ending at \swiftT+10 sec. To further probe for any extended emission (EE), we constructed a light curve with S/N criteria with 5 or 10-second binning, which uncovers faint emission persisting beyond the main emission phase. This softer and longer tail emission extends up to $\sim$ \swiftT+300 sec with a \tninty duration of 57.20 $\pm$ 8.57 sec \citep{2024GCN.37355....1M}. We further compared GRB 240825A's 25–50 keV and 50-100 keV fluence to a sample of \swift-BAT and BATSE GRBs, finding its fluence ($\sim$ 10$^{-5}$ erg cm$^{-2}$) to be high (see Figure \ref{FIG1}).

\subsection{X-ray afterglow}

The \swift X-Ray Telescope (XRT; \citealt{2005SSRv..120..165B}) initiated observations of the GRB 240825A field at 15:54:22.9 UT, approximately 83.1 seconds after the \swift-BAT trigger. The XRT identified a bright, uncatalogued, and fading X-ray source at RA = 22h 58m 16.58s, Dec = +01d 01' 31.1" (J2000), with a positional uncertainty of 12.3 arcseconds (90\% containment), and further an astrometrically corrected position was determined at RA = 22h 58m 17.28s, Dec = +01d 01' 36.3" (J2000), with a refined uncertainty of 2.7 arcseconds (90\% confidence). This position lies 75 arcseconds from the BAT onboard location \citep{2024GCN.37290....1E}. Temporal and spectral data for the X-ray afterglow were retrieved from the \textit{Swift} XRT GRB light curve and spectrum repositories \citep{2007A&A...469..379E, 2009MNRAS.397.1177E}, provided by the UK Swift Science Data Centre\footnote{\url{https://www.swift.ac.uk/}}. The XRT spectrum in the 0.3--10 keV energy range was fitted with an absorbed power-law model, using both a fixed Galactic absorption model (\texttt{phabs}) and a free absorption model (\texttt{zphabs}) to constrain contributions from the host galaxy (see \citealt{2024A&A...683A..55C, 2023arXiv231216265G} for detailed methodology).

\subsection{Ultra-Violet and optical observations}

\subsubsection{Swift-UVOT}

The {\it Swift}/UVOT \citep{2005SSRv..120...95R} began observations of the field of GRB 240825A \citep{2024GCN.37296....1K} 77 sec after the \swift-BAT trigger. The source counts were extracted using a source region of 5 arcsec radius. Background counts were extracted using a 20 arcsec circular region located in a source-free region. The afterglow count rates were converted to magnitudes using the UVOT photometric zero points \citep{poole, bre11}. To improve the S/N ratio, the count rates in each filter were binned using $\Delta t/t$ =0.2.
The UVOT detector is less sensitive in a few small patches\footnote{\url{https://heasarc.gsfc.nasa.gov/docs/heasarc/caldb/swift/docs/uvot/uvotcaldb_sss_02b.pdf}} for which a correction has not yet been determined. Therefore, we determine if the afterglow falls on any of these patches. We find that during the $white$, $u$, and UV filters during the first 1300s, the afterglow does fall on one of these patches. We have therefore excluded the UV data affected by this, but we include $white$, and $u$ band data since it provides information on the early time behavior of the afterglow; as such, caution should be used when interpreting the early $white$ and $u$ band data. The \textit{Swift}/UVOT observations of GRB 240825A, detailed in Table~\ref{tab:combined_photometry} of the appendix.

\subsubsection{OSN and BOOTES}

We conducted optical afterglow observations of GRB 240825A using the 1.5m telescope at the Observatorio de Sierra Nevada (OSN) in Spain and the 0.6m Dolores P\'erez-Ram\'irez robotic telescope (DPRT) at the BOOTES-6 station in South Africa (as part of the BOOTES Global Network, \citealt{2023NatAs...7.1136C, 2023FrASS..10.2887H}). OSN observations began approximately 8.4 hours (T-\swiftT = 30176 sec) post-trigger, covering the $B$, $V$, $I$, and $R$ bands. The OSN data were reduced with a custom pipeline that included bias subtraction, flat-field correction using flat frames, and background subtraction to account for varying sky conditions. Aperture photometry, optimized for the afterglow's point-spread function, was calibrated against nearby Pan-STARRS1 stars to derive AB magnitudes, reported without corrections for Galactic or host extinction (see Table \ref{tab:combined_photometry} of the appendix). Complementary $R$-band observations were obtained with the 0.6m BOOTES-6/DPRT starting approximately 2.5 hours (T-\swiftT = 9063 sec) post-trigger. BOOTES-6 data were reduced with bias and flat-field corrections and initially calibrated to the Vega magnitude system. These magnitudes were converted to the AB system using a zero-point offset of 0.21 mag.

\subsubsection{GRB IKI-FuN observations}

We continued monitoring the afterglow using a range of telescopes from the IKI Gamma-Ray Burst Follow-up Network (IKI GRB-FuN network \citep{Volnova21}) and additional facilities. Optical observations were conducted with the 1.5-meter AZT-22 telescope at the Maidanak Astronomical Observatory (MAO), operated by the Ulugh Beg Astronomical Institute in Uzbekistan. Further R-band imaging was obtained with the Zeiss-1000 telescope, located at the Special Astrophysical Observatory (SAO) in Russia. Additional observations were carried out using the RC-35 (Kitab) and the Zeiss-1000 telescopes. All our optical data obtained from the IKI GRB-FuN network were processed using NOAO's \textsc{iraf} software package\footnote{\url{https://iraf-community.github.io/} }. Results of the observations are presented in Table~\ref{tab:combined_photometry} of the appendix.

\subsection{Late time Imaging and Spectroscopy: Search for SN and host galaxy}

\begin{figure*}[ht]
    \centering
    \includegraphics[width = 0.75\textwidth]{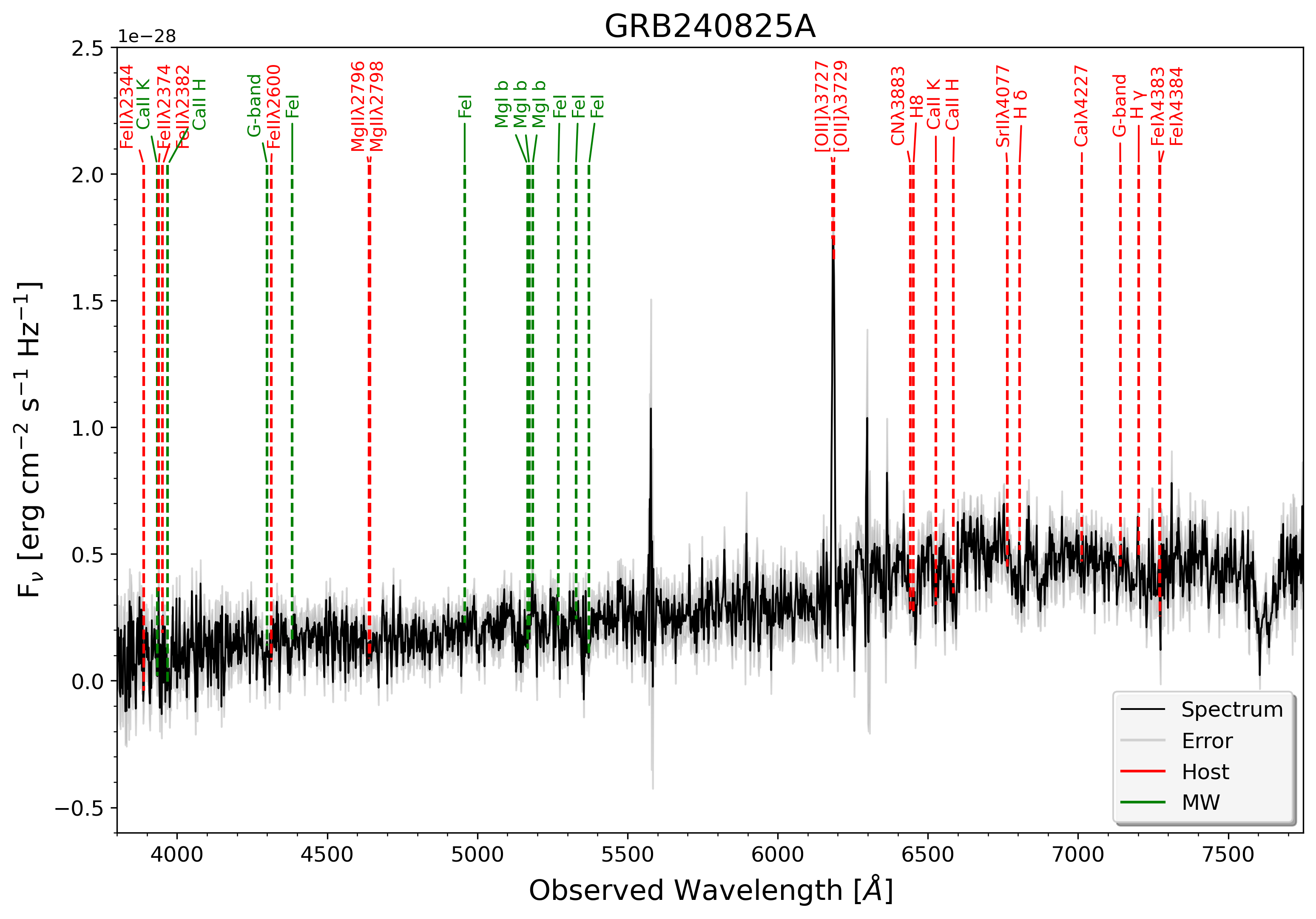}
\caption{Optical spectrum of the host galaxy of GRB~240825A obtained with the 10.4\,m GTC. The spectrum (black) is overlaid with the corresponding error spectrum (gray). Features associated with the host galaxy are marked in red, while foreground (Milky Way) features are shown in green. Detected lines include [O\,\textsc{ii}]~$\lambda\lambda$3727, 3729, Ca\,\textsc{ii}~H, Ca\,\textsc{ii}~K, and several Fe\,\textsc{ii}, Mg\,\textsc{ii}, and Balmer lines, which were used to determine a redshift of $z = 0.659$ for the host galaxy.}
\label{spectrum}
\end{figure*}

\subsubsection{10.4m GTC spectroscopy observations and redshift determination}

We performed spectroscopic observations of the field of GRB~240825A using the Optical System for Imaging and low-Intermediate-Resolution Integrated Spectroscopy (OSIRIS) mounted on the 10.4\,m GTC, located at the Roque de los Muchachos Observatory, La Palma, Canary Islands. Observations were conducted on 2024 October 9 and 10 under good atmospheric conditions, with airmass ranging from 1.13 to 1.25. We used the R1000B grism, which covers the wavelength range of 3630--7500\,\AA\ with a spectral resolving power of $R \sim 1000$. 

Data reduction was performed using standard IRAF routines and complementary Python-based tools. The two-dimensional spectroscopic frames were bias-subtracted, flat-field corrected, and cleaned of cosmic rays using the L.A.~Cosmic algorithm \citep{2001PASP..113.1420V}. One-dimensional spectra were extracted using an optimal aperture centered on the galaxy trace after subtracting the background via a low-order polynomial fit to spatial regions adjacent to the target. Wavelength calibration was achieved using comparison arc lamps (Ne, Ar, and Hg), and flux calibration was performed using a spectrophotometric standard star (Feige110) observed on the same nights. Finally, we accounted for slit losses ($\sim$29$\%$) scaling the spectrum to the photometric magnitude measured in the r'-band (see detailed methodology \citep{2024A&A...692A...3S, 2021A&A...646A..50H, 2022MNRAS.511.1694G}).

The reduced and calibrated spectrum (Figure~\ref{spectrum}) exhibits a prominent [O\,\textsc{ii}]~$\lambda\lambda$3727, 3729 emission lines associated with the host galaxy, as well as several absorption features including Ca\,\textsc{ii}~H, Ca\,\textsc{ii}~K, the G-band, Ca\,\textsc{i}$\lambda$4227, Sr\,\textsc{ii}$\lambda$4077, CN$\lambda$3883, and multiple Fe\,\textsc{ii}, Fe\,\textsc{i} and Mg\,\textsc{ii} lines. These lines were consistently identified across both nights, yielding a redshift measurement of $z = 0.659$ for the host galaxy of GRB~240825A. Our redshift measurement is consistent with the value measured by the VLT equipped with the X-shooter spectrograph \citep{2024GCN.37293....1M}. We also noted several foreground (Milky Way) features (see Figure~\ref{spectrum}) in our spectrum.

\subsubsection{10.4m GTC NIR and optical panchromatic observations}

\begin{deluxetable*}{lcccc}
\tablecaption{Host galaxy photometric observations of GRB 240825A. \label{tab:host_observations}}
\tablehead{
\colhead{Instrument/Survey} & \colhead{Filter/Band} & \colhead{T-\swiftT (sec)} & \colhead{Exp. (sec)} & \colhead{AB Mag $\pm$ error}
}
\startdata
\multicolumn{5}{c}{\textit{GTC Optical Imaging}} \\
GTC & $r$ & 9004750 & 540 & $22.45 \pm 0.07$ \\
GTC & $g$ & 9005075 & 960 & $23.76 \pm 0.11$ \\
GTC & $i$ & 9005502 & 540 & $22.03 \pm 0.05$ \\
GTC & $z$ & 9006216 & 400 & $21.63 \pm 0.12$ \\ \hline
\multicolumn{5}{c}{\textit{GTC NIR Imaging}} \\
GTC & $K$ & 1668561 & 1260 & $21.10 \pm 0.06$ \\
GTC & $J$ & 1671315 & 910 & $21.77 \pm 0.10$ \\
GTC & $H$ & 1673259 & 1134 & $21.50 \pm 0.09$ \\
GTC & $K$ & 2365811 & 1260 & $21.14 \pm 0.06$ \\
GTC & $J$ & 2368461 & 910 & $21.81 \pm 0.14$ \\ \hline
\multicolumn{5}{c}{\textit{DESI Legacy Surveys (DR10)}} \\
DESI & $g$ & \nodata & \nodata & $23.607 \pm 0.063$ \\
DESI & $r$ & \nodata & \nodata & $22.463 \pm 0.027$ \\
DESI & $i$ & \nodata & \nodata & $21.875 \pm 0.051$ \\
DESI & $z$ & \nodata & \nodata & $21.627 \pm 0.032$ \\
DESI & $W1$ & \nodata & \nodata & $20.577 \pm 0.102$ \\
DESI & $W2$ & \nodata & \nodata & $21.692 \pm 0.965$ \\
DESI & $W3$ & \nodata & \nodata & $18.141 \pm 1.322$ \\ \hline
\multicolumn{5}{c}{\textit{CAHA Optical Imaging}} \\
CAHA & $g$ & 2961761 & 750 & $23.53 \pm 0.20$ \\
CAHA & $r$ & 2960729 & 720 & $22.71 \pm 0.14$ \\
CAHA & $i$ & 2959660 & 720 & $21.78 \pm 0.15$ \\
CAHA & $z$ & 2958492 & 700 & $21.87 \pm 0.29$ \\ 
\enddata
\tablecomments{AB magnitudes are reported without corrections for Galactic or host extinction. For GTC observations, T-\swiftT represents the time since the GRB trigger, and Exp. is the exposure time. DESI Legacy Surveys data are from DR10, with no reddening correction applied.}
\end{deluxetable*}

We performed near-infrared (NIR) and optical imaging observations of the host galaxy of GRB 240825A using the 10.4\,m GTC, equipped with the OSIRIS instrument for optical bands and the Espectr\'ografo Multiobjeto Infra-Rojo (EMIR) instrument for NIR bands. Optical imaging was conducted in the $g$, $r$, $i$, and $z$ filters, starting approximately 104 days after the GRB trigger. NIR observations in the $J$, $H$, and $K$ bands were obtained spanning a temporal range of $\sim$ 19.31 to 27.41 days post-trigger. The raw data were processed using the standard pipeline, which included bias subtraction, flat-fielding, and sky subtraction tailored to the varying background levels in each band. Photometry was performed using aperture photometry with a radius optimized for the host's point-spread function (see \citealt{2021MNRAS.505.4086G} for detailed methodology). Optical AB magnitudes were calibrated against Pan-STARRS standard stars, while NIR magnitudes were calibrated using 2MASS standards. These magnitudes are reported without corrections for Galactic or host extinction (see Table \ref{tab:host_observations}).

\subsubsection{DESI optical infrared observations}

Complementing GTC host observations, we extracted multi-band photometry from the Dark Energy Spectroscopic Instrument (DESI) Legacy Surveys Data Release 10 (DR10), covering the $g$, $r$, $i$, $z$, $W1$, $W2$, and $W3$ bands, using the survey's public catalog. For the data analysis of DESI data, we performed aperture photometry with a radius optimized to the host's point-spread function, calibrated against Pan-STARRS standard stars for optical bands ($g$, $r$, $i$, $z$) and WISE standards for infrared bands ($W1$, $W2$, $W3$). The resulting magnitudes were derived without applying reddening corrections and are summarized in Table \ref{tab:host_observations}. We noted that GTC optical measurements remain consistent with DESI values.

\subsubsection{2.2m CAHA optical observations}

We also obtained observations with the 2.2\,m telescope at the Calar Alto Observatory in Almería, Spain, using the Calar Alto Faint Object Spectrograph (CAFOS). Imaging was performed in the Sloan $g'$, $r'$, $i'$, and $z'$ filters on two epochs: 2024 September 28 and 2024 November 24. However, only the data from the first night were usable due to poor weather conditions during the second epoch. The first observation began at 21:31 UT, corresponding to approximately 34 days post-burst. A faint detection of the GRB host galaxy was noted in the stacked images (see Table \ref{tab:host_observations}). The raw data were reduced using standard procedures. Absolute photometric calibration was performed using nearby field stars from the Pan-STARRS1 catalog.

\subsubsection{LBT optical observations}

We observed the field of GRB 240825A with the twin Large Binocular Camera (LBCs) mounted on the LBT at Mt. Graham, AZ, USA, in the Sloan $r'$ and $z'$ bands at three epochs (Program: IT-2024B-023, PI E. Maiorano): 2024-09-12 ($\sim$17.59 days post-burst), 2024-09-28 ($\sim$33.70 days post-burst), and 2024-11-09 ($\sim$75.52 days post-burst), with varying exposure times ranging from 30 to 60 minutes per filter to optimize S/N ratios. LBT data were reduced using the standard pipeline \citep{2014A&A...570A..11F}, which includes bias subtraction and flat-fielding, bad pixel and cosmic ray masking, astrometric calibration, and coaddition. We used HOTPANTS-based image subtraction to search for associated SN (see section \ref{SN}). We calibrated the LBT images against Pan-STARRS field stars and uncorrected for foreground Galactic extinction. 

\subsection{Radio observations}

As part of the multi-wavelength coverage of GRB 240825A, radio observations from the Atacama Large Millimeter/submillimeter Array (ALMA; 97.5 GHz, \citep{2024GCN.37314....1L}), the Very Large Array (VLA; 8.6 GHz, \citep{2024GCN.37322....1P}), MeerKAT (1.3 GHz, \citep{2024GCN.37353....1P}), and the Australia Telescope Compact Array (ATCA; multiple frequencies, \citep{2024GCN.37388....1G}) were included. These measurements were reported in public GCN circulars.

\section{Results}
\label{results}

\subsection{Prompt emission temporal characteristics}

\begin{figure}[ht]
    \centering
        \includegraphics[width = 0.47\textwidth]{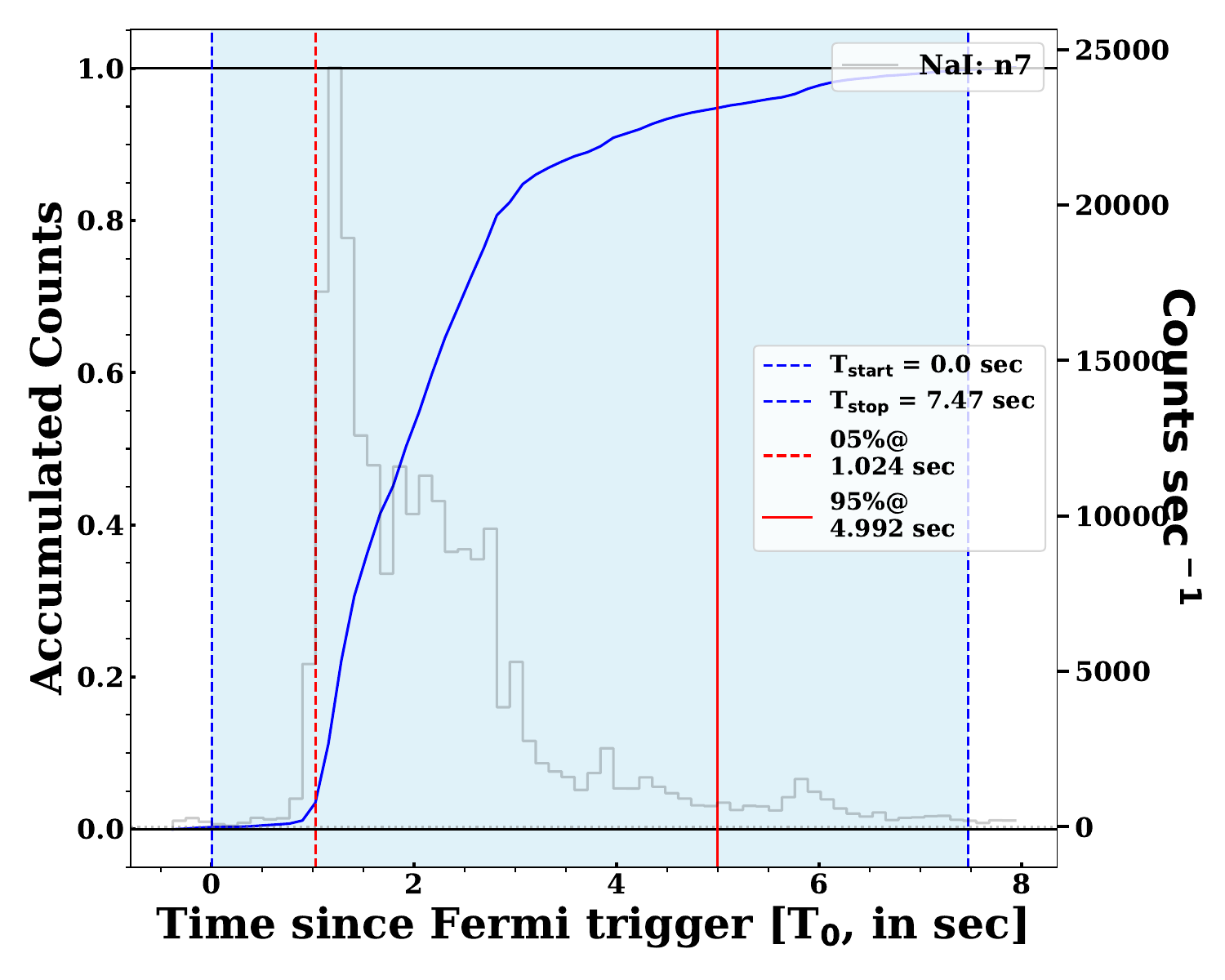}
    \caption{Cumulative count rate and \tninty calculation for GRB 240825A in \fermi-GBM 50–300 keV depicting the cumulative count rate of GRB 240825A as observed by the \fermi-GBM in the 50–300 keV energy band, with time measured relative to the GBM trigger. The \tninty duration, representing the interval containing 90\% of the burst's fluence, is calculated as 3.968 sec, defined by the 5\% accumulation level at 1.024 sec (red dashed line) and the 95\% level at 4.992 sec (red solid line), with a start time of 0.0 sec and stop time of 7.47 sec (blue dashed lines). The background-subtracted cumulative profile, shown in blue, illustrates the burst's temporal evolution, with the \tninty value $\sim$ 2 sec suggesting an ambiguous classification. The background-subtracted GBM light curve is shown in grey.}
    \label{FIG2_T90_Fermi}
\end{figure}

The duration of GRB 240825A depends strongly on the instrument, energy band, and treatment of the weak soft tail. In our \fermi-GBM analysis, the bright hard main pulse in the 50–300 keV band has \tninty = 3.968 $\pm$ 0.091 sec (see Figure \ref{FIG2_T90_Fermi}). This value should be interpreted as the duration of the main hard prompt-emission episode. Independent high-energy observations support a similarly short main-emission phase: \cite{2025ApJ...985L..30W} reported \tninty = 4.96$^{+0.09}_{-0.13}$ sec for GECAM-B in 40–6000 keV, \tninty = 2.15$^{+0.25}_{-0.40}$ sec for GECAM-C in 6–6000 keV, and \tninty = 1.90$^{+0.15}_{-0.25}$ sec for Insight-HXMT/HE in 30–1000 keV. These measurements indicate that the bright hard prompt episode was relatively short.

In contrast, \swift-BAT is more sensitive to softer emission and detects a weak tail following the main pulse. If this tail is included as a prompt emission, the BAT \tninty duration becomes 57.20 $\pm$ 8.57 sec. If the weak tail is instead interpreted as early afterglow and excluded, the BAT main-emission duration is much shorter, \tninty = 7.42 $\pm$ 0.16 sec in 15-350 \keV \citep{2025ApJ...985L..30W}. Additionally, \kw detected the total duration of main emission of \thisgrb $\sim$ 2 sec, followed by a weaker emission and softer emission lasts until $\sim$ \fermiT + 25 sec \citep{2024GCN.37302....1F} and \AstroSat Cadmium Zinc Telluride Imager (CZTI) measured the \tninty duration of 6.2$^{+0.9}_{-0.8}$ sec \citep{2024GCN.37298....1J}. These discrepancies highlight the limitations of using a fixed \tninty criteria and point to the importance of probabilistic classification techniques—such as Gaussian Mixture Models (GMM)—to more robustly distinguish GRB populations across instruments \citep{2016MNRAS.462.3243Z}.

\begin{figure}[ht]
    \centering
    \includegraphics[width = 0.47\textwidth]{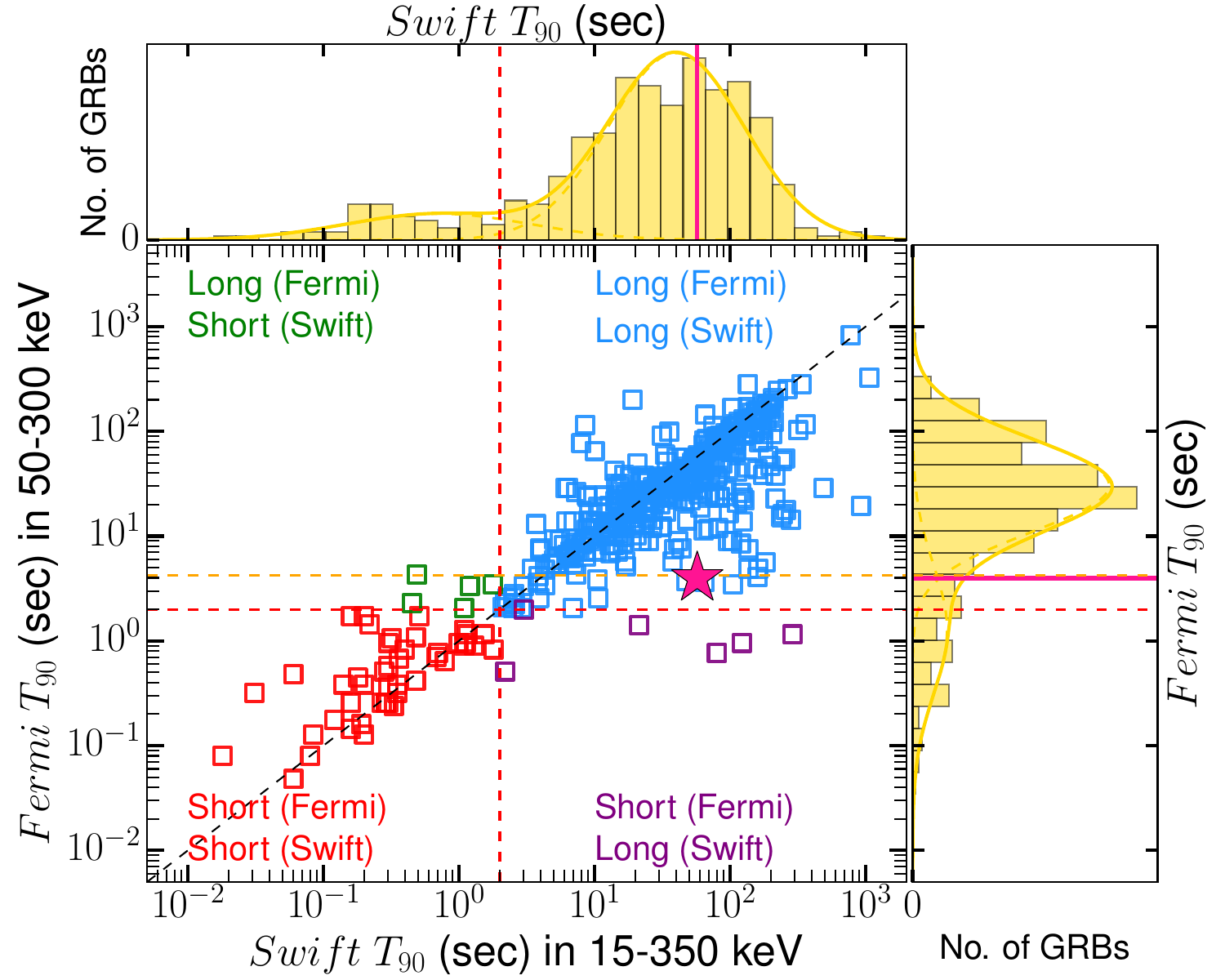}
    \caption{Scatter plot of \swift ~\tninty versus \fermi ~\tninty durations for GRBs detected by both instruments. GRBs are classified into four categories: pure long (both \tninty $> 2$ sec, blue squares), pure short (both \tninty $\leq 2$ sec, red squares), \fermi long–\swift short (\fermi ~\tninty $> 2$ sec, \swift ~\tninty $< 2$ sec, green squares), and \swift long–\fermi short (\swift ~\tninty $> 2$ sec, \fermi ~\tninty $< 2$ sec, purple squares). GRB 240825A is highlighted in pink star with error bars (\swift ~\tninty = 57.20 $\pm$ 8.57 sec, \fermi ~\tninty = 3.968 $\pm$ 0.091 sec). The dashed black line indicates y = x, red dashed lines mark \tninty = 2 sec, and the orange dashed line marks \tninty = 4.2 sec obtained from the fourth Fermi Catalog \citep{2020ApJ...893...46V}. Histograms along the top and right axes show the logarithmic distributions of \swift and \fermi ~\tninty values, respectively, with fitted bimodal Gaussian Mixture Models (solid gold line) and individual short (dashed gold, lower mean) and long (dashed gold, higher mean) components. Vertical/horizontal lines indicate GRB 240825A's \tninty values. The GMM fits reveal distinct short and long GRB populations, with probabilities of GRB 240825A being short calculated as 0.3 \% for \swift and 48.3 \% for \fermi.}
    \label{Fig4}
\end{figure}

We analyzed the \tninty durations\footnote{\url{https://user-web.icecube.wisc.edu/~grbweb_public/}} of GRBs jointly observed by BAT (15–350 keV) and GBM (50–300 keV) to investigate their classification into short and long categories (see Figure \ref{Fig4}). Using a GMM, we fitted bimodal log-normal distributions to the log$_{10}$(\tninty) values for both instruments. The matched GRBs were classified into four categories: Pure Long (both \tninty $>$ 2 sec), Pure Short (both \tninty $\leq$ 2 sec), \fermi Long–\swift Short (\fermi $> 2$ sec, \swift $\leq$ 2 sec), and \swift Long–\fermi Short (\swift $> 2$ sec, \fermi $\leq$ 2 sec), represented by blue, red, green, and purple markers, respectively. For GRB 240825A, with \swift ~\tninty = 57.20 $\pm$ 8.57 sec and \fermi ~\tninty = 3.968 $\pm$ 0.091 sec, we calculated the probability of being a short GRB \citep{2019MNRAS.485.5294P} using the GMM posterior probabilities, finding a low probability (0.3 \%) for \swift (consistent with a long GRB) and a high probability (48.3 \%) for \fermi due to its proximity to the classification boundary. This analysis highlights the differences in \tninty measurements between instruments, which arise not only from differences in energy sensitivity ranges (\swift: 15–350 keV; \fermi: 50–300 keV), but also from variations in effective area as a function of energy—both of which impact detection sensitivity.

\begin{figure*}
    \centering
    \includegraphics[width = 0.47\textwidth]{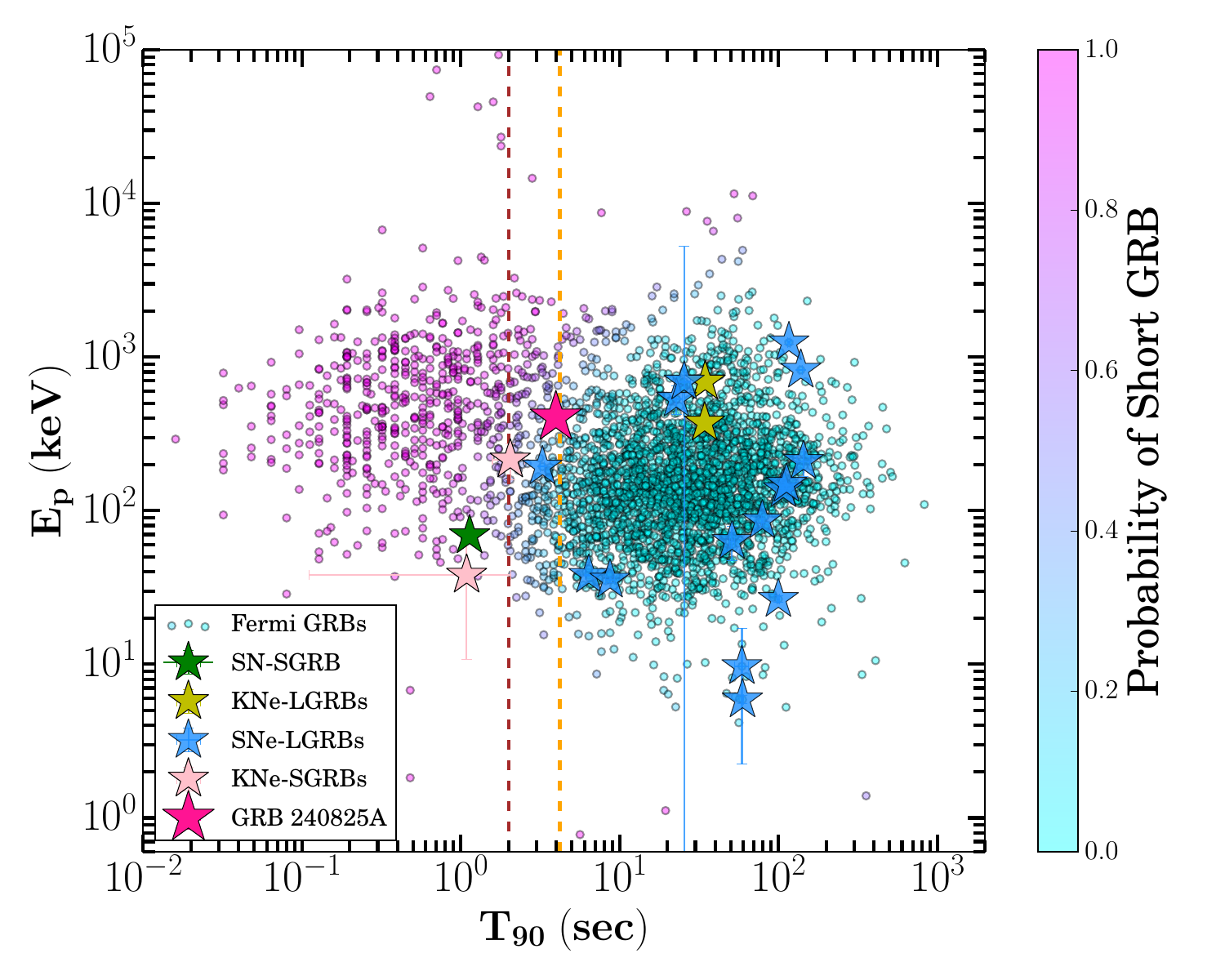}
    \includegraphics[width = 0.47\textwidth]{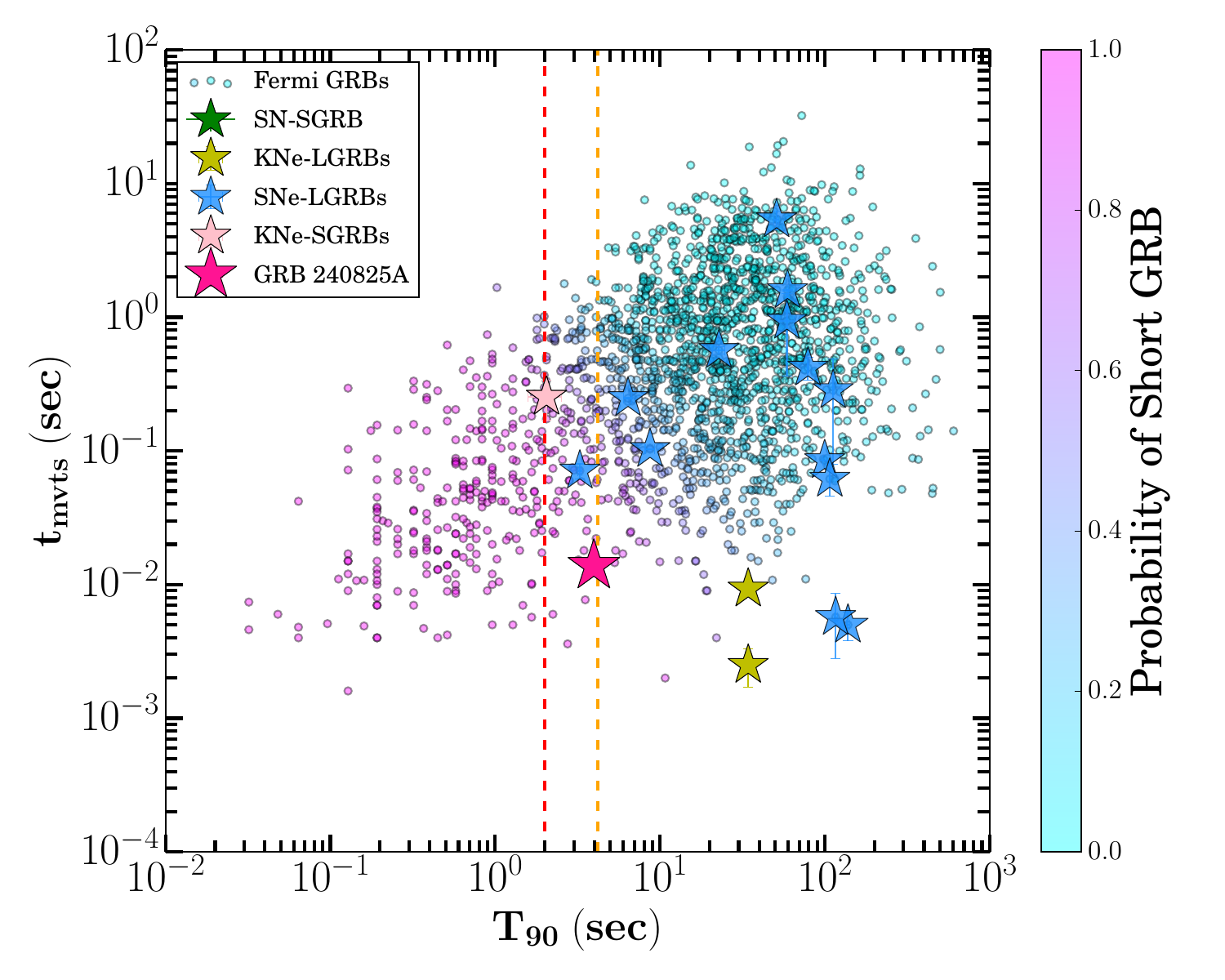}
    \includegraphics[width = 0.47\textwidth]{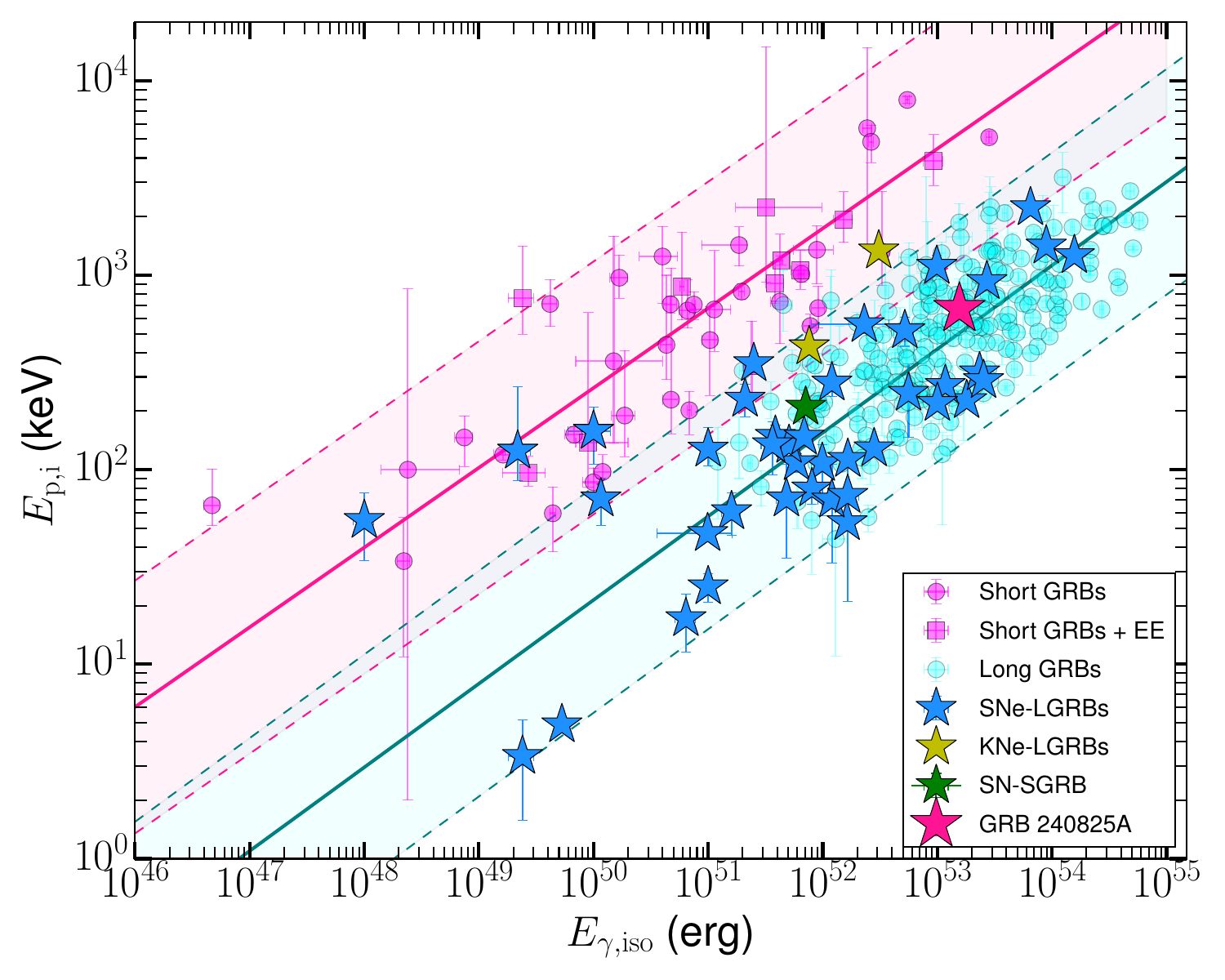}
        \includegraphics[width = 0.47\textwidth]{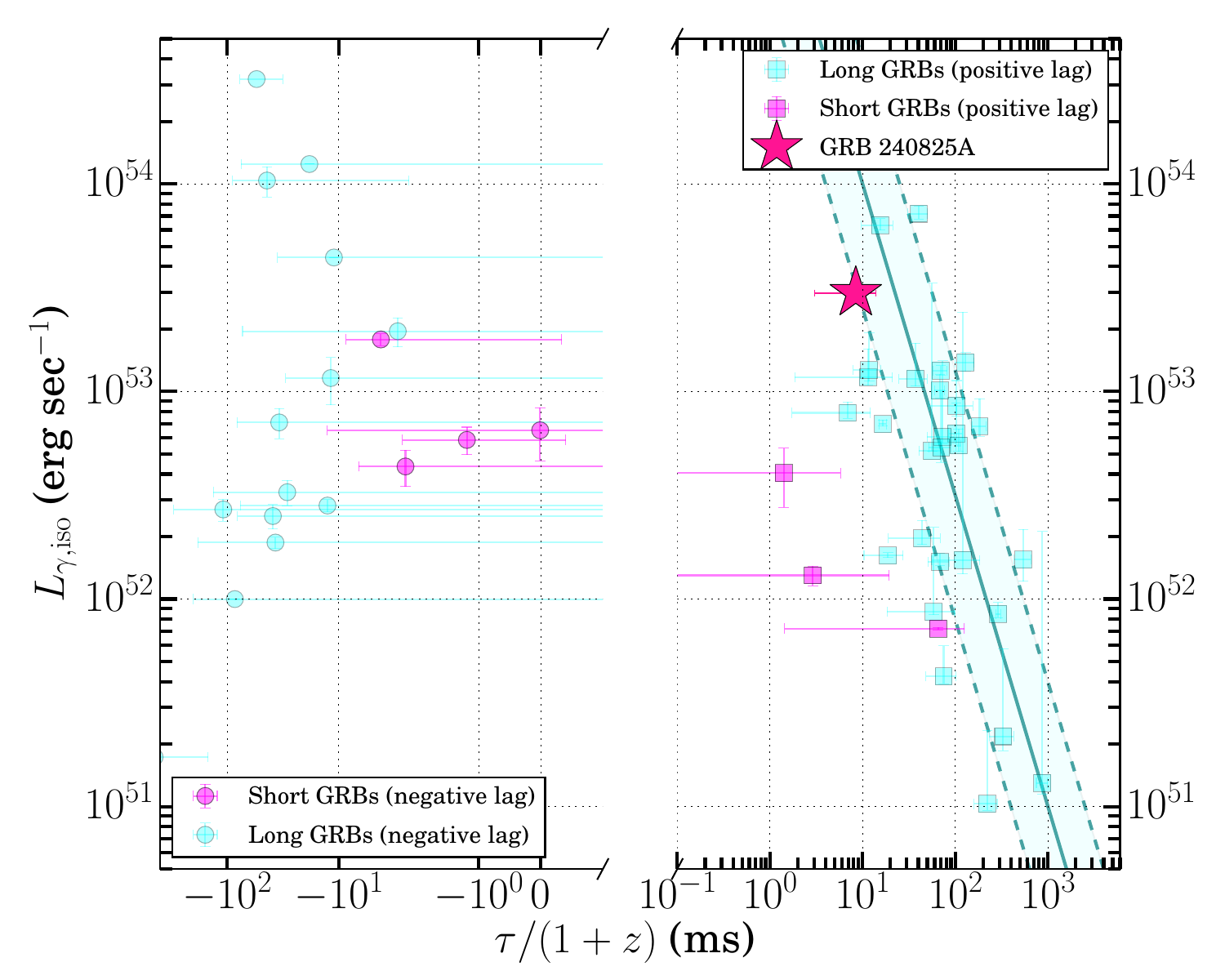}
    \caption{Prompt emission empirical correlations: Top-left: Scatter plot of peak energy (\Ep) versus \tninty duration for a sample of GRBs detected by \fermi-GBM. Top-right: Log-log scatter plot of minimum variability timescale (MVT, in seconds) versus \tninty duration (in seconds) for GRBs detected by \fermi-GBM. Red dashed lines mark \tninty = 2 sec, and the orange dashed line marks \tninty = 4.2 sec, respectively. Bottom-left: Log-log scatter plot of peak energy (\Ep) versus isotropic-equivalent energy ($E_{\gamma, iso}$) for GRBs obtained from \cite{2020MNRAS.492.1919M}. The dashed lines represent the Amati relation for collapsar (cyan) and merger (pink) GRBs \citep{2020MNRAS.492.1919M}. GRB 240825A, with an estimated rest-frame $E_p \approx $ 671.57 keV and $E_{\gamma, iso} \approx 1.6 \times 10^{53}$ erg, falls within the collapsar GRB cluster, indicating its classification as a collapsar GRB. Bottom-right: GRB 240825A in the lag-luminosity plane $\log(\tau)$ vs. $\log(L_\text{iso})$, compared to a sample of \swift-BAT GRBs, marginally following the LGRB lag (rest frame)-luminosity relation \citep{2010ApJ...711.1073U}.}
        \label{traditional_classification}
\end{figure*}

\subsection{Prompt emission empirical correlations: Classical classification of \thisgrb}

In this section, we classify GRB 240825A using standard prompt emission diagnostics, including the time-integrated peak energy–\tninty duration (\Ep-\tninty) correlation, the minimum variability timescale–\tninty duration (\mvts-\tninty) relation, the Amati relation, and the lag-luminosity correlation. 

\subsubsection{\texorpdfstring{\Ep--\tninty}{Ep--T90}}

The \Ep-\tninty (peak energy--duration) parameter space is a well-established diagnostic tool for distinguishing between merger and collapsar GRBs \citep{2000ApJS..126...19P}, particularly using \fermi-GBM observations \citep{2021ApJ...913...60P, 2024ApJ...972..166G}. To investigate the classification of GRB~240825A, we compiled a sample of GRBs from the \fermi-GBM catalog\footnote{\url{https://heasarc.gsfc.nasa.gov/W3Browse/fermi/fermigbrst.html}} and extracted their \tninty and \Ep (404.81$^{+9.02}_{-8.69}$ keV) values (peak energy from time-integrated spectral fits, Gupta et al. in prep). We applied a two-component GMM in the log \tninty--log \Ep space to statistically separate the population into merger and collapsar GRBs. This probabilistic approach suggests a refined classification beyond the conventional hard threshold at \tninty = 2 sec, providing a likelihood estimate for each burst. GRB~240825A is located near the boundary between the two populations, with a probability of 71.6 \% of belonging to the collapsar GRB class. In Figure~\ref{traditional_classification}, we show the position of GRB~240825A on the \Ep-\tninty diagram, along with other GRBs associated with SNe and KNe, for context and comparison.

\subsubsection{\texorpdfstring{\mvts-\tninty}{Mvts--T90}}

The minimum variability timescale (\mvts or MVT) of a GRB is the shortest timescale over which coherent temporal variations in the light curve can be observed. When the GRB light curve exhibits multiple pulses, the MVT typically corresponds to the timescale of the shortest significant pulse. It has been shown that MVT, in combination with the burst duration (\tninty), can effectively distinguish between collapsar and merger GRB populations \citep{2023A&A...671A.112C}. We compiled a sample of \fermi GRBs with \mvts measurements from \citet{Golkhou+15tvar} and additional literature sources \citep{2023ApJ...954L...5V}, and constructed the \mvts-\tninty distribution in log-log space. A two-component GMM was applied to this dataset to statistically separate the GRB population. GRB~240825A was found to have an MVT of $13.8 \pm 1.6$~ms utilizing GBM observations, as derived via the wavelet decomposition technique of \citet{Golkhou+15tvar}. Its location on the \mvts-\tninty plane (see Figure~\ref{traditional_classification}, GRB~240825A along with other KN/SN connected GRBs) places it near the boundary of the merger and collapsar GRB populations, with a calculated merger GRB probability of 95.3\%. 





\subsubsection{Amati relation}

The correlation between the intrinsic peak energy ($E_{\rm p,i}$) and the isotropic-equivalent energy ($E_{\gamma,\mathrm{iso}}$), known as the Amati relation, has long served as an empirical discriminator between collapsar and merger GRBs \citep{2002A&A...390...81A, 2013MNRAS.430..163Q}. Collapsar GRBs tend to follow a tight correlation in the $E_{\rm p,i}$–$E_{\gamma,\mathrm{iso}}$ plane, while merger GRBs, including those with EE, typically fall above this track. To assess the nature of GRB~240825A, we compiled a sample of GRBs from the literature \citep{2020MNRAS.492.1919M}, including merger GRBs with and without EE, collapsar GRBs, and those with confirmed SN or KN associations \citep{2022Natur.612..232Y, 2022ApJ...931L...2W, 2025NSRev..12E.401S}. We then plotted these events in the $E_{\rm p,i}$–$E_{\gamma,\mathrm{iso}}$ plane and overlaid the best-fit Amati relations for collapsar and merger GRBs based on previous studies \citep{2020MNRAS.492.1919M}. GRB~240825A ($E_{\gamma,\mathrm{iso}}$ = 1.56 $\times$ 10 $^{53}$ erg), shown in Figure~\ref{traditional_classification}, lies within the $2\sigma$ scatter of the collapsar GRB correlation, overlapping with the region occupied by SN-associated long GRBs (SNe-LGRBs). This placement suggests that GRB~240825A is more consistent with the collapsar GRB population in terms of energetics.

\subsubsection{Lag-luminosity relation}

The lag-luminosity relation, which connects the spectral lag ($\tau$) and the isotropic peak luminosity ($L_{\gamma,\mathrm{iso}}$), has been widely employed as a diagnostic to distinguish between merger and collapsar GRBs \citep{2000ApJ...534..248N}. Collapsar GRBs typically exhibit positive spectral lags (where higher-energy photons arrive earlier than lower-energy ones) and follow a strong anti-correlation between lag and luminosity \citep{2002ApJ...579..386N}. In contrast, merger GRBs often show negligible or even negative lags, deviating from the collapsar GRB track in this parameter space \citep{2010ApJ...711.1073U, 2015MNRAS.446.1129B}. We collected a sample of GRBs with measured lags and luminosities, including both positive and negative lag populations for merger and collapsar GRBs \citep{2010ApJ...711.1073U, 2015MNRAS.446.1129B, 2023MNRAS.519.3201C}. 

We calculated the spectral lag for GRB 240825A using the cross-correlation function (CCF) between the 15--25 keV and 50--100 keV light curves from the \swift-BAT, focusing on the time interval from 0.67 to 7.58 seconds post-burst. Employing a robust CCF computation with \texttt{numpy.correlate} and Monte Carlo error estimation, we fitted the CCF peak with an asymmetric Gaussian model via Markov Chain Monte Carlo (MCMC) sampling, yielding a spectral lag of $\sim$  $0.014 \pm 0.009$ seconds (68\% confidence), indicating that the softer 15--25 keV band lags the harder 50--100 keV band. To contextualize GRB 240825A within the lag-luminosity relation, we estimated its isotropic peak luminosity ($L_\text{iso}$) utilizing a peak flux derived from GBM data (Gupta et al. in prep). As shown in Figure~\ref{traditional_classification}, GRB~240825A (pink star) satisfies the collapsar GRB lag–luminosity relation, with a rest-frame lag of $\tau/(1+z) \approx 8.44$~ms and $L_{\gamma,\mathrm{iso}} \approx 2.98 \times 10^{53}$~erg~sec$^{-1}$.

\subsection{Fermi and ASIM timing analysis of GRB 240825A}

We searched for quasi-periodic oscillations (QPOs, \citealt{2025MNRAS.538L.100C}) in the prompt light curves using Power Density Spectra (PDS) from Atmosphere-Space Interactions Monitor-High Energy Detector (ASIM-HED, \citealt{2019arXiv190612178N}) and \fermi-GBM (see Figure \ref{timing1} and Table \ref{Joint_timinganalysis_Table} in the Appendix). We noticed low-significance peaks at around 6\,Hz and 20\,Hz, with single-trial significance of $0.3, 1.3\,{\sigma}$ for ASIM-HED and with an F-test probability of 0.56, 0.60 for \fermi-GBM. Given these low significance levels, we consider the QPO evidence inconclusive and do not further use it in the progenitor classification. 


\begin{figure*}[ht]
    \centering
    \includegraphics[width = 0.8\textwidth]{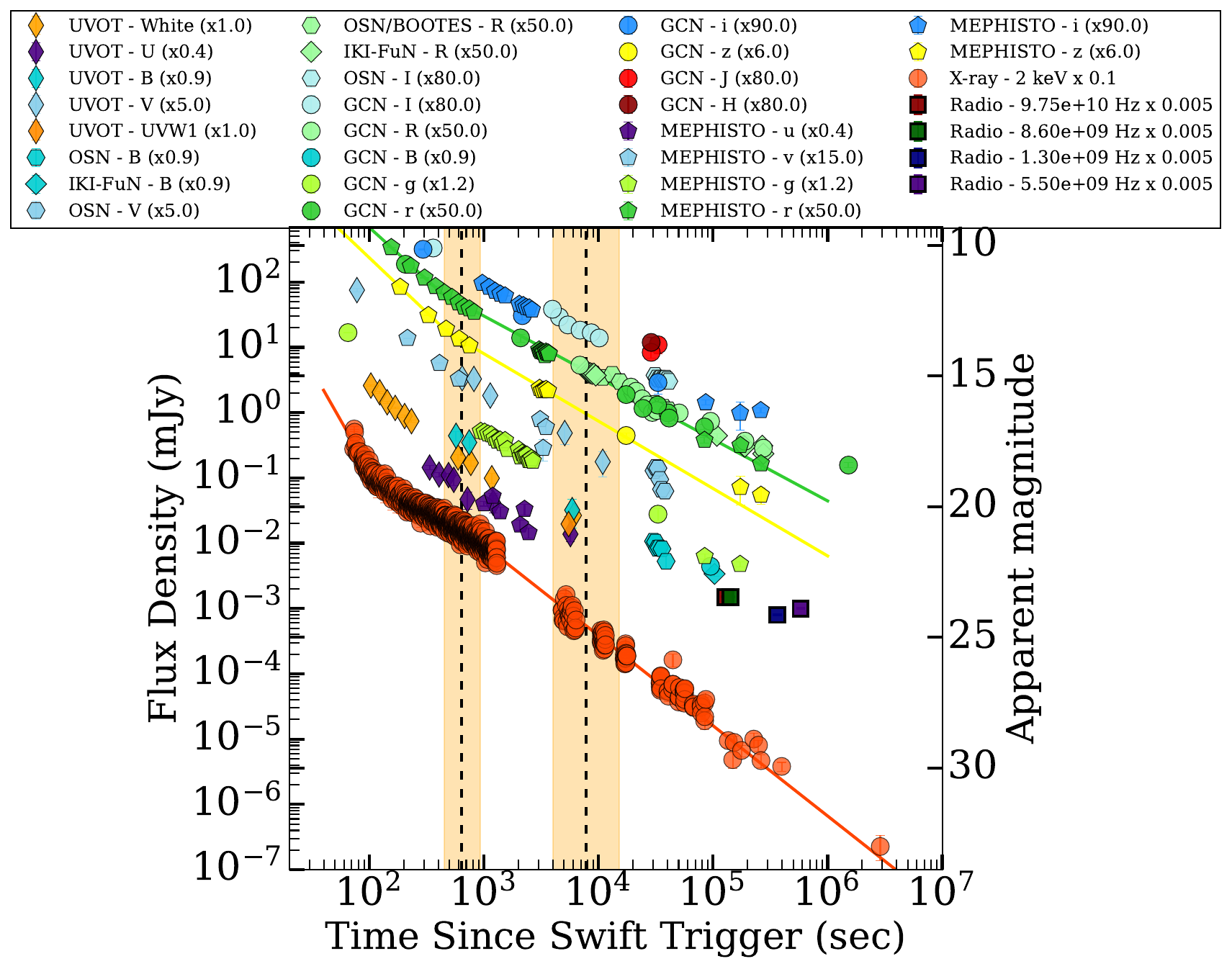}
    \caption{Multi-wavelength light curves of the GRB 240825A afterglow. Solid lines in different colors represent best-fit power-law or broken power-law models applied to the observed data points. Optical data have been corrected for Galactic extinction. Vertical orange shaded bands mark the time intervals used for constructing the spectral energy distributions, with vertical black dashed lines indicating the central times of the SEDs.}
    \label{afterglowlc}
\end{figure*}

\subsection{Afterglow temporal and spectral evolution}

\begin{table*}
\centering
\caption{Summary of the temporal fit parameters for the optical and X-ray afterglow light curves of GRB 240825A.}
\label{tab:temporal_fit}
\begin{tabular}{l l c c}
\hline
\hline
Band & Model & Temporal index & Break time (sec) \\
\hline
Optical & Broken power law 
& $\alpha_{O1} = -1.60 \pm 0.01$ 
& $t_{\rm opt} = 425.95 \pm 6.70$ \\

Optical & 
& $\alpha_{O2} = -0.94 \pm 0.01$ 
& -- \\
\hline
X-ray, 2 keV & Double broken power law 
& $\alpha_{X1} = -3.11 \pm 0.25$ 
& $t_{X1} = 115.94 \pm 2.42$ \\

X-ray, 2 keV & 
& $\alpha_{X2} = -0.97 \pm 0.02$ 
& -- \\

X-ray, 2 keV & 
& $\alpha_{X3} = -1.38 \pm 0.02$ 
& $t_{X2} = 836.65 \pm 40.10$ \\
\hline
\end{tabular}
\end{table*}

We conducted the analysis of the optical r-band light curve of GRB 240825A using data from \cite{2025ApJ...979...38C, 2024GCN.37287....1I, 2024GCN.37373....1S, 2024GCN.37293....1M, 2024GCN.37295....1B, 2024GCN.37300....1L, 2024GCN.37306....1W, 2024GCN.37335....1G}, focusing on the first $10^5$ seconds to avoid contamination from the host galaxy. The optical light curve was fitted using a broken power-law function (see Table \ref{tab:temporal_fit}). The observed transition from a steeper decay ($\alpha_{O1} = -1.60$) to a shallower decay ($\alpha_{O2} = -0.94$) is consistent with a transition from the reverse shock-dominated phase to the forward shock-dominated phase (e.g., \citealt{1999ApJ...520..641S}). Given the prompt duration $T_{90} = 3.96$ seconds (\fermi-GBM), the break time at $\sim426$ seconds satisfies the condition $t_{\text{dec}} \gg T_{90}$, indicating the thin-shell case \citep{2025arXiv250702806W}. In this regime, the reverse shock is Newtonian and the temporal decay of optical emission (in the slow-cooling regime and for $\nu > \nu_m^{\rm RS}$) follows $\alpha = -\frac{27p + 7}{35}$ \citep{2005ApJ...628..315Z, 2026arXiv260628267G}. Solving for $p$, we find $p \approx 1.81$, a physically reasonable value for GRB ejecta and consistent with other GRB reverse shock studies \citep{2021MNRAS.505.4086G}. On the other hand, the shallow post-break decay index ($\alpha_{O2} = -0.94$) is consistent with the expected range for forward shock emission in a constant-density interstellar medium, where the temporal decay index is related to the electron energy distribution index, $p$, via $\alpha = (3p-3)/4$ for a slow-cooling regime, i.e, $\nu_{\rm opt}$ $<$ $\nu_{\rm c}$ $<$ $\nu_{\rm x-ray}$ or $\nu_{\rm opt}$ $<$ $\nu_{\rm x-ray}$ $<$ $\nu_{\rm c}$ \citep{1998ApJ...497L..17S}. Assuming $\alpha_{O2} = -0.94$, this implies $p \approx 2.25$ \citep{2025arXiv250702806W}.

The X-ray afterglow light curve at 2 keV was best described by a double broken power-law model (see Table \ref{tab:temporal_fit}). We also conducted a spectral analysis of the XRT data during the early steep decay phase (t $<$ 118 seconds). The spectrum was modeled with a power-law function, yielding an X-ray spectral index of $\beta = -1.28^{+0.10}_{-0.09}$. This steep decay phase is consistent with high-latitude emission, commonly interpreted as the tail of the prompt GRB emission, where photons from off-axis regions arrive at the observer with a delay due to geometric effects (e.g., \citealt{2000ApJ...541L..51K}). For non-thermal emission described by a single power-law, the temporal decay index is theoretically related to the spectral index by the relation $\alpha = -2 + \beta$. Given the measured spectral index $\beta = -1.28^{+0.10}_{-0.09}$, the expected temporal decay index is approximately -3.28$^{+0.10}_{-0.09}$, which is in agreement with the observed value of $\alpha_{X1} = -3.11 \pm 0.25$. 

\begin{figure}
    \centering
    \includegraphics[width = 0.45\textwidth]{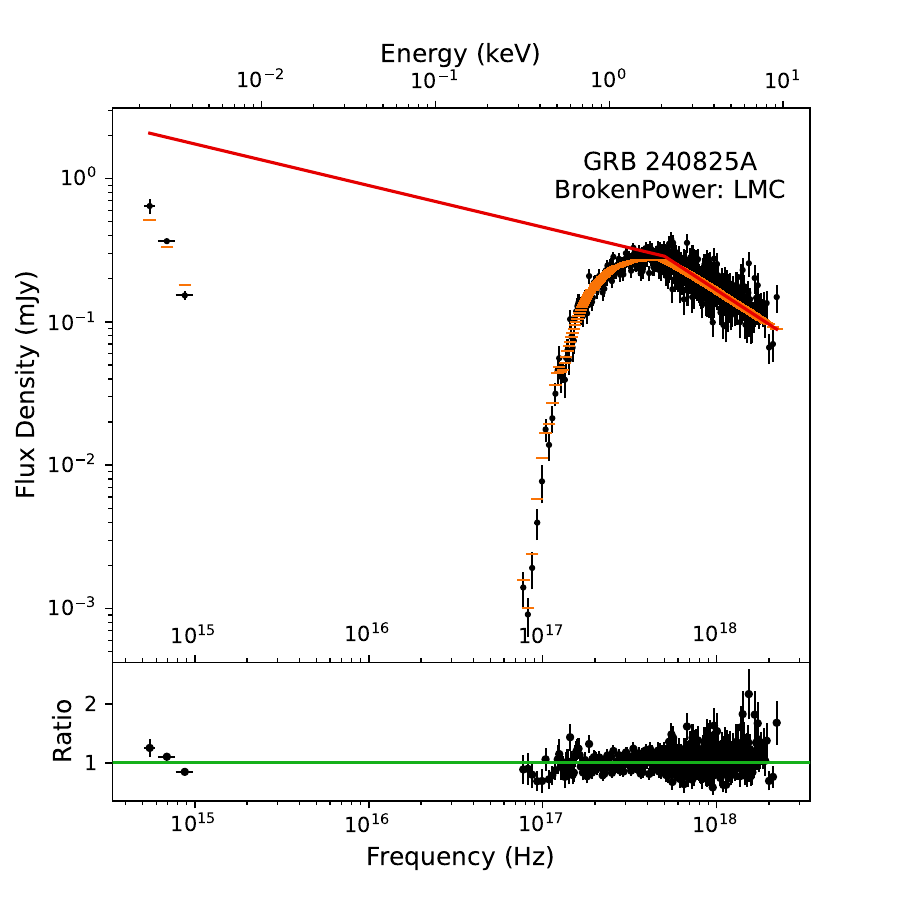}
    \includegraphics[width = 0.45\textwidth]{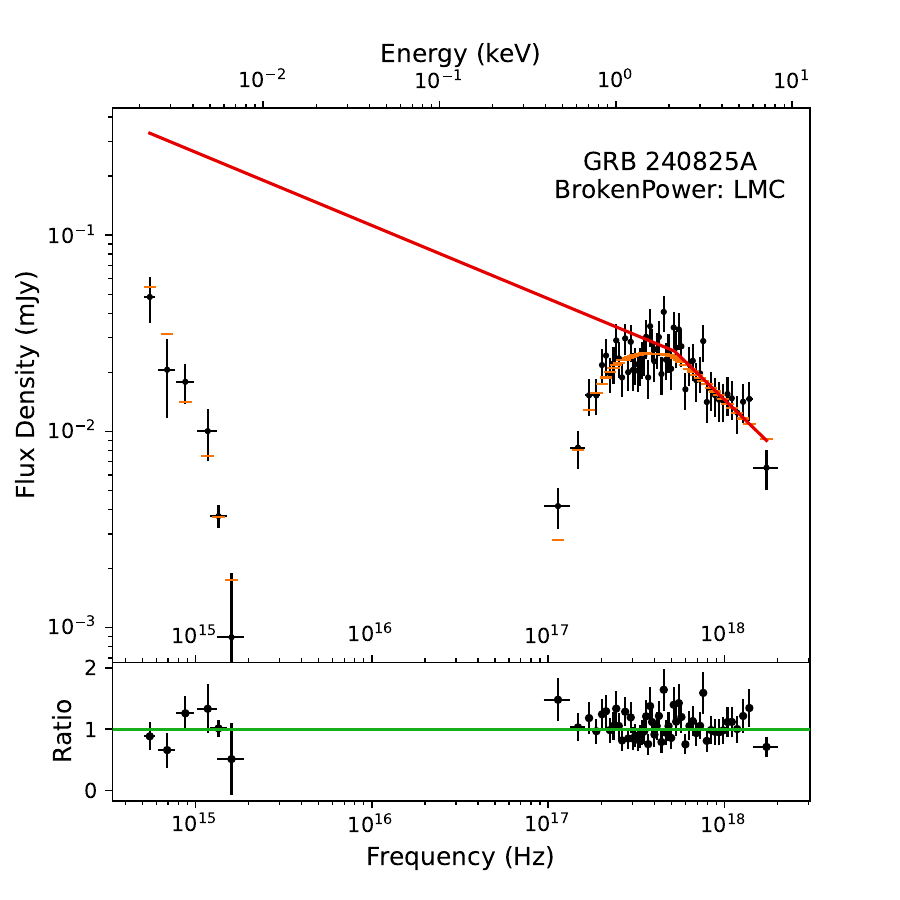}
    \caption{Combined X-ray and UV/optical SEDs for GRB 240825A. Top panel: SED 1, centered at \swiftT+640 s, constructed from UVOT data in the $u$, $b$, and $v$ filters observations and contemporaneous XRT data. Bottom panel: SED 2, centered at \swiftT+ 7800\,s, constructed using UVOT data in the $u$, $b$, $v$, $uvw1$, $uvw2$, and $uvm2$ filters observations and XRT data. Both SEDs were optimally fitted with a broken power-law model, with the Large Magellanic Cloud extinction law providing the best fit. The host galaxy absorbing column density ($N_{\rm H, z}$) was fixed at 8.68 $\times 10^{22}$ cm$^{-2}$ in both SEDs. Observed and model data points are shown in black and orange, respectively. The solid red line represents the best-fit model corrected for absorption and extinction. The lower sub-panel in each figure displays the data-to-model ratio, demonstrating the quality of the fit across the energy range. The broken power law provides a phenomenological description of the broadband curvature; its fitted break energy should not necessarily be interpreted as a sharply defined synchrotron cooling frequency.}
    \label{SED}
\end{figure}

\begin{table*}
\centering
\caption{Results of simultaneous UV/optical and X-ray spectral fits for GRB 240825A using MW, LMC, and SMC extinction laws with single power-law (POW) or broken power-law (BKP) continuum models. For the BKP, the high-energy spectral index was fixed according to $\beta_2 = \beta + 0.5$. Columns list the low-energy spectral index ($\beta$), color excess ($E(B-V)$), break energy ($E_{bk}$) for broken power-law models, fit statistic $\chi^2$, degrees of freedom (dof), and null hypothesis probability. In every fit, host galaxy absorbing column density ($N_{\rm H, z}$) was fixed at 8.68 $\times 10^{22}$ cm$^{-2}$, as determined from the XRT-only spectrum at the epoch corresponding to SED 1, approximately \swiftT+640\,sec. Thus, the fits assume a time-independent host-galaxy absorbing column density.} 
\label{tab:sedfits}
\begin{tabular}{@{}cccccccc}
\hline
SED & Model & Dust & $\beta$ & $E(B-V)$ (mag) & $E_{bk}$ (keV) & $\chi^2$ (dof) & Null Hypothesis Probability \\
\hline
\multicolumn{8}{c}{SED 1 (\swiftT+640\,sec)} \\
\hline
640\,s & POW & MW  & $1.52 \pm 0.01$ &  $0.46 \pm 0.01$ & -- & 642.23 (459) & 3.12e-08 \\
640\,s & POW & LMC & $1.52 \pm 0.01$ &  $0.45 \pm 0.01$ & -- & 636.96 (459) & 6.75e-08 \\
640\,s & POW & SMC & $1.52 \pm 0.01$ &  $0.47 \pm 0.01$ & -- & 637.65 (459) & 6.12e-08 \\
640\,s & BKP & MW  & $1.29 \pm 0.02$ &  $0.26 \pm 0.02$ & $2.06 \pm 0.10$ & 479.35 (458) & 2.37e-01 \\
640\,s & BKP & LMC & $1.29 \pm 0.02$ &  $0.27 \pm 0.02$ & $2.07 \pm 0.10$ & 482.00 (458) & 2.11e-01 \\
640\,s & BKP & SMC & $1.29 \pm 0.02$ &  $0.26 \pm 0.02$ & $2.06 \pm 0.10$ & 480.31 (458) & 2.27e-01 \\
\hline
\multicolumn{8}{c}{SED 2 (\swiftT+7800\,sec)} \\
\hline
7800\,s & POW & MW  & $1.60 \pm 0.04$ &  $0.62 \pm 0.04$ & -- & 59.32 (56) & 3.56e-01 \\
7800\,s & POW & LMC & $1.49 \pm 0.03$ &  $0.40 \pm 0.03$ & -- & 70.21 (56) & 9.59e-02 \\
7800\,s & POW & SMC & $1.41 \pm 0.03$ &  $0.31 \pm 0.03$ & -- & 90.19 (56) & 2.56e-03 \\
7800\,s & BKP & MW  & $1.50 \pm 0.07$ &  $0.56 \pm 0.06$ & $2.79 \pm 0.62$ & 50.48 (55) & 6.48e-01 \\
7800\,s & BKP & LMC & $1.37 \pm 0.04$ & $0.35 \pm 0.03$ & $2.15 \pm 0.28$ & 44.82 (55) & 8.35e-01 \\
7800\,s & BKP & SMC & $1.28 \pm 0.04$ &  $0.24 \pm 0.02$ & $1.81 \pm 0.30$ & 48.12 (55) & 7.33e-01 \\
\hline
\end{tabular}
\end{table*}

We further generated two joint spectral energy distributions (SEDs) using contemporaneous UVOT and XRT data, centered at \swiftT+640\,s (SED 1) and \swiftT+7800\,s (SED 2). We followed the methodology described by \citet{schady07, sch10} to construct the SEDs, and both SEDs were analyzed using \textsc{XSPEC} (version 12.12.0), following the procedures in \citet{schady07,sch10}. We tested two continuum models: a simple power-law (POW) and a broken power-law (BKP). For the BKP model, the high-energy spectral index was fixed according to $\beta_2=\beta+0.5$, as expected for a synchrotron cooling transition. Each model incorporated two dust and gas components to account for Galactic and host galaxy extinction and photoelectric absorption, using the \textit{phabs}, \textit{zphabs}, and \textit{zdust} models in \textsc{XSPEC}. The Galactic components were fixed to the reddening values from \citet{schlegel} and column density from \citet{kalberla}. For the host galaxy, we tested three extinction laws: Small Magellanic Cloud (SMC), Large Magellanic Cloud (LMC), and Milky Way (MW) extinction laws. To reduce the degeneracy between intrinsic low-energy spectral curvature and photoelectric absorption, we did not allow the host galaxy absorbing column to vary independently between the two epochs. We first determined the host galaxy column density from the XRT-only spectrum at the epoch corresponding to SED~1, approximately $T_0+640$~s. The XRT-only fit gave N$_{\rm H,z}=8.68 \times10^{22}\ {\rm cm}^{-2}$. We then fixed $N_{\rm H,z}$ at this value in all broadband fits to both SEDs and for all continuum and extinction models. Our analysis therefore assumes a time-independent line-of-sight absorbing column. 

The fits were evaluated using the reduced chi-squared relative to degrees of freedom ($\chi^2$/dof), and the null hypothesis probability. For both SEDs, BKP provided the best fit (see Table \ref{tab:sedfits}). For SED~1, the single power-law fits are statistically unacceptable for all three extinction laws, with null-hypothesis probabilities of order $10^{-8}$. In contrast, the broken power-law fits are acceptable, with null-hypothesis probabilities ranging from 0.21 to 0.24. The improvement is substantial for all three extinction laws. For example, for the LMC extinction law, the fit statistic improves from $\chi^2/{\rm dof}=636.96/459$ for the single power law to $\chi^2/{\rm dof}=482.00/458$ for the BKP. At the second epoch, the BKP also gives a lower $\chi^2$ than the corresponding single power-law model for all three extinction laws. 
The best-fitting model among those tested for SED~2 (higher number of UVOT filters) is the BKP with LMC extinction, for which $\chi^2/{\rm dof}=44.82/55$ and the null-hypothesis probability is 0.835. The corresponding single power law LMC fit gives $\chi^2/{\rm dof}=70.21/56$. Adopting the LMC extinction law consistently at both epochs, the BKP parameters are $\beta_1=1.29\pm0.02,\qquad E(B-V)=0.27\pm0.02\ {\rm mag},\qquad E_{\rm bk}=2.07\pm0.10\ {\rm keV}$ for SED~1, and $\beta_1=1.37\pm0.04,\qquad E(B-V)=0.35\pm0.03\ {\rm mag},\qquad E_{\rm bk}=2.15\pm0.28\ {\rm keV}$ for SED~2. 

We emphasize that the BKP model is a phenomenological description of the broadband curvature. A physical synchrotron spectrum is expected to exhibit a smooth transition around the cooling frequency rather than an infinitesimally sharp break, and this curvature may extend over a broad frequency interval \citep{2002ApJ...568..820G, 2014ApJ...780...82U, 2009MNRAS.394.2164V}. Consequently, the fitted value of $E_{\rm bk}$ can depend on the assumed continuum and extinction law and should not necessarily be identified with a sharply defined synchrotron cooling frequency. This qualification is also supported by the temporal behavior. At approximately $T_0+640$~s, the post-break optical and contemporaneous X-ray temporal indices are similar, $\alpha_{\rm O,2}=-0.94\pm0.01$ and $\alpha_{\rm X,2}=-0.97\pm0.02$. A cooling break cleanly
separating the optical and X-ray bands would generally predict different temporal indices in the two bands. Furthermore, if the fitted \(E_{\rm bk}\sim2\) keV were identified directly with the synchrotron cooling frequency, the expected temporal evolution would not be reproduced. In a constant-density medium, \(\nu_c\propto t^{-1/2}\), so the break should move to lower energy by a factor of \(\sim(7800/640)^{-1/2}\approx0.29\) between the two SED epochs. In a wind-like medium, the expected evolution is in the opposite direction, \(\nu_c\propto t^{1/2}\) and an increase by
a comparable factor would be expected. Instead, the fitted LMC break energies remain approximately constant within the uncertainties: $E_{\rm bk}=2.07\pm0.10$~keV and $E_{\rm bk}=2.15\pm0.28$~keV. We therefore interpret $E_{\rm bk}$ as an empirical measure of the broadband curvature sampled primarily by the XRT band, rather than as a robust direct measurement of $\nu_c$. Under this interpretation, the cooling frequency is likely close to or above the X-ray band, but the present data do not allow us to determine its exact location robustly.


The broadband afterglow is broadly consistent with synchrotron emission, but no single standard closure relation simultaneously explains all temporal and spectral indices. This tension may indicate a complex afterglow evolution, potentially driven by a two-component jet model (e.g., a narrow core and wider cocoon), a wavelength-dependent model, a transition in the external medium density profile (e.g., from wind to ISM), evolving microphysical parameters, or a structured outflow. Alternatively, the circumburst environment may follow a non-trivial density profile, parameterized as $\rho \propto r^{-k}$ with $0 \leq k \leq 2$, which has been shown to affect the temporal and spectral indices \citep{2013NewAR..57..141G}. Further multiwavelength observations, including radio and high-energy data, are essential to refine these models and elucidate the physical processes governing the afterglow of GRB 240825A.

\subsection{Host galaxy properties}

Host galaxy properties provide crucial context for distinguishing between collapsar and merger GRBs, offering insights into their progenitor systems \citep{Fruchter2006, Savaglio2009, Berger2014, Fong2013}. In our analysis, we utilize the host galaxy properties to probe the likely progenitor and classify the physical origin of GRB 240825A.

\begin{figure}[ht]
    \centering
    \includegraphics[width = 0.47\textwidth]{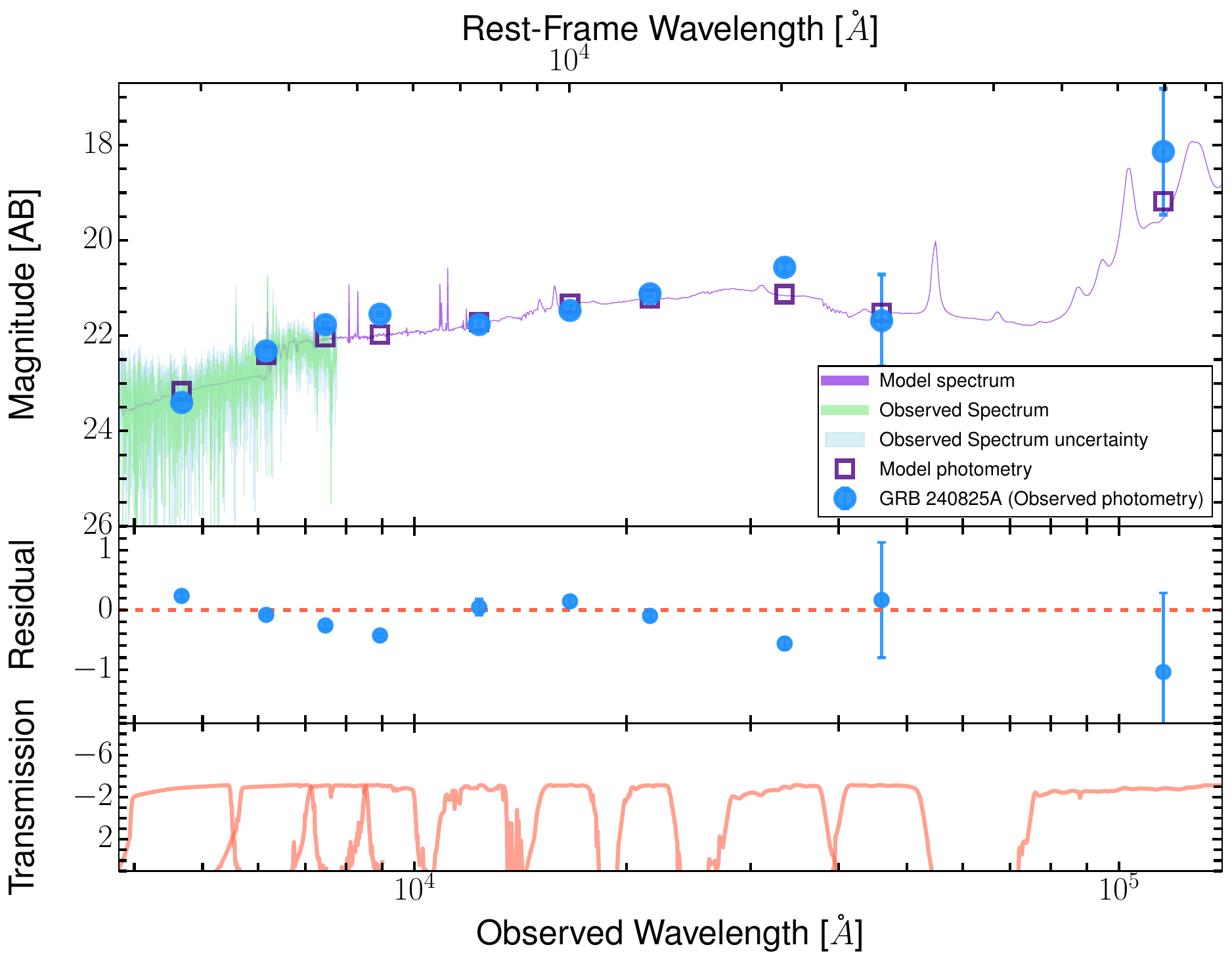}
    \caption{Best-fit spectral energy distribution model for the host galaxy of GRB 240825A using \textit{Prospector}. The sky blue points represent observed photometry from 10.4\,m GTC and DESI Legacy Surveys in the $g$, $r$, $i$, $z$, $J$, $H$, $K$, $W1$, $W2$, and $W3$ bands. Residuals between the observed and modeled fluxes are shown in the middle panel, with a dashed line marking the residuals equal to zero. The transmission of each filter is shown in the lower panel.}
    \label{host_sed}
\end{figure}

\subsubsection{Host SED analysis} 

We conducted a comprehensive SED analysis of the host galaxy of GRB 240825A using the \textit{Prospector} (see Figure \ref{host_sed}). The SED was constructed using the GTC spectrum and broadband photometry spanning optical to mid-infrared wavelengths, including deep imaging from the 10.4\,m GTC in the $g$, $r$, $i$, $z$, $J$, $H$, and $K$ bands, and archival data from the DESI Legacy Surveys in the $W1$, $W2$, and $W3$ bands. Data were corrected for galactic extinction using the \cite{2011ApJ...737..103S} dust maps, while dust attenuation was fixed to the afterglow-derived line-of-sight value within \textit{Prospector}. We employed the \textit{parametric\_sfh} model\footnote{\url{https://prospect.readthedocs.io/en/latest/sfhs.html}} in \textit{Prospector}, assuming a delayed-$\tau$ star formation history (SFR$(t) \propto (t / \tau^2) e^{-t / \tau}$), and sampled the parameter space using Markov Chain Monte Carlo (MCMC) methods. Key parameters included stellar mass ($M_\star$), metallicity ($Z$), galaxy age ($t_{\mathrm{age}}$), and the e-folding time ($\tau$). The corner plot (see Figure \ref{host_sed_corner} of the appendix) displays posterior distributions for parameters. The best-fit model yields a stellar mass of $\log (M_\star / M_\odot) \approx 10.01$ and a star formation timescale of $\tau \approx 0.62$ Gyr. The star formation rate (SFR), derived from the model, is approximately 3.18 M$_{\odot} \mathrm{yr}^{-1}$, consistent with a star formation history for a massive galaxy. 

\begin{figure}
    \centering
    \includegraphics[width = 0.45\textwidth]{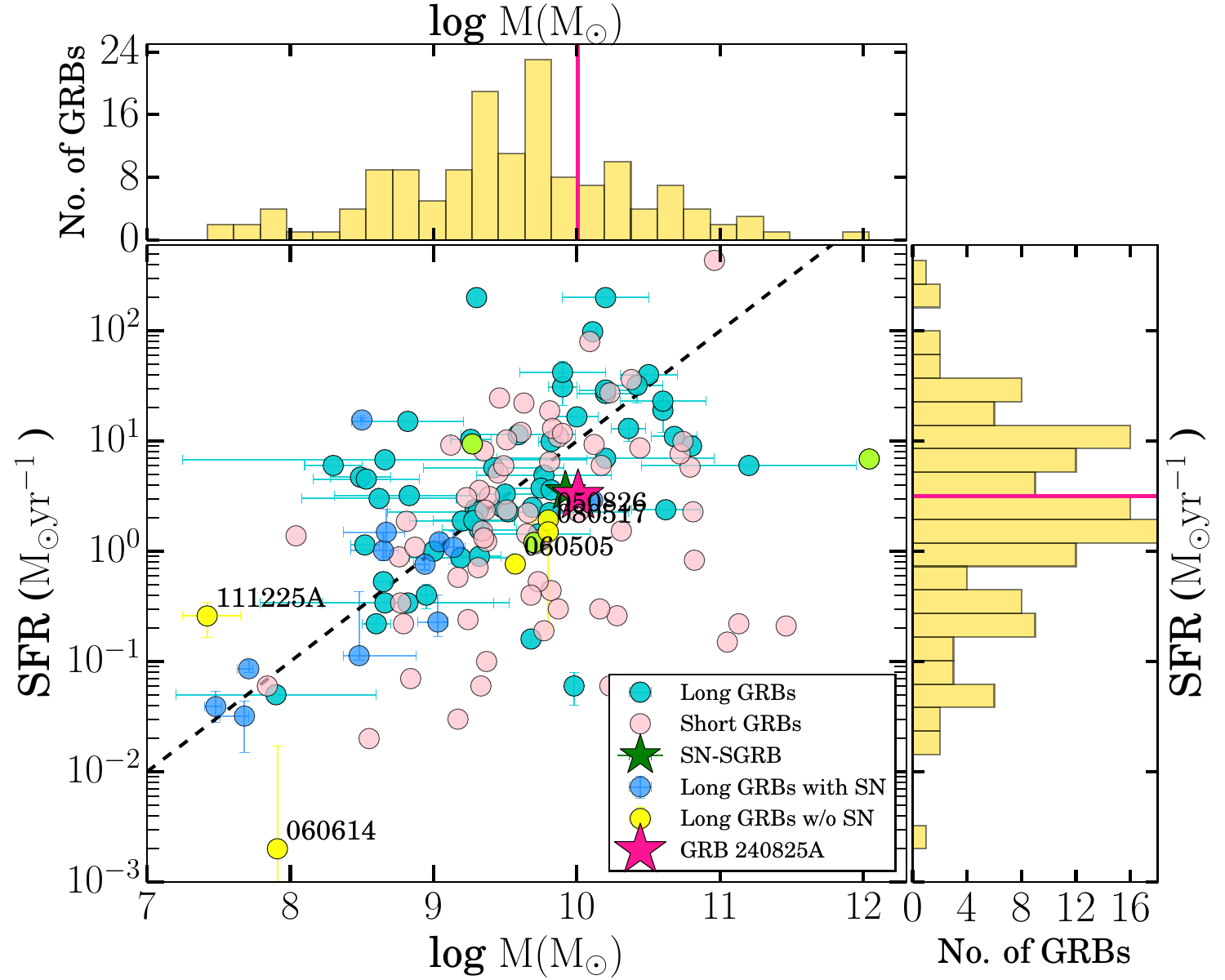}
        \includegraphics[width = 0.45\textwidth]{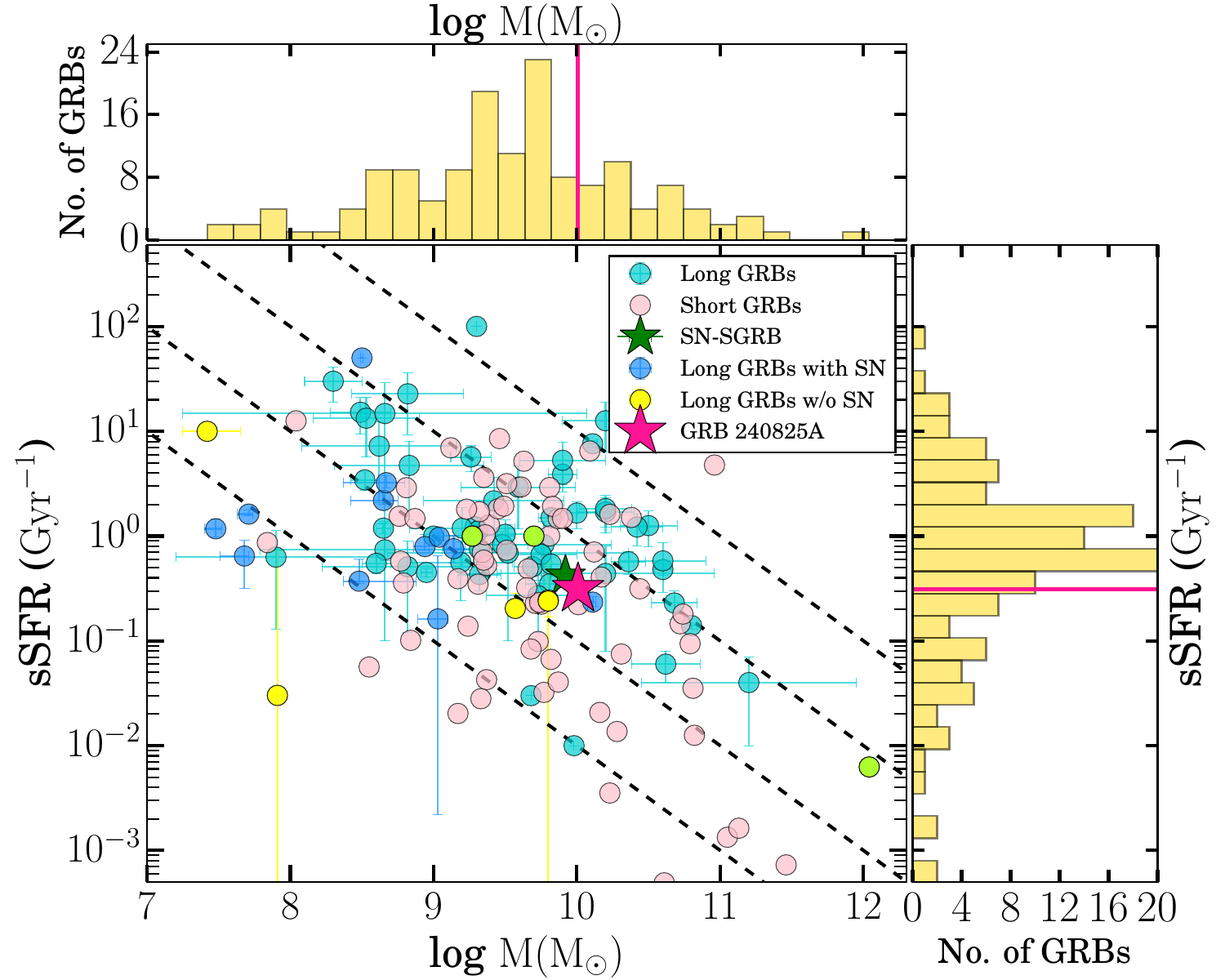}
    \caption{This figure illustrates the host galaxy properties of GRB 240825A compared to collapsar (depicted with cyan circles) and merger (depicted with pink circles) GRBs taken from \cite{2006AIPC..836..540S, Savaglio2009, 2021MNRAS.503.3931T, 2022ApJ...940...57N, 2022JApA...43...82G}. Top Panel: Stellar mass (log $\mathrm{M}/M_{\odot}$) versus star formation rate ($\mathrm{SFR}/M_{\odot} \mathrm{yr}^{-1}$). Histograms of stellar mass and SFR are shown along the top X-axis and right Y-axis, respectively. A dashed black line denotes a specific star formation rate of 1 Gyr$^{-1}$, representing the main sequence of star-forming galaxies. Bottom Panel: Stellar mass versus specific star formation rate ($  \mathrm{sSFR}/\mathrm{yr}^{-1}$), with histograms of stellar mass and sSFR along the top X-axis and right Y-axis, respectively. In both panels, GRB 240825A's host galaxy is highlighted with a pink star, and its position in the histograms is marked by vertical pink lines. Dashed black lines denote star formation rates (sSFR) of 0.1, 1, 10, 100 M$_{\odot} \mathrm{yr}^{-1}$.}
    \label{host_comparison}
\end{figure}

\begin{figure*}
    \centering
    \includegraphics[width = 0.9\textwidth]{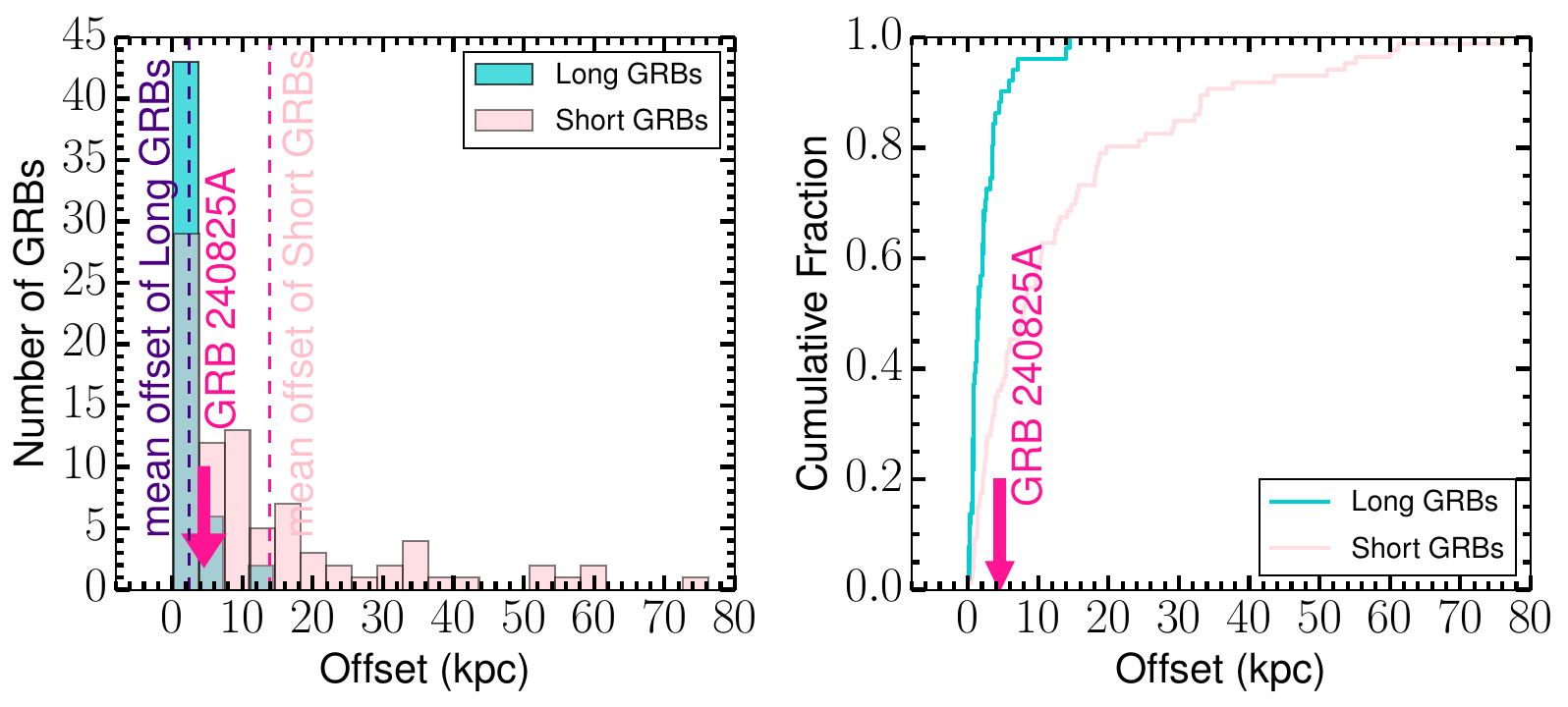}
    \caption{Left panel: Histogram of physical offsets (in kpc) for collapsar GRBs (cyan) and merger GRBs (light pink), with the mean offset of GRB 240825A (4.54 kpc) marked by a pink arrow. The mean offsets are 2.426 kpc for collapsar GRBs and 13.838 kpc for merger GRBs, with medians of 1.451 kpc and 7.820 kpc, respectively. Right panel: Cumulative fraction of offsets for collapsar GRBs (cyan) and merger GRBs (light pink), with the position of GRB 240825A (4.54 kpc) indicated by a pink arrow. The offset of GRB 240825A, derived from the host position (RA 22:58:17.23, Dec +01:01:36.48) and VLA position (RA 22:58:17.27, Dec +01:01:36.78, \citealt{2024GCN.37322....1P}) at ($z$ = 0.659), suggests an intermediate location between typical collapsar and merger GRB populations, hinting at a potential off-axis massive star progenitor scenario.}
    \label{host_offset}
\end{figure*}

\subsubsection{Comparative analysis of GRB 240825A host galaxy properties with Collapsar and Merger GRB hosts}

Using the \textit{Prospector} tool, we derived the stellar mass, star formation rate (SFR), and specific star formation rate (sSFR) of the host galaxy of GRB 240825A to contextualize its properties relative to other GRB hosts and infer its progenitor environment. The host's parameters (see Figure \ref{host_sed_corner} of the appendix) are consistent with a moderately star-forming dusty galaxy at its redshift ($z$ = 0.659). GRB 240825A's host is typical to the main sequence of star-forming galaxies in the log $\mathrm{SFR}$ versus log $\mathrm{M}/M_{\odot}$ and log $\mathrm{sSFR}$ versus log $\mathrm{M}/M_{\odot}$ planes (see Figure \ref{host_comparison}). These properties—a relatively massive host, subsolar metallicity, and ongoing star formation—are consistent with typical hosts of long-duration GRBs \citep{Savaglio2009, 2022JApA...43...82G}. 

\subsubsection{Host offset and progenitor implications} 

The projected physical offset between a GRB afterglow and its host galaxy center provides key insights into the nature of its progenitor system. Collapsar GRBs, believed to originate from the collapse of massive stars, are typically found near the star-forming regions of their hosts and thus exhibit smaller offsets. In contrast, merger GRBs, which are thought to result from compact object mergers, often show larger offsets due to natal kicks and long inspiral times \citep[e.g.,][]{Fong2010, Berger2014}. For GRB 240825A, we derive a projected offset of $4.54$ kpc based on the angular separation ($0.64$ arcseconds) between the VLA-detected afterglow position (RA: 22:58:17.27, Dec: +01:01:36.78, \citealt{2024GCN.37322....1P}) and the host galaxy centroid (RA: 22:58:17.23, Dec: +01:01:36.48). In the context of known GRB populations \citep{2002AJ....123.1111B, 2016ApJ...817..144B, 2022ApJ...940...56F, 2022MNRAS.515.4890O}, this offset lies between the typical ranges for merger and collapsar GRBs. Specifically, collapsar GRBs have a mean (median) offset of $2.43$ kpc ($1.45$ kpc), while merger GRBs exhibit significantly larger mean (median) values of $13.84$ kpc ($7.82$ kpc). Figure~\ref{host_offset} shows the offset distribution for both GRB classes and the cumulative distribution comparison. The offset of GRB 240825A is more consistent with the upper tail of collapsar GRB offsets, suggesting a probable massive star origin.

\section{Discussion} 
\label{sec:Discussion}

\subsection{Rest-frame Energetics: \texorpdfstring{$E_{\gamma,\mathrm{iso}}$ and $E_{\rm p,i}$}{Egamma,iso and Ep,i}}

\begin{figure}[ht]
    \centering
    \includegraphics[width=0.9\linewidth]{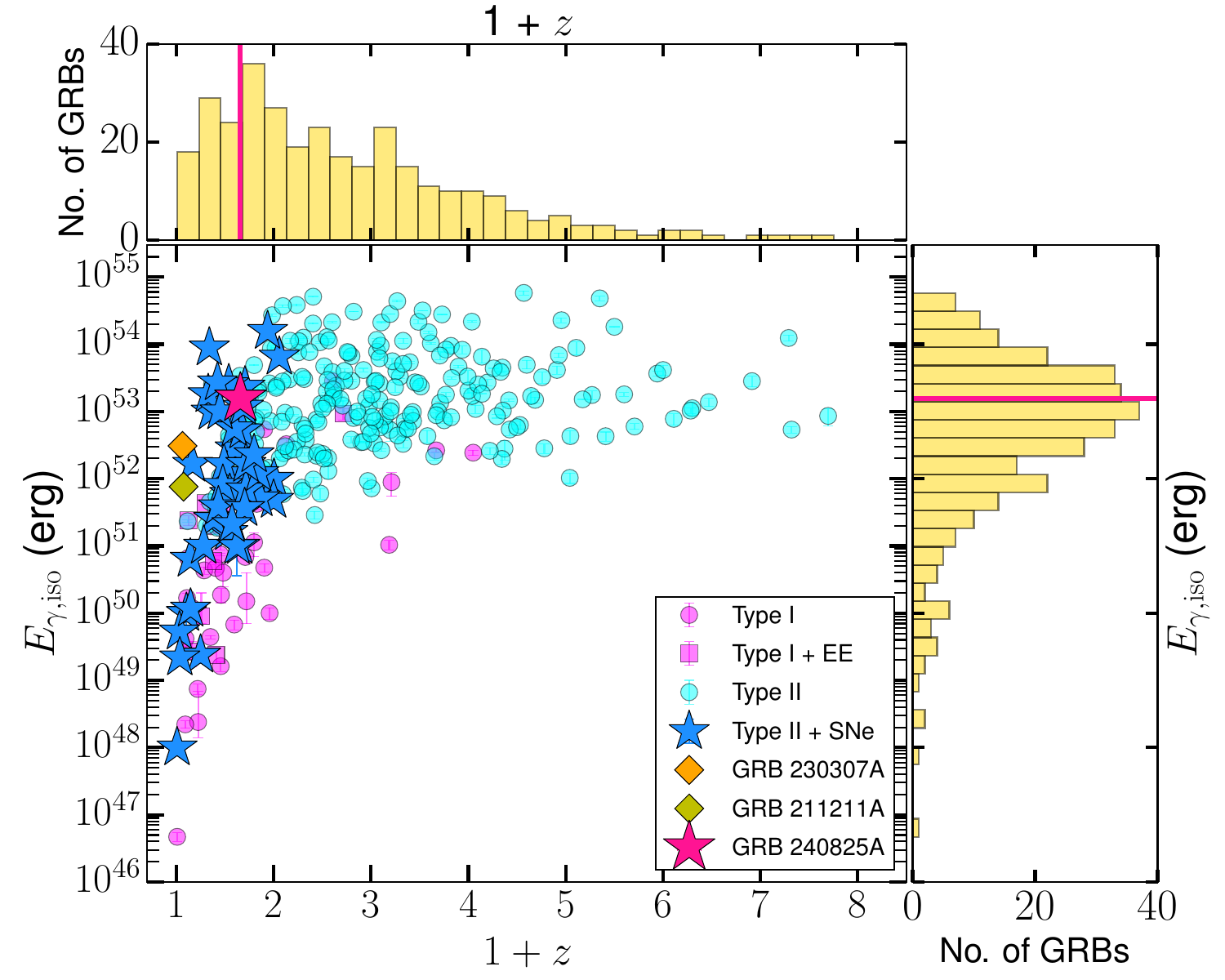}
     \includegraphics[width=0.9\linewidth]{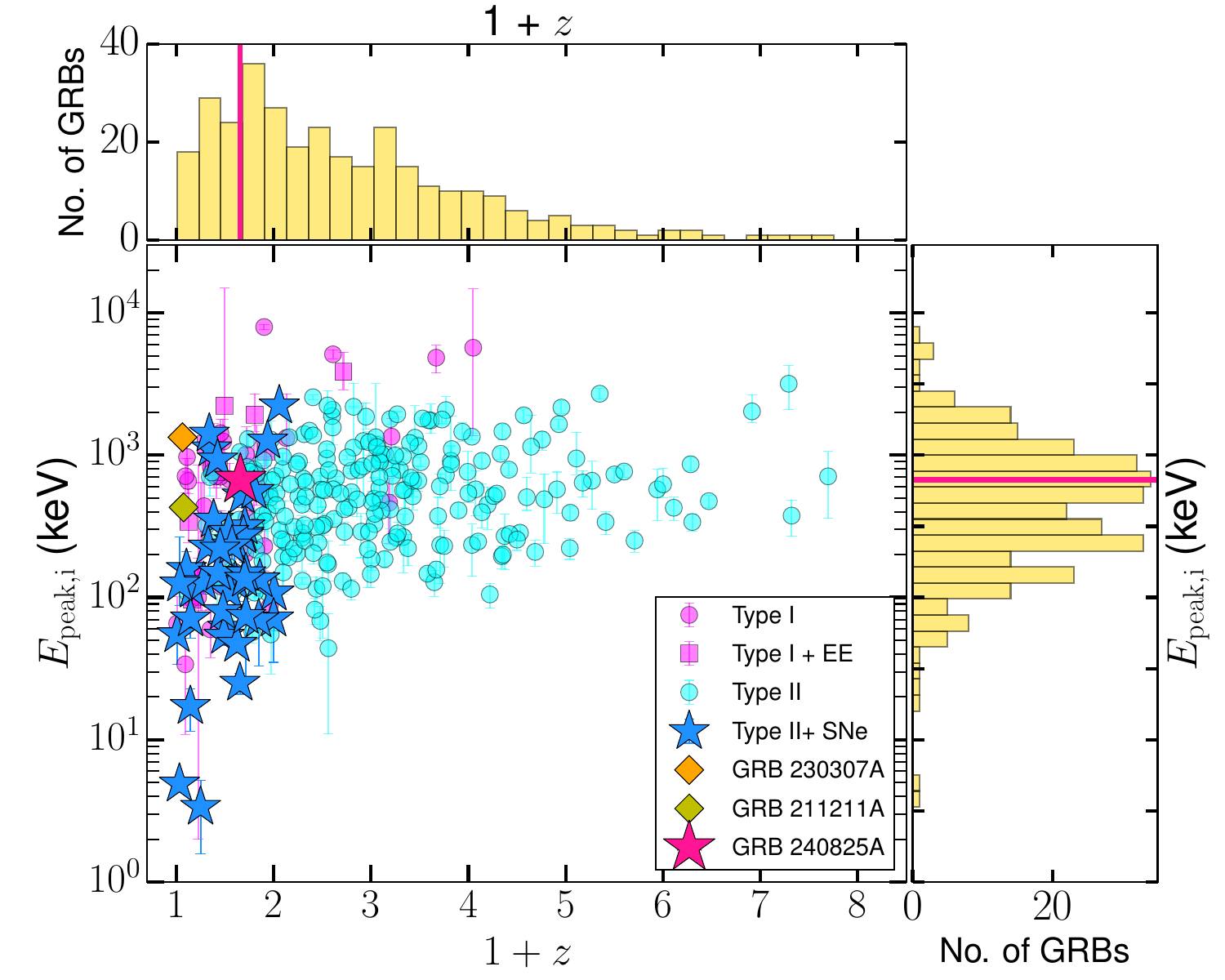}
    \caption{Top panel: Isotropic-equivalent energy $E_{\gamma,\mathrm{iso}}$ plotted against redshift for a sample of classified GRBs from \citet{2020MNRAS.492.1919M}. GRB~240825A (pink star) is shown alongside GRB~211211A and GRB~230307A for comparison. Merger GRBs (Type I), merger GRBs with extended emission (Type I+EE), collapsar GRBs (Type II), and collapsar GRBs with confirmed SNe (Type II+SNe) are color-coded. Bottom panel: Rest-frame spectral peak energy $E_{\rm p,i}$ as a function of redshift, with the same GRB sample. In these planes, GRB~240825A (pink star) appears consistent with typical collapsar GRBs.}
    \label{fig:eiso_z}
\end{figure}

Understanding the redshift evolution of prompt emission properties such as the isotropic-equivalent gamma-ray energy ($E_{\gamma,\mathrm{iso}}$) and rest-frame peak energy ($E_{\rm p,i}$) is critical for classifying GRBs and probing their progenitor systems. Collapsar GRBs, associated with collapsars, typically exhibit higher $E_{\gamma,\mathrm{iso}}$ and lower $E_{\rm p,i}$ values over a broad redshift range, while merger GRBs, often originating from compact binary mergers, tend to occupy a distinct region in this parameter space with lower energetics and higher spectral hardness. We compared the location of GRB~240825A in the $E_{\gamma,\mathrm{iso}}$ vs $(1+z)$ and $E_{\rm p,i}$ vs $(1+z)$ planes against a well-classified GRB population from \cite{2020MNRAS.492.1919M, 2025NSRev..12E.401S, 2022Natur.612..232Y}. As shown in Figure~\ref{fig:eiso_z}, GRB~240825A (pink star) appears consistent with typical collapsar GRBs. 

\begin{figure}[ht]
    \centering
    \includegraphics[width=0.95\linewidth]{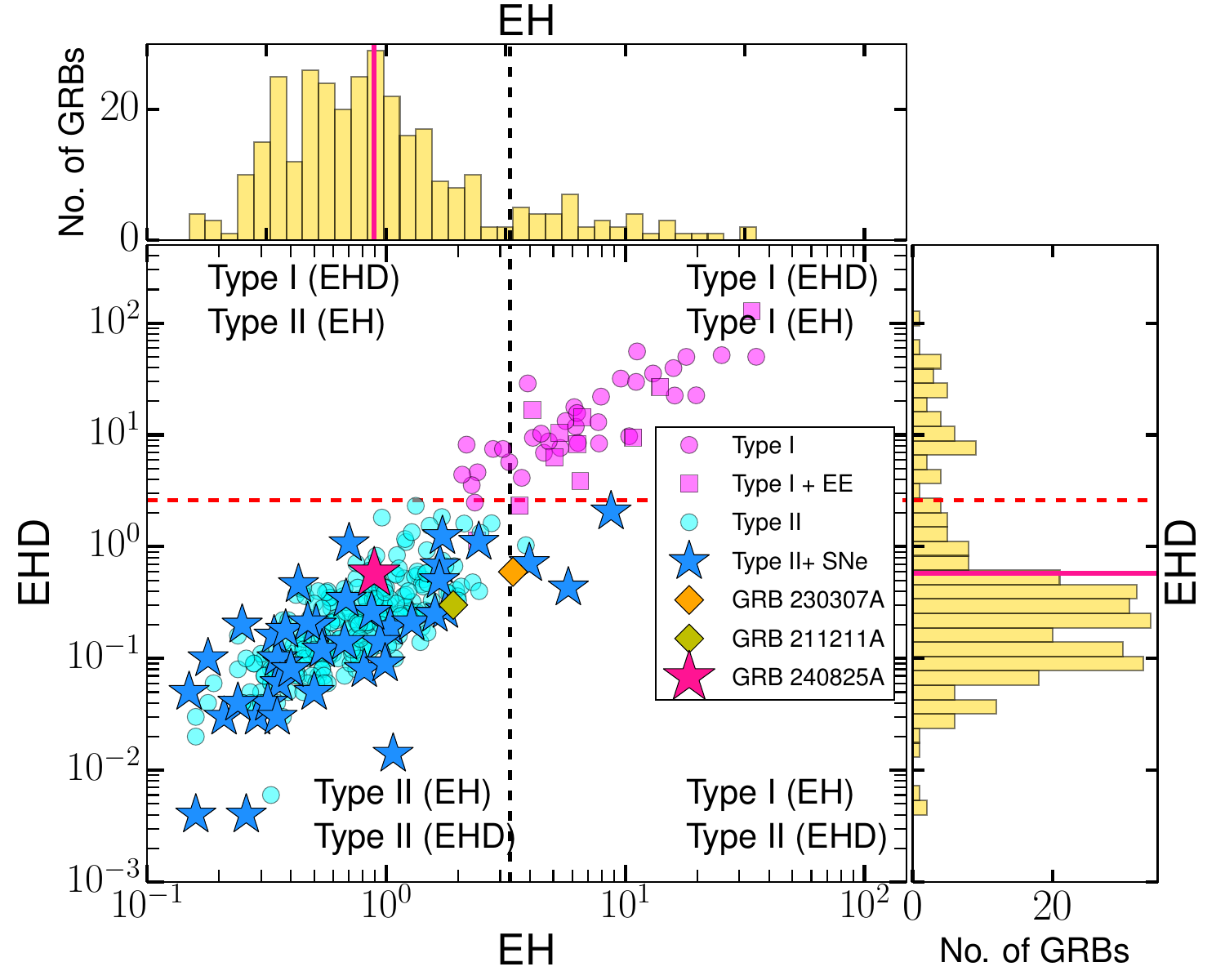}
    \caption{EHD vs. EH distribution of gamma-ray bursts from \citet{2020MNRAS.492.1919M}, showing the distinction between Type I, Type II, those with extended emission (Type I+EE), and Type II+SNe. GRB~240825A (pink star) is shown alongside GRB~211211A and GRB~230307A for comparison. GRB~240825A lies close to the classical type II GRBs region. The vertical black and red horizontal dashed lines show EHD= EH, respectively.}
    \label{fig:eh_ehd}
\end{figure}

\subsection{GRB 240825A: New classification schemes and possible progenitor}

In recent times, some authors have proposed physically motivated classifications of GRBs to complement the traditional duration-based scheme. For example, \cite{2020MNRAS.492.1919M} introduced a robust classification scheme based on two parameters: energy-hardness (EH) and energy-hardness-duration (EHD). These parameters provide a valuable diagnostic tool for GRB classification, especially when traditional duration- or spectral-based criteria prove ambiguous. In the EH--EHD parameter space, collapsar GRBs (Type II) tend to occupy regions with low EH-EHD, while merger GRBs (Type I) cluster at higher values of both parameters. In Figure~\ref{fig:eh_ehd}, we compare GRB~240825A against this framework. Our analysis places the burst in a region overlapping with type II GRBs (see Figure \ref{fig:eh_ehd}). We also analyzed GRB 211211A and GRB 230307A (long merger events, \citealt{2025NSRev..12E.401S, 2022Natur.612..232Y}) following the same parameter spaces and noted that GRB 211211A is consistent with type II GRBs; however, for GRB 230307A, the EH parameter suggests that the burst possibly belongs to type I GRBs.

\begin{figure}[ht]
\centering
\includegraphics[width=0.5\textwidth]{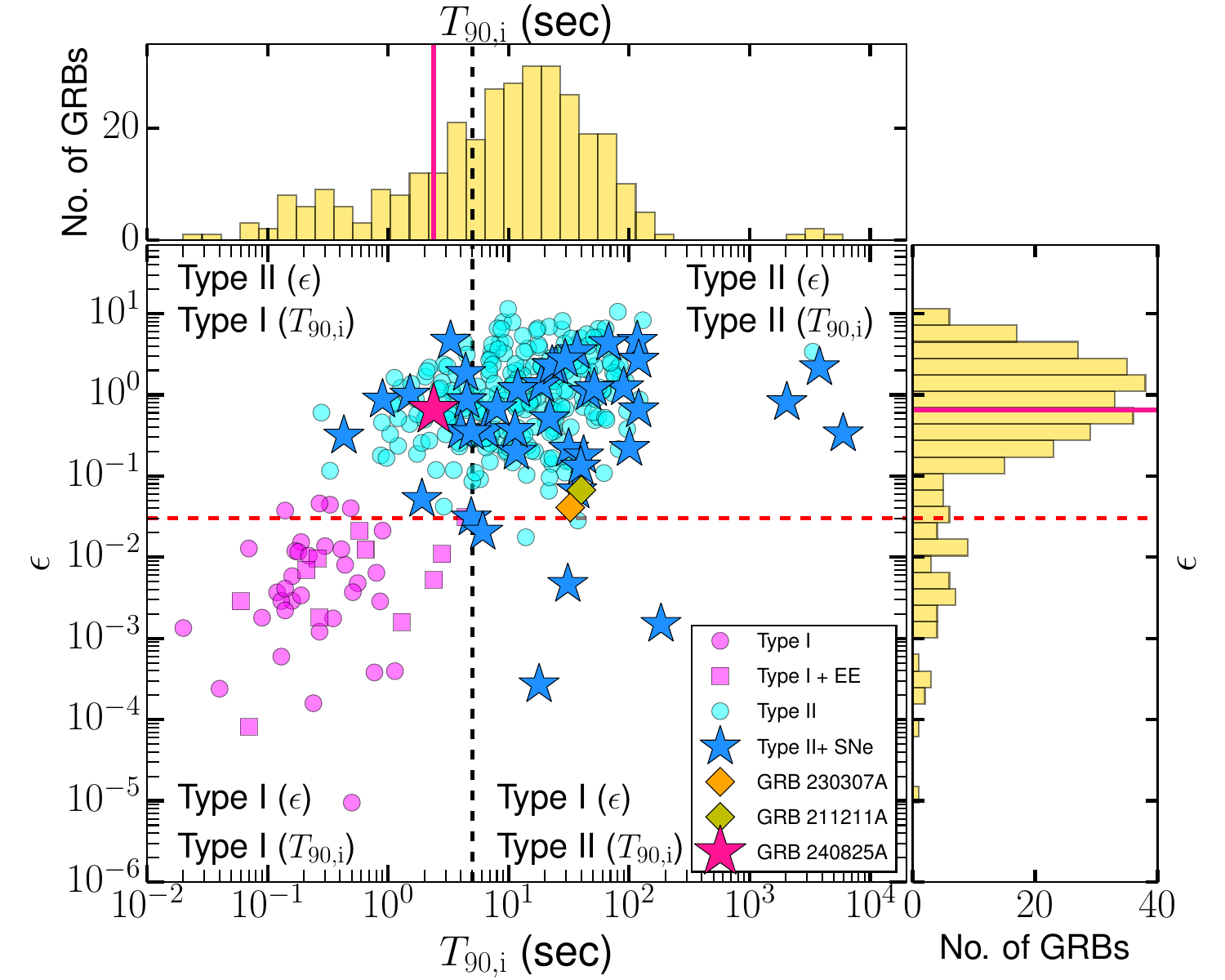}
\caption{Rest-frame duration $T_{90,i}$ vs. $\varepsilon$ for GRB 240825A (pink star), compared with the population of GRBs classified using the $\varepsilon$ method \citep{2010ApJ...725.1965L}. The plot shows distinct regions occupied by Type I (merger-origin) and Type II (collapsar-origin) GRBs, with special populations such as Type I with EE and Type II associated with SNe (II+SNe) indicated. GRB 240825A falls within the high-$\varepsilon$, but short-duration region, clustering with classical Type I bursts ($T_{90,i}$) and Type II GRBs ($\varepsilon$), supporting a hybrid origin. GRB 230307A (marked in orange) and GRB 211211A (marked in yellow) are overlaid, demonstrating their positions relative to the classical GRB populations. The black and red dashed lines show the $T_{90,i}$ = 5 sec, and $\varepsilon$ = 0.03, respectively.}
\label{fig:t90_epsilon}
\end{figure}

An important tool for understanding the physical origin of GRBs lies in the classification based on their prompt emission properties. However, the traditional scheme does not always correlate well with the underlying progenitor systems, namely collapsars (Type II) and compact object mergers (Type I). To address this, \citet{2010ApJ...725.1965L} introduced a new phenomenological classification method based on the parameter $\varepsilon \equiv E_{\gamma,\mathrm{iso},52} / E_{p,z,2}^{5/3}$, where $E_{\gamma,\mathrm{iso},52}$ is the isotropic gamma-ray energy in units of $10^{52}$ erg and $E_{p,z,2}$ is the rest-frame spectral peak energy in units of 100 keV. This parameter shows a clear bimodal distribution with a division at $\varepsilon \sim 0.03$, effectively distinguishing Type I and Type II GRBs. In Figure~\ref{fig:t90_epsilon}, we plot GRB 240825A in the $\log T_{90,i}$--$\log \varepsilon$ plane, alongside other GRBs (obtained from \citealt{2020MNRAS.492.1919M, 2025NSRev..12E.401S, 2022Natur.612..232Y}) classified using this scheme. GRB 240825A lies in the high-$\varepsilon$, but short-duration region, clustering with classical Type I bursts ($T_{90,i}$) and Type II GRBs ($\varepsilon$), supporting a hybrid origin (merger from $T_{90,i}$ and collapsar from $\varepsilon$).

\subsection{Machine learning-based progenitor classification of GRB 240825A}

\begin{figure}[!ht]
\centering
\includegraphics[width=0.5\textwidth]{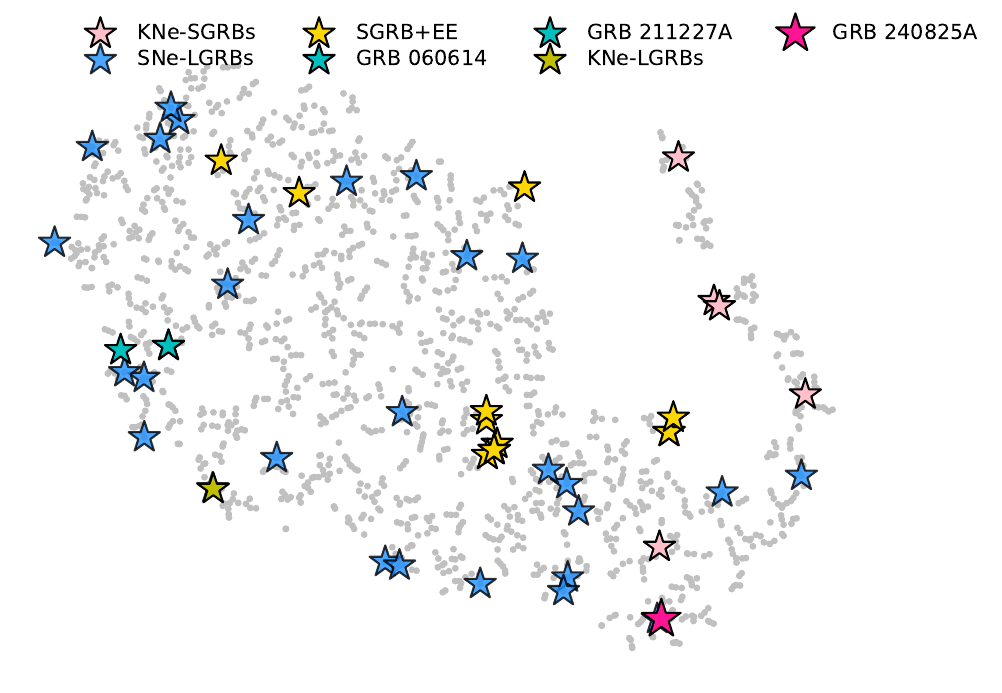}
\caption{t-SNE classification map of \swift GRBs using normalized, Fourier-transformed prompt light curves as in \citet{2020ApJ...896L..20J}. GRB 240825A is marked with a pink star and lies near the type-L (long) GRB cluster. Non-special GRBs are shown as small gray points. Light pink stars represent ``Known Mergers", blue stars denote ``Known Collapsar" (e.g., GRB 190829A), and green stars, the long mergers GRB 060614 and GRB 211227A.}
\label{fig:tsne_map}
\end{figure}

Recent machine-learning approaches provide useful, but not definitive, ways to compare GRBs in multidimensional prompt-emission parameter spaces. One such technique is the t-distributed stochastic neighbor embedding (t-SNE), which performs unsupervised dimensionality reduction on the high-dimensional prompt emission light curves to reveal intrinsic clustering in a lower-dimensional space \citep{2020ApJ...896L..20J}. This method, applied to the full \swift GRB catalog, results in an unambiguous separation into two well-defined groups that broadly correspond to collapsar and merger GRBs, respectively. The classification is purely empirical and based on the global similarity of light curves, retaining rich temporal and spectral structure. We caution that this map should not be interpreted as a clean physical separation between collapsar and merger progenitors. As shown in Figure~\ref{fig:tsne_map}, the t-SNE projection most clearly separates the population of classical short GRBs from the broader long-duration population. Several hybrid or unusual events, including short GRBs with extended emission and long GRBs with KN signatures, overlap with standard long GRBs. This indicates that duration and broad light-curve morphology remain important drivers of the embedding. Figure~\ref{fig:tsne_map} presents the position of GRB 240825A on the t-SNE map along with a sample of \swift-detected GRBs. The burst is consistent with type-L (long) GRBs.

\begin{figure}[ht]
\centering
\includegraphics[width=0.5\textwidth]{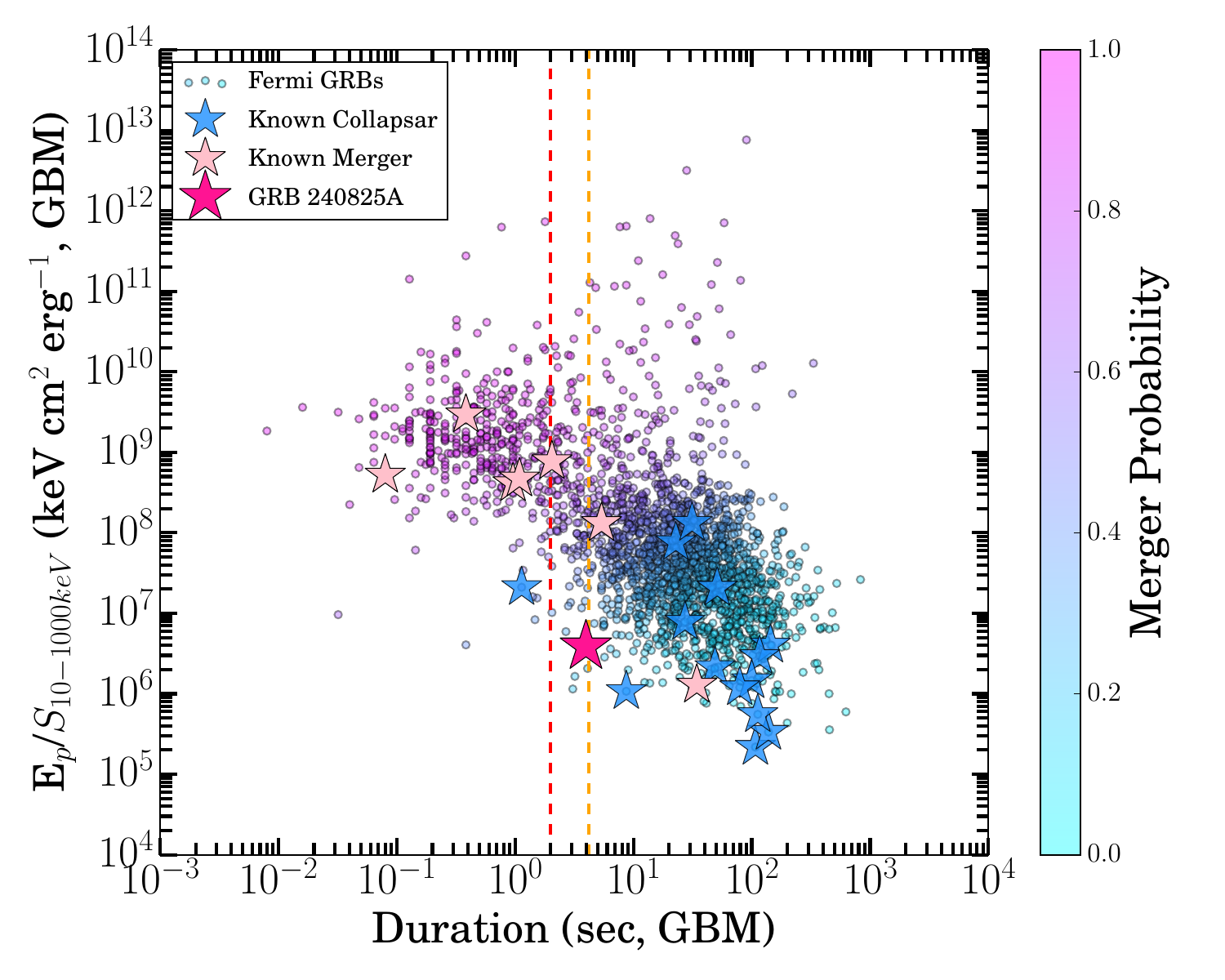}
\caption{ML-classification of GRB 240825A (shown with pink star) in the plane of burst duration (\tninty) versus spectral peak energy over fluence ($E_p/S$), following the SVM-based method of \cite{2024ApJ...974..120N}. The color scale represents the predicted merger probability, ranging from 1 (merger-like) to 0 (collapsar-like). Known collapsars (blue stars) and known mergers (pink stars) are also shown for reference. GRB 240825A falls in a region with intermediate $T_{90}$ and moderate $E_p/S$, with a classifier prediction of $\sim$76\% probability for a collapsar origin. The vertical orange and red lines show a 4.2 sec duration boundary from the fourth Fermi Catalog \citep{2020ApJ...893...46V} and \tninty= 2 sec, respectively.} 
\label{fig:svm_classification}
\end{figure}

Furthermore, \cite{2024ApJ...974..120N} demonstrated that machine learning methods can classify GRB progenitors using prompt emission observables, moving beyond simplistic duration-based schemes. In particular, the ratio of the spectral peak energy ($E_p$) to fluence ($S$) in the 10–1000 keV band, plotted against the burst duration ($T_{90}$), serves as a physically motivated proxy to distinguish progenitor types, as collapsars and compact object mergers tend to populate different regions in this plane. This methodology was adopted in a probabilistic framework using a support vector machine (SVM) with a radial basis function kernel, trained on a sample of GRBs with known progenitor associations. We applied this classifier to GRB 240825A and obtained a probability of $\sim 76\%$ for a collapsar origin and $\sim 24\%$ for a merger origin. The fourth Fermi Catalog suggested a boundary of 4.2 seconds for classifying GRBs \citep{2020ApJ...893...46V}. We noted that GRB 240825A has a duration slightly below this \fermi-GBM threshold. Following the classification scheme proposed by \citet{2024ApJ...974..120N}, GRB 240825A would likely be considered a short collapsar similar to GRB 200826A (see Figure~\ref{fig:svm_classification}, \citealt{2021NatAs...5..917A}). 

\subsection{Search for associated Supernova with GRB 240825A} 
\label{SN}

\begin{figure*}[ht]
    \centering
    \includegraphics[width = 0.325\textwidth]{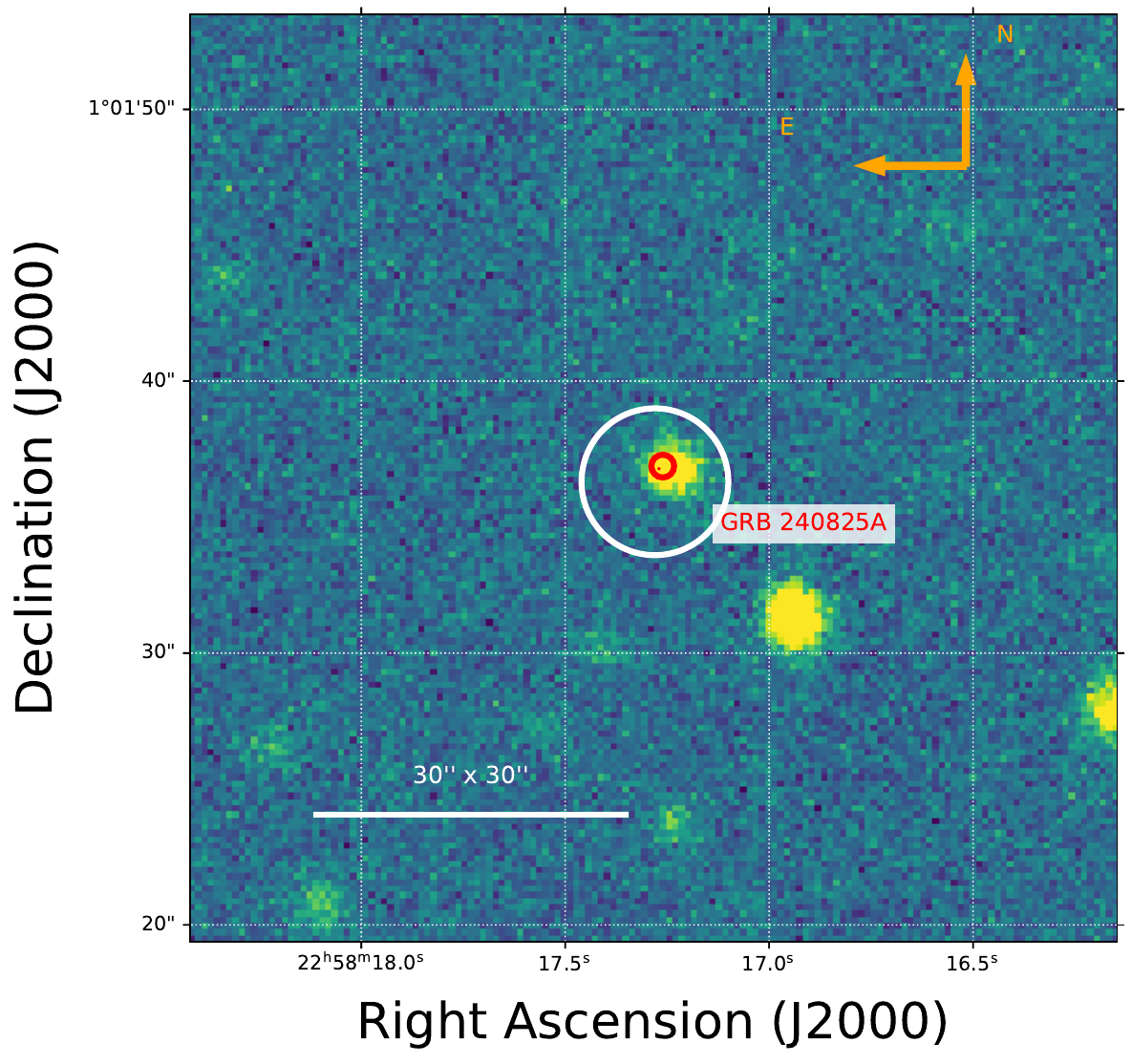}
        \includegraphics[width = 0.325\textwidth]{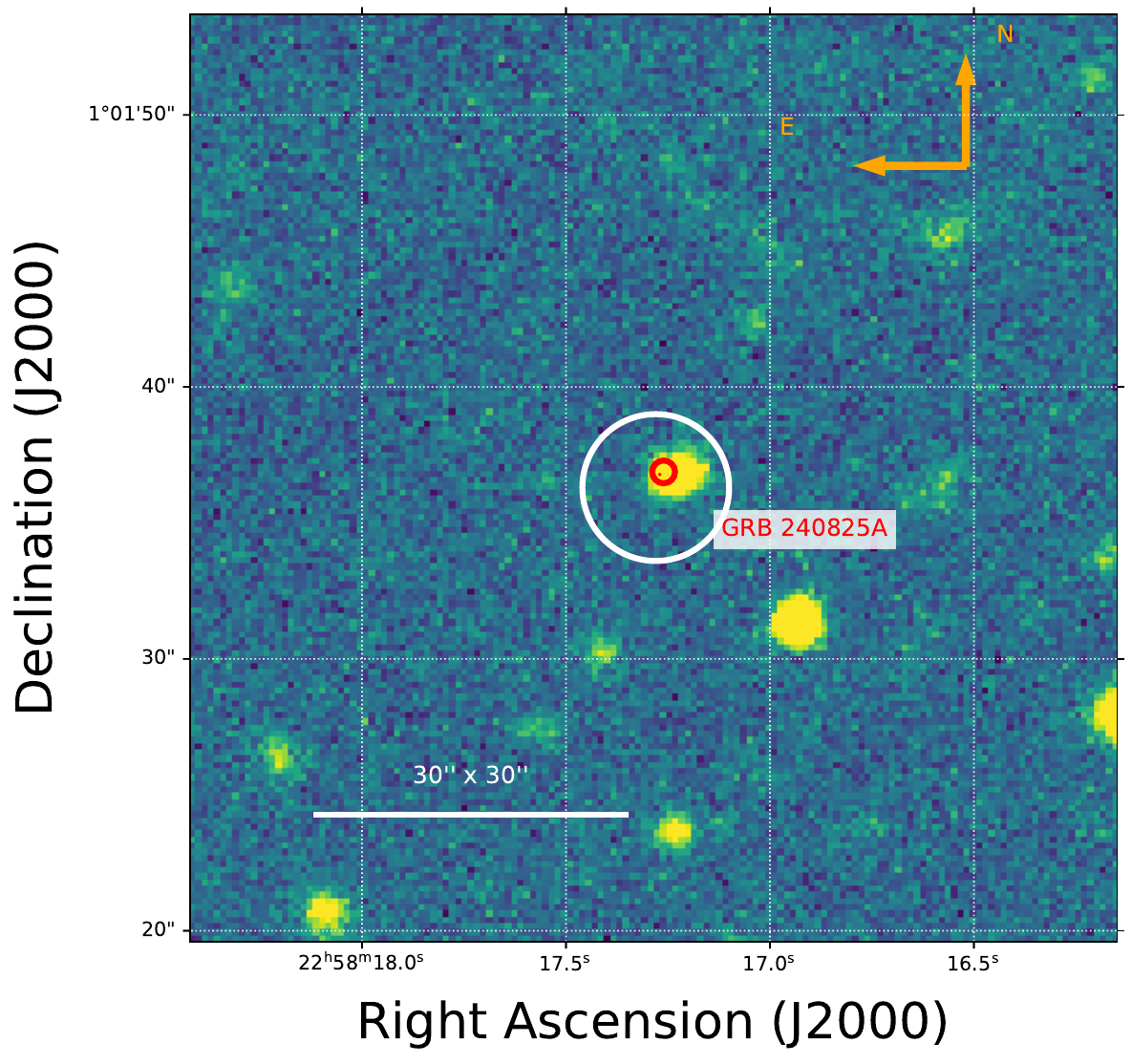}
    \includegraphics[width = 0.325\textwidth]{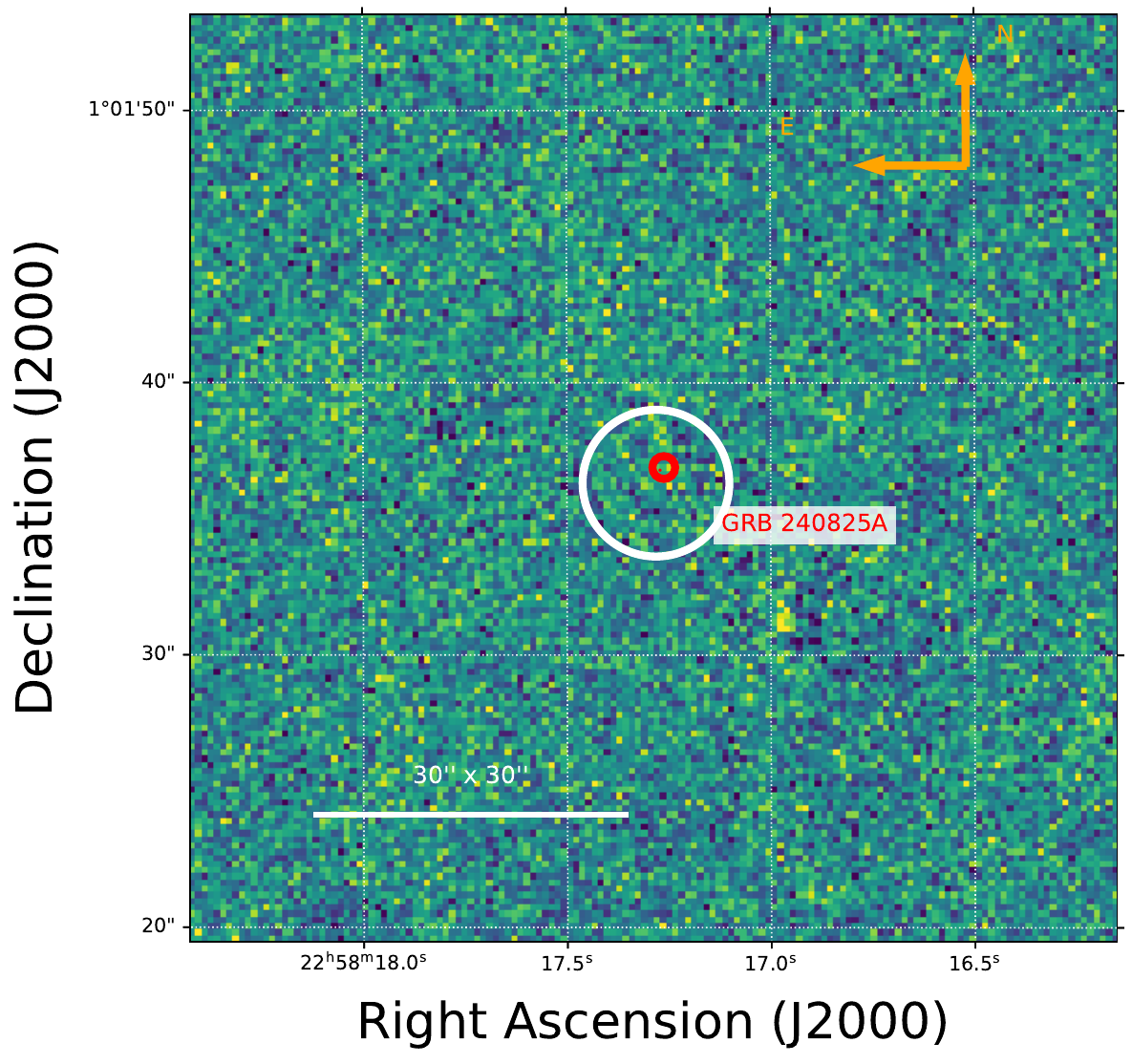}
    \caption{Lack of bright SN associated with GRB 240825A. {\emph{Left:} LBT $r'$-band image of the GRB 240825A field on 2024 Sept 28 ($\sim$ 33.70 days post-burst). \emph{Middle:} LBT $r'$-band on 2024 Nov 9 ($\sim$ 75.52 days) used as the subtraction template. \emph{Right:} Difference image (Sept 28 $-$ Nov 9) produced with AutoPhot/HOTPANTS, showing no point-like residual at the GRB afterglow position (red circle). The white and red circles show the uncertainty in the position measured by \swift-XRT and \swift-UVOT. Images are oriented with north up and east to the left. The scale bar corresponds to $10''$.}
}
\label{SN_search}
\end{figure*}

GRB 240825A exhibits a duration of approximately 4 seconds and lies at a redshift of $z = 0.659$, placing it relatively nearby in the cosmological context. The burst released a substantial fluence of $\sim 10^{-4}~\mathrm{erg~cm^{-2}}$, indicative of a highly energetic event and consistent with the Amati relation of collapsar GRBs. Furthermore, its X-ray afterglow luminosity is also consistent with those observed in collapsar GRBs, and significantly more luminous than typical Type I GRBs of merger origin \citep{2024GCN.37536....1R}. These observational signatures support the classification of GRB 240825A as a likely collapsar event. Its properties are broadly consistent with the binary-driven hypernova (BdHN I) scenario (e.g., \citealt{2024ApJ...966..219B}), indicating a core-collapse progenitor and an associated SN \citep{2024GCN.37536....1R}. These properties of GRB 240825A motivated us to search for an accompanying SN signature. We conducted a comprehensive search for SN associated with the long-duration burst GRB 240825A, located at a redshift of $z = 0.659$, using optical (using the LBC mounted on the Large Binocular Telescope \citealt{2008A&A...482..349G}), UV (\swift-UVOT), and near-infrared (using EMIR mounted on the 10.4\,m GTC Telescope) observations. Optical observations were performed in the Sloan $r'$ and $z'$ bands across three epochs: 2024-09-12 ($\sim$ 17.59 days post-burst), 2024-09-28 ($\sim$ 33.70 days post-burst), and 2024-11-09 ($\sim$ 75.52 days post-burst). With a time dilation factor of $1 + z = 1.659$ due to the redshift, these epochs correspond to rest-frame times of approximately 10.60, 20.31, and 45.52 days, respectively, aligning with the typical rise and decay phases of SN-GRB light curves \citep{2017AdAst2017E...5C}. Furthermore, \swift-UVOT performed a deep late-time search for SN emission in the U-band up to $\sim$ 33.14 days post-burst (see Table \ref{tab:uvot_sn}). We also utilized GTC NIR observations in the $J$, $H$, and $K$ bands, spanning $\sim$ 19.31–27.41 days post-trigger ($\sim$ 11.6–16.5 days rest frame), a period optimal for detecting SN emission near its peak \citep{2017A&A...605A.107C}. The choice of $r'$, $z'$, $J$, $H$, and $K$ bands leverages their red and near-infrared coverage, ideal for detecting redshifted optical emission from a distant source at $z = 0.659$, where rest-frame ultraviolet and blue light are shifted into these wavelengths.

To identify a potential SN signature superimposed on the host galaxy's light, we employed image subtraction using the AutoPhot software package \citep{2022A&A...667A..62B}, which implements the HOTPANTS algorithm for optimal difference imaging \citep{2015ascl.soft04004B}. This technique aligns a science image (containing the potential transient) with a template image (where the transient has faded), scales them to match in flux, and subtracts the template to remove static sources like the host galaxy. For GRB 240825A, the third epoch (2024-11-09, 75.52 days post-burst) served as the template, as any SN would have significantly faded by this late time (45.52 days rest frame), leaving primarily the host galaxy emission. This template was subtracted from the first (2024-09-12) and second (2024-09-28) epoch LBT images in both $r'$ and $z'$ bands. 

The resulting difference images revealed no significant residual flux at the precise position of GRB 240825A, indicating the absence of a detectable SN component. For the GTC NIR data, the second epoch ($\sim$ 27.41 days) was subtracted from the first epoch ($\sim$ 19.31 days) in $J$ and $K$ bands. No significant residual flux was detected in any difference images at the GRB position (derived from XRT and UVOT uncertainty circles), further indicating the absence of an SN signature. Given the non-detection of a transient source, we calculated upper limits on the brightness of any potential SN. These limits represent the faintest magnitudes detectable at a 3$\sigma$ confidence level above the background noise in the difference images. For the LBT optical data, the limits are: on 2024-09-12, $r' > 25.0$ mag, $z' > 24.8$ mag; on 2024-09-28, $r' > 24.5$ mag, $z' > 23.8$ mag; and on 2024-11-09, $r' > 25.9$ mag, $z' > 25.0$ mag (AB system). For the GTC NIR data, limits at the first epoch ($\sim$ 19.31–27.41 days) are $J > 23.4$ mag, $H > 23.3$ mag, and $K > 23.1$ mag (AB system). These values reflect the sensitivity of the observations, with variations likely due to differences in seeing conditions, exposure times, or sky brightness across epochs. Furthermore, using the UVOT data, we attempted a deep late-time search for emission from a SN potentially associated with the GRB; however, no evidence of any SN was detected up to 23.4 mag (AB) in the U-band. We limit our search to the U band, since it is the most sensitive of UVOT's filters, and this GRB has the most late-time coverage in the U filter. We summed the last 3 observations and then calculated the upper limits using {\tt uvotsource}  (see Table \ref{tab:uvot_sn}). To obtain even deeper limits, we then continued to sequentially add the prior segment's U-band exposure. 

\begin{figure}
    \centering
    \includegraphics[width = 0.47\textwidth]{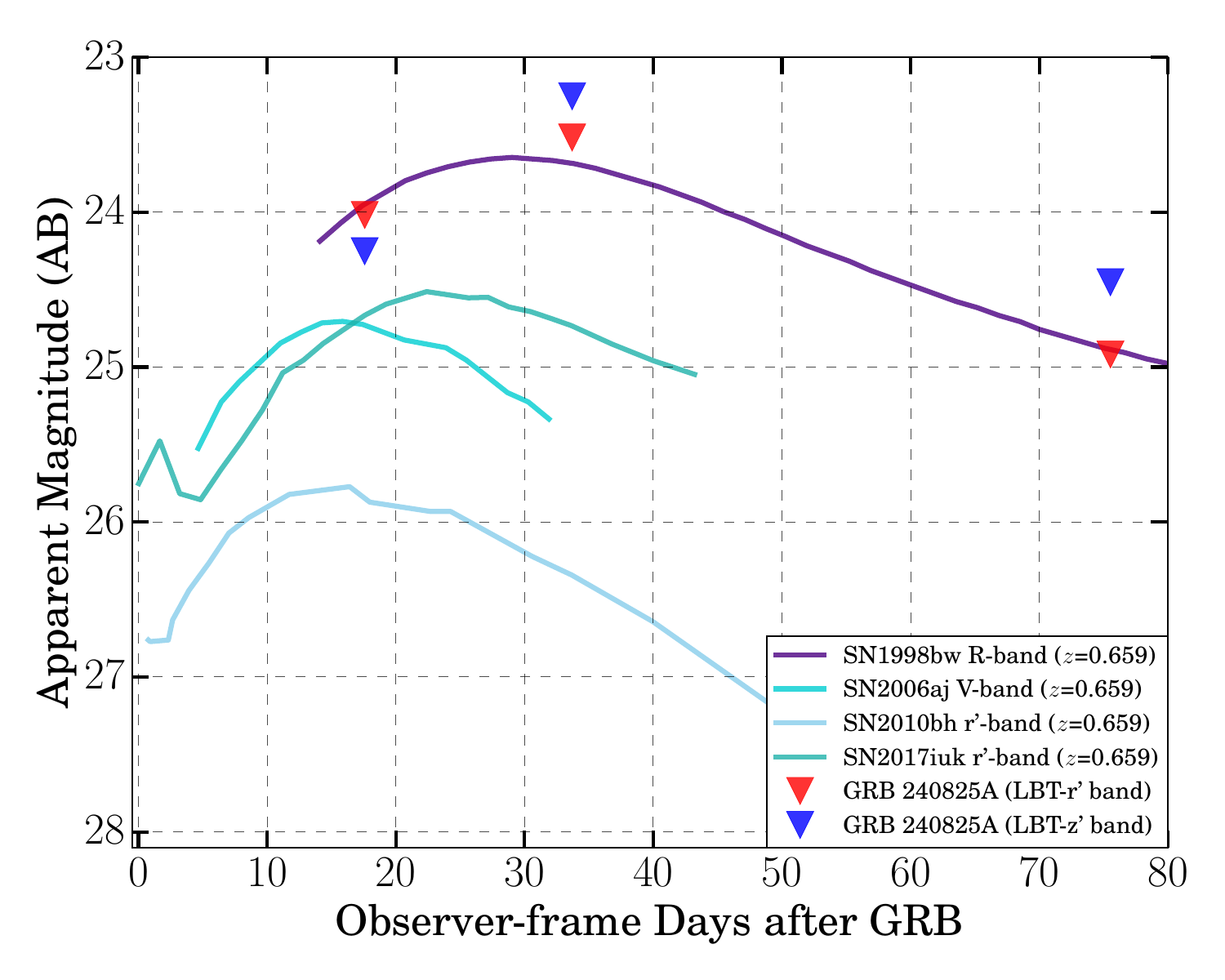}
    \caption{Comparison of our deep optical limits (extinction corrected) for GRB 240825A with well-studied GRB-associated SNe. Optical deep limits in the $r$ and $z$ bands from LBT observations of GRB 240825A, compared to SN1998bw, SN2006aj, SN2010bh, and SN2017iuk \citep{2025A&C....5200954F} scaled to the redshift of GRB 240825A ($z = 0.659$).}
    \label{SN_comparison}
\end{figure}

To interpret these non-detections, we compared our upper limits to the light curves of well-characterized SN-GRBs: SN1998bw (associated with GRB 980425, \citealt{1998Natur.395..670G}), SN2006aj (GRB 060218, \citealt{2006Natur.442.1008C}), SN2010bh (GRB 100316D, \citealt{2011MNRAS.411.2792S}), and SN2017iuk (GRB 171205A, \citealt{2019Natur.565..324I}). These light curves were adjusted to the redshift of GRB 240825A ($z = 0.659$) by applying a time dilation factor of 1.659 and a distance modulus correction to account for luminosity distance differences. The LBT limits do not rule out the full range of known GRB-SNe. In particular, the limits at 33.70 days and 75.52 days observer-frame lie above most of the comparison light curves. The earliest epoch at $\sim 17.59$ days observer-frame provides the most useful constraint ($r^{\prime}>25.0$ AB mag) and mildly disfavors a very luminous SN 1998bw-like component at that phase (see light curve comparison in Figure \ref{SN_comparison}). However, fainter or more slowly evolving events, including SN2022xiw/GRB 221009A \citep{2023ApJ...949L..39S} and SN2026fgk/GRB 260310A \citep{2026arXiv260518973O}, would remain consistent with our measurements. The non-detection also suggests either a low-luminosity SN component below our detection limits or a GRB without a detectable SN, potentially indicating diverse physical mechanisms or progenitor properties \citep{2010Natur.463..513S}. Our NIR limits ($J>23.4$, $H>23.3$, and $K>23.1$ AB mag at $\sim 19.3$--27.4 days after the burst) obtained using GTC provide useful ancillary non-detections at redder wavelengths, where dust extinction is reduced. These limits further strengthen the constraint against a bright SN. 



For the non-detection of an SN associated with GRB 240825A, observed using the LBT with an r-band magnitude limit of $m_r > 25.0$ at approximately 17.59 days post-burst (observer frame), we derived an extinction-corrected absolute magnitude upper limit of $M_r > -18.68$ mag (with k-correction). Applying a bolometric correction of $-0.43$ to account for flux outside the observed band \citep{2016MNRAS.458.2973P}, we obtained a bolometric absolute magnitude limit of $M_{\text{bol}} > -19.11$ (with k-correction), corresponding to a bolometric luminosity upper limit of $L_{\text {bol}} < 1.33 \times 10^{43} \, \text{erg/sec}$. Assuming that any accompanying SN would be powered by the radioactive decay chain ${}^{56}\mathrm{Ni} \rightarrow {}^{56}\mathrm{Co} \rightarrow {}^{56}\mathrm{Fe}$ (e.g., \citealt{1982ApJ...253..785A, 2008MNRAS.383.1485V, 2021A&A...646A..50H}), we used the standard heating function, which relates the bolometric luminosity to the nickel mass via $L_{\text{bol}}(t) = \frac{M_{\text{Ni}}}{\rm M_\odot} \cdot \left[ (\varepsilon_{\text{Ni}} - \varepsilon_{\text{Co}}) e^{-t / t_{\text{Ni}}} + \varepsilon_{\text{Co}} e^{-t / t_{\text{Co}}} \right]$ \citep{2017suex.book.....B}, where $t$ is the rest-frame time since explosion. Converting the observer frame time to rest frame using $t_{\text{rest}} = t_{\text{obs}} / (1 + z) = 17.59 / 1.659 \approx 10.60 \, \text{days}$, and using standard values $\varepsilon_{\text{Ni}} = 7.9 \times 10^{43} \, \text{erg/sec}$, $\varepsilon_{\text{Co}} = 1.45 \times 10^{43} \, \text{erg/sec}$, $t_{\text{Ni}} = 8.8 \, \text{days}$, and $t_{\text{Co}} = 111.3 \, \text{days}$ \citep{2022PASP..134e4201B}, we estimate an upper limit on the synthesized nickel mass of $M_{\rm Ni} \lesssim 0.41\,{\rm M_\odot}$. This limit is consistent with SN~1998bw which had an $M_{\rm Ni} \approx 0.4\,{\rm M_\odot}$ \citep{1998Natur.395..672I} while SN~2003dh had a lower value of $M_{\rm Ni} \approx 0.35\,{\rm M_\odot}$ \citep{2003ApJ...599L..95M}. As the amount of the nickel synthesized determines the peak of an SN powered by the radioactive decay chain ${}^{56}\mathrm{Ni} \rightarrow {}^{56}\mathrm{Co} \rightarrow {}^{56}\mathrm{Fe}$, the inferred $M_{\rm Ni}$ suggests that any accompanying SN to GRB 240825A must be faint.

\begin{figure}[ht]
\centering
\includegraphics[width=0.48\textwidth]{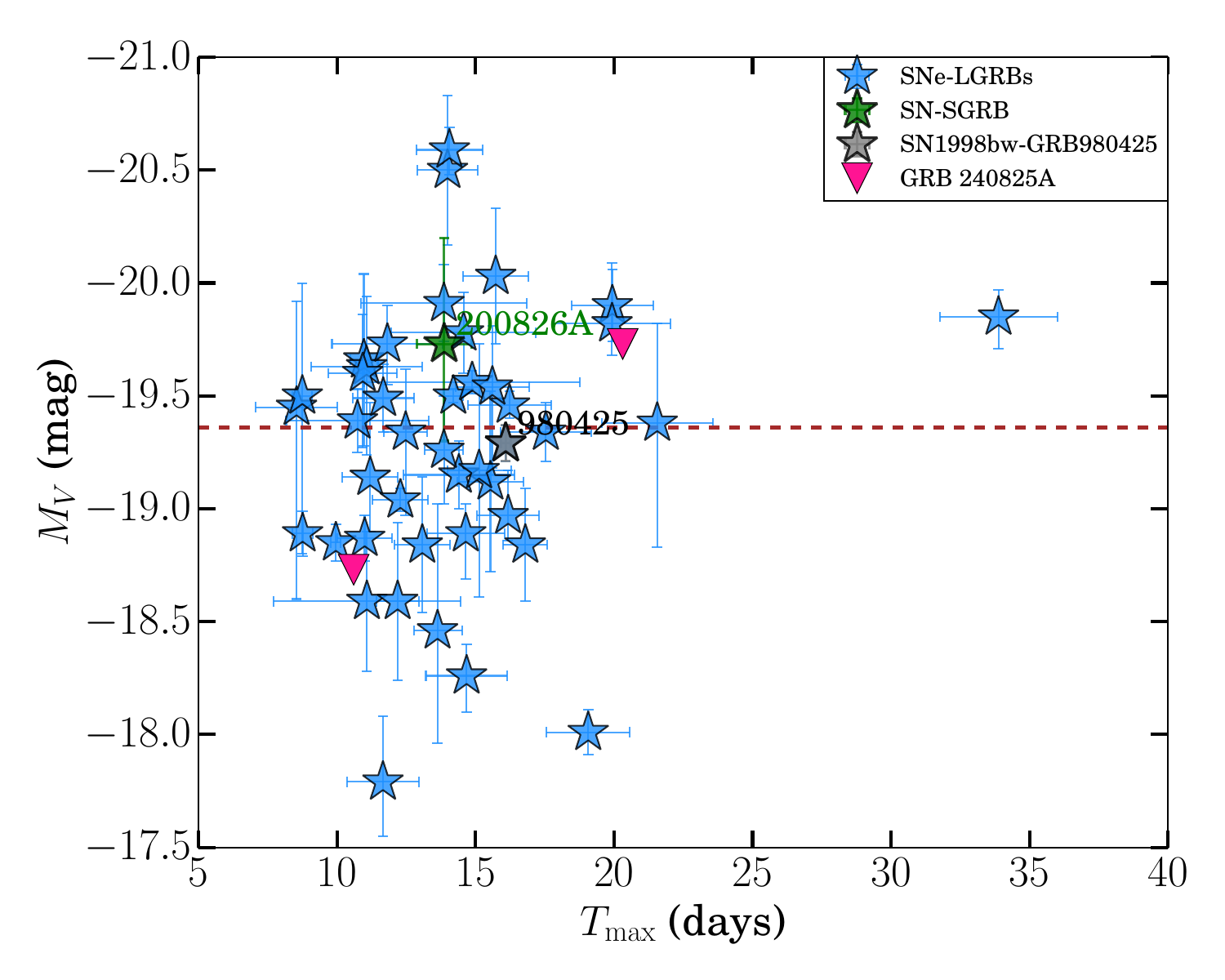}
\caption{Distribution of peak absolute $V$-band magnitude ($M_V$) versus rest-frame peak time ($T_{\rm max}$) for GRB-associated SN. Blue stars represent SNe-LGRBs, green stars denote SN-SGRB (GRB 200826A), and the grey star marks SN 1998bw associated with GRB 980425. The pink triangles indicate the GRB 240825A upper limits (derived from our $z$-band observations) at the actual observing epochs: 17.59 and 33.70 days in the observer frame, corresponding to 10.60 and 20.31 days in the rest frame, respectively. The earlier 17.59-day limit may be interpreted as a constraint on a fast/early-peaking GRB-SN, whereas the 33.70-day limit is more relevant for a broader, SN1998bw-like event whose observed peak would be expected closer to $\sim 30$ days after the burst. Data points for comparison are taken from \citet{2024AstL...50..701B}, and the brown dashed line marks the median peak brightness of the SN-LGRB population.
}
\label{fig:mv_tmax}
\end{figure}

To search for a potential SN associated with GRB 240825A, we obtained late-time photometry in the observer-frame $z$-band, which probes the rest-frame $V$-band at the burst redshift ($z = 0.659$). Based on our non-detections, we derive upper limits on the absolute magnitude. For rest-frame SN peak times of 10.60 days (observer-frame 17.59 days) and 20.31 days (observer-frame 33.70 days), the limits correspond to $M_V > -18.73$ and $M_V > -19.73$, respectively. The 17.59-day limit is the deepest optical constraint and could be relevant if the associated SN were fast-evolving and peaked at an early rest-frame time of $\sim 10$ days, as seen in some rapidly evolving GRB-SNe. In this case, the non-detection would imply that such a fast SN was fainter than $M_V \simeq -18.73$. However, this epoch is earlier than the expected observer-frame peak of an SN1998bw-like template at $z=0.659$. For a broader, SN1998bw-like light curve, the 33.70-day limit is the more appropriate comparison, and this constraint is substantially milder. We placed these values of GRB 240825A in the distribution of peak absolute $V$-band magnitude ($M_V$) versus rest-frame peak time ($T_{\rm max}$) for GRB-associated SN. This constraint is shown in Figure~\ref{fig:mv_tmax} (we emphasize that, because no SN component was detected, the peak time of any associated SN is not directly measured. Therefore, the limits should be interpreted as epoch-dependent upper limits), alongside the GRB-SN sample from \citet{2024AstL...50..701B}. Our observations show that no bright SN component was detected at either an early/fast-SN epoch or near the expected peak epoch of a broader GRB-SN. The non-detection therefore argues against a luminous, unobscured SN peaking near the observed epochs, but it does not exclude a normal, moderately faint, or dust-obscured GRB-SN.

\begin{figure}[ht]
    \centering
    \includegraphics[width = 0.48\textwidth]{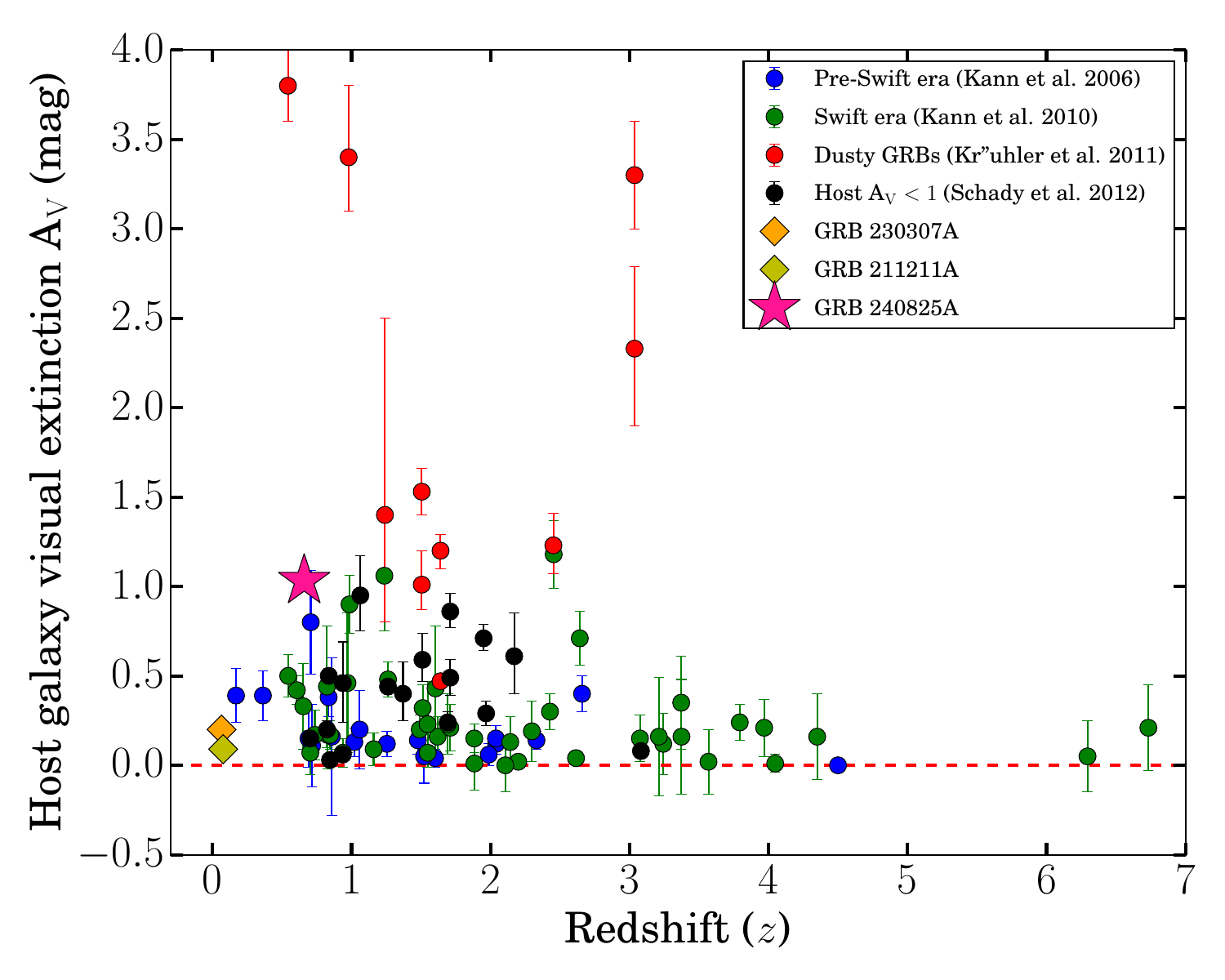}
    \caption{The figure illustrates the relationship between host galaxy visual extinction (A$_V$) and redshift ($z$) for GRB 240825A alongside various GRB samples, with the horizontal red dashed line indicating A$_V$ = 0. Our analysis reveals that GRB 240825A exhibits a significantly elevated A$_V$ at its redshift relative to the broader GRB population.}
    \label{Host_DUST}
\end{figure}

We explored the possible cause (such as the local environment) of the non-detection of the associated SN for GRB 240825A. The visual extinction (A$_V$), defined as the attenuation of light in the visual band due to dust along the line of sight, serves as a critical indicator of the dust content and density of the ISM. Based on the afterglow analysis using UVOT and XRT data, we derive a color excess of (E(B-V) = $0.35\pm0.03$ mag) and a visual extinction of (A$_V$ $\sim$ 1.03 mag) for LMC, indicating significant dust attenuation along the line of sight of GRB 240825A. In this study, we investigate the A$_V$ as a function of redshift ($z$) for GRB 240825A, comparing it with a well-established sample of GRBs from prior works \citep{2006ApJ...641..993K, 2010ApJ...720.1513K, 2011A&A...534A.108K, 2012A&A...537A..15S}. This high LOS extinction places GRB~240825A among the more dust-obscured events at its redshift relative to the broader GRB population, suggesting an unusually dust-rich host galaxy (see Figure \ref{Host_DUST}). Notably, despite extensive follow-up observations in the NIR, UV, and optical bands, no SN counterpart was detected for GRB 240825A, which could be attributed to this substantial dust obscuration. An intrinsically faint SN also remains consistent with the observational limits. Across the GRB sample, we observe a general trend of decreasing A$_V$ with increasing redshift, likely driven by selection biases inherent in high-redshift observations \citep{2010ApJ...720.1513K, 2022JApA...43...82G, 2024A&A...683A..55C}. 

\section{Conclusion} 
\label{sec:Conclusion}

In this study, we have performed a detailed, multiwavelength investigation of GRB~240825A—a unique gamma-ray burst that challenges the traditional dichotomy between short and long GRBs. Classical diagnostics such as its high fluence, energetics, positive spectral lag, placement within the $E_{\mathrm{p,i}}$--$E_{\gamma,\mathrm{iso}}$ (Amati) relation, and a massive, dusty, star-forming host favor an interpretation more consistent with collapsar-driven long GRBs. To further refine its classification, we employed advanced machine-learning classifiers (e.g., t-SNE, support vector machines) and modern physical metrics such as the energy-hardness-duration parameter, and  $\varepsilon = E_{\gamma,\mathrm{iso},52} / E_{p,z,2}^{5/3}$. The SVM classifier assigns a 76\% probability of collapsar origin, and the EH/EHD and $\epsilon$ parameters place GRB 240825A firmly in the Type II (collapsar) region. 

Late-time photometric and spectroscopic follow-up with the 10.4\,m GTC, LBT, and \textit{Swift}/UVOT revealed no detectable SN component down to deep limits ($m_r > 25.0$, $m_J > 23.4$, $m_H > 23.3$, and $m_K > 23.1$~AB mag), despite a relatively nearby redshift of $z = 0.659$. The deepest 17.59 days LBT limit constrains a possible fast/early-peaking SN, while the 33.70 days limit is more relevant for a broader SN1998bw-like event. Taken together, the observations rule out only 
a sufficiently bright SN peaking near the observed epochs. The absence of bright SN features, especially those similar to SN~1998bw, raises two possibilities: either the SN was intrinsically faint (low luminosity SN with absolute magnitude M$_{V}$ fainter than -18.5 mag), or it was heavily extinguished by dust in the host environment. We therefore interpret GRB 240825A as a collapsar-favored GRB with no detected luminous SN along a dusty sightline.


In summary, GRB~240825A is an archetypal example of a GRB that lies near the boundary of current classification schemes. This burst reinforces the need for a unified framework that combines prompt emission physics, afterglow behavior, host galaxy properties, and contextual progenitor signatures. Future studies, particularly those leveraging high-cadence multiwavelength follow-up and statistical learning techniques, will be essential to identify and interpret such events within the broader landscape of GRB phenomenology. Additionally, JWST mid-infrared follow-up could potentially detect dust-reprocessed SN emission and definitively resolve the nature of such events.

\begin{acknowledgments}
We sincerely thank the referee for their thorough and valuable suggestions, which have significantly improved the quality of our manuscript. RG was sponsored by the National Aeronautics and Space Administration (NASA) through a contract with ORAU. The views and conclusions contained in this document are those of the authors and should not be interpreted as representing the official policies, either expressed or implied, of the National Aeronautics and Space Administration (NASA) or the U.S. Government. The U.S. Government is authorized to reproduce and distribute reprints for Government purposes notwithstanding any copyright notation herein. AR acknowledges support by PRIN-MIUR 2017 (grant 20179ZF5KS). AJCT acknowledges support from the Spanish Ministry project PID2023-151905OB-I00 and Junta de Andaluc\'ia grant P20\_010168 and from the Severo Ochoa grant CEX2021-001131-S funded by MCIN/AEI/10.13039/501100011033. Based on observations collected at the Observatorio de Sierra Nevada (operated by IAA-CSIC) and at the Centro Astronómico Hispano en Andalucía (CAHA) at Calar Alto (proposal 24B-2.2-010), operated jointly by Junta de Andaluc\'ia and Consejo Superior de Investigaciones Cient\'ificas (IAA-CSIC). Also based on observations made with the Gran Telescopio Canarias (GTC) (proposal GTCMULTIPLE7B-24B 104-MULTIPLE-7/24A), installed at the Spanish Observatorio del Roque de los Muchachos of the Instituto de Astrofísica de Canarias, on the island of La Palma. This work is partly based on data obtained with the instrument OSIRIS, built by a Consortium led by the Instituto de Astrofísica de Canarias in collaboration with the Instituto de Astronomía of the Universidad Autónoma de México. OSIRIS was funded by GRANTECAN and the National Plan of Astronomy and Astrophysics of the Spanish Government. MCG acknowledges support from the Spanish Ministry project PID2023-149817OB-C31. The programme of development within Priority-2030 is acknowledged for supporting the research at UrFU. ASIM is a mission of ESA’s SciSpace programme for scientific utilization of the {\it ISS} and non-{\it ISS} space exploration platforms and space environment analogues. ASIM and the ASIM Science Data Centre are funded by ESA and by national grants of Denmark, Norway and Spain. This research has used data obtained through the HEASARC Online Service, provided by the NASA-GSFC, in support of NASA High Energy Astrophysics Programs. This work made use of data supplied by the UK Swift Science Data Centre at the University of Leicester. The LBT is an international collaboration of the University of Arizona, Italy (INAF: Istituto Nazionale di Astrofisica), Germany (LBTB: LBT Beteiligungsgesellschaft), The Ohio State University, representing also the University of Minnesota, the University of Virginia, and the University of Notre Dame. LH acknowledges support from Research Ireland grant 19/FFP/6777. DT acknowledges support from the National Research Foundation of Korea (NRF) grant, RS-2024-00343729, funded by the Korea government (MSIT).

\end{acknowledgments}

\facilities{Swift(BAT, XRT, and UVOT), Fermi, ASIM, GTC, LBT, OSN, BOOTES, CAHA, IKI-FuN}

\software{astropy \citep{2013A&A...558A..33A,2018AJ....156..123A,2022ApJ...935..167A}, 
          XSPEC \citep{1996ASPC..101...17A}, 
          Source Extractor \citep{1996A&AS..117..393B}
          }

\appendix
\restartappendixnumbering

\section{Figure and Table}

\startlongtable
\begin{deluxetable*}{llrrrrl}
\tabletypesize{\scriptsize}
\tablecaption{Optical/UV follow-up observations of GRB 240825A.
\label{tab:combined_photometry}}
\tablehead{
\colhead{Telescope} &
\colhead{Filter} &
\colhead{$T-T_0$} &
\colhead{Exp.} &
\colhead{Mag.} &
\colhead{Mag. Err.} &
\colhead{System} \\
\colhead{} &
\colhead{} &
\colhead{(sec)} &
\colhead{(sec)} &
\colhead{} &
\colhead{} &
\colhead{}
}
\startdata
Swift-UVOT & WH   & 102.99740    & 10.00000    & 15.612 & 0.036 &  AB \\
Swift-UVOT & WH   & 122.99740    & 10.00000    & 15.871 & 0.037 &  AB \\
Swift-UVOT & WH   & 142.99740    & 10.00000    & 16.253 & 0.041 &  AB \\
Swift-UVOT & WH   & 167.99740    & 15.00000    & 16.470 & 0.035 &  AB \\
Swift-UVOT & WH   & 202.99740    & 20.00000    & 16.781 & 0.034 &  AB \\
Swift-UVOT & WH   & 232.87380    & 9.87640     & 16.976 & 0.052 &  AB \\
Swift-UVOT & WH   & 594.91400    & 9.88150     & 18.365 & 0.104 &  AB \\
Swift-UVOT & WH   & 767.51850    & 9.87600     & 18.574 & 0.118 &  AB \\
Swift-UVOT & WH   & 1170.87090   & 9.87590     & 19.161 & 0.206 &  AB \\
Swift-UVOT & WH   & 6095.52310   & 99.89210    & 20.586 & 0.165 &  AB \\
Swift-UVOT & V    & 77.82920     & 5.00000     & 13.625 & 0.068 &  AB \\
Swift-UVOT & V    & 644.85540    & 9.87690     & 16.997 & 0.261 &  AB \\
Swift-UVOT & V    & 817.11690    & 9.88140     & 17.025 & 0.266 &  AB \\
Swift-UVOT & V    & 1134.47400   & 96.11000    & 17.661 & 0.298 &  AB \\
Swift-UVOT & V    & 5069.72090   & 99.88650    & 19.098 & 0.424 &  AB \\
Swift-UVOT & V    & 10883.36850  & 453.36470   & 20.196 & 0.538 &  AB \\
Swift-UVOT & V    & 53432.26540  & 3390.49850  & $>$20.059 & --  & AB \\
Swift-UVOT & V    & 81813.87990  & 3405.69650  & $>$20.121 & --   & AB \\
Swift-UVOT & B    & 570.52030    & 9.88160     & 17.393 & 0.185 &  AB \\
Swift-UVOT & B    & 743.22400    & 9.87600     & 17.650 & 0.220 & AB \\
Swift-UVOT & B    & 1146.45580   & 9.87660     & 18.845 & 0.795 &  AB \\
Swift-UVOT & B    & 5890.87900   & 99.88650    & 20.240 & 0.616 &  AB \\
Swift-UVOT & U    & 335.41420    & 30.00000    & 17.767 & 0.087 &  AB \\
Swift-UVOT & U    & 405.41420    & 40.00000    & 18.035 & 0.087 &  AB \\
Swift-UVOT & U    & 490.41420    & 45.00000    & 18.057 & 0.083 &  AB \\
Swift-UVOT & U    & 545.29720    & 9.88300     & 18.240 & 0.210 &  AB \\
Swift-UVOT & U    & 718.32840    & 9.88170     & 18.990 & 0.357 &  AB \\
Swift-UVOT & U    & 1204.12500   & 92.42510    & 19.120 & 0.382 &  AB \\
Swift-UVOT & U    & 5685.01030   & 99.89210    & 20.318 & 0.279 &  AB \\
Swift-UVOT & U    & 40602.84590  & 329.04030   & $>$21.426 & --    & AB \\
Swift-UVOT & U    & 143988.61540 & 8738.43570  & $>$22.226 & --   & AB \\
Swift-UVOT & U    & 178393.99880 & 9192.96970  & $>$22.249 & --    & AB \\
Swift-UVOT & U    & 246282.46700 & 20264.49860 & $>$22.406 & --    & AB \\
Swift-UVOT & U    & 377364.80130 & 37202.24030 & $>$23.069 & --    & AB \\
Swift-UVOT & U    & 425514.30050 & 692.72390   & 22.834 & 1.586 &  AB \\
Swift-UVOT & UVW1 & 5480.11240   & 99.88650    & 20.929 & 0.383   & AB \\
Swift-UVOT & UVW1 & 34758.21760  & 408.60420   & $>$22.214 & --   & AB \\
Swift-UVOT & UVM2 & 5274.65740   & 99.88650    & 21.945 & 0.884 &  AB \\
Swift-UVOT & UVM2 & 11449.02620  & 107.17380   & 22.714 & 3.395 &  AB \\
Swift-UVOT & UVW2 & 4865.04370   & 99.88100    & $>$21.426 & --  & AB \\
Swift-UVOT & UVW2 & 6301.23740   & 99.88650    & 22.303 & 1.101 & AB \\
Swift-UVOT & UVW2 & 44520.64640  & 266.27400   & 23.028 & 1.960 & AB \\
Swift-UVOT & UVW2 & 86962.40790  & 2930.34870  & $>$23.389 & -- & AB \\ \hline
OSN & $B$ & 30176 & 300 & 21.43 & 0.17  & AB \\
OSN & $B$ & 31419 & 300 & 21.46 & 0.17  & AB \\
OSN & $B$ & 32663 & 300 & 21.70 & 0.19  & AB \\
OSN & $B$ & 33904 & 300 & 21.68 & 0.22  & AB \\
OSN & $B$ & 35769 & 600 & 21.72 & 0.14  & AB \\
OSN & $B$ & 38871 & 900 & 22.20 & 0.24  & AB \\
OSN & $V$ & 30486 & 300 & 20.54 & 0.15  & AB \\
OSN & $V$ & 31762 & 300 & 20.41 & 0.12  & AB \\
OSN & $V$ & 32970 & 300 & 20.44 & 0.12  & AB \\
OSN & $V$ & 34210 & 300 & 20.87 & 0.18  & AB \\
OSN & $V$ & 35455 & 300 & 21.25 & 0.24  & AB \\
OSN & $V$ & 37944 & 900 & 21.32 & 0.09  & AB \\
OSN & $I$ & 31103 & 300 & 19.83 & 0.07  & AB \\
OSN & $I$ & 32348 & 300 & 19.93 & 0.08  & AB \\
OSN & $I$ & 33589 & 300 & 19.98 & 0.08  & AB \\
OSN & $I$ & 34830 & 300 & 19.94 & 0.08  & AB \\
OSN & $I$ & 36077 & 300 & 19.99 & 0.07  & AB \\
OSN & $I$ & 37320 & 300 & 19.98 & 0.07  & AB \\
OSN & $I$ & 38560 & 300 & 19.94 & 0.08  & AB \\
OSN & $I$ & 39801 & 300 & 20.00 & 0.08  & AB \\
OSN & $I$ & 41041 & 300 & 20.06 & 0.09 & AB \\
OSN & $R$ & 30796 & 300 & 20.22 & 0.11 & AB \\
OSN & $R$ & 32037 & 300 & 20.48 & 0.10  & AB \\
OSN & $R$ & 33282 & 300 & 20.33 & 0.10 & AB \\
OSN & $R$ & 34521 & 300 & 20.38 & 0.09  & AB \\
OSN & $R$ & 35767 & 300 & 20.46 & 0.10 &  AB \\
OSN & $R$ & 37013 & 300 & 20.37 & 0.09 &  AB \\
OSN & $R$ & 38250 & 300 & 20.47 & 0.12 &  AB \\
OSN & $R$ & 39493 & 300 & 20.52 & 0.12 &  AB \\
OSN & $R$ & 40730 & 300 & 20.47 & 0.10 &  AB \\ \hline
BOOTES-6 & $R$ & 9063  & 600  & 19.30 & 0.18 &  AB \\
BOOTES-6 & $R$ & 10871 & 1200 & 19.44 & 0.18 &  AB \\
BOOTES-6 & $R$ & 13159 & 1200 & 19.32 & 0.10 &  AB \\
BOOTES-6 & $R$ & 15386 & 1200 & 19.60 & 0.10 &  AB \\
BOOTES-6 & $R$ & 17379 & 1200 & 20.07 & 0.11 &  AB \\
BOOTES-6 & $R$ & 19853 & 1680 & 20.08 & 0.11 &  AB \\ \hline
AZT-22 & $R$ & 7827   & 300  & 18.96 & 0.03 &  Vega \\
AZT-22 & $R$ & 8151   & 300  & 19.04 & 0.03 &  Vega \\
AZT-22 & $R$ & 8476   & 300  & 19.07 & 0.03 &  Vega \\
AZT-22 & $R$ & 8800   & 300  & 19.09 & 0.03 &  Vega \\
AZT-22 & $R$ & 9125   & 300  & 19.08 & 0.03 &  Vega \\
AZT-22 & $R$ & 9449   & 300  & 19.16 & 0.03 &  Vega \\
AZT-22 & $R$ & 108002 & 3000 & 21.46 & 0.08 &  Vega \\
AZT-22 & $B$ & 102975 & 4200 & 22.59 & 0.09 &  Vega \\
AZT-22 & $R$ & 185045 & 4200 & 21.85 & 0.07 & Vega \\
AZT-22 & $R$ & 268488 & 5100 & 21.85 & 0.07 &  Vega \\ \hline
RC-35 (Kitab) & Clear & 17876  & 3600 & $>$18.30 & --    & Vega \\ \hline
Zeiss-1000 (TSHAO) & $R$ & 84755  & 3600 & $>$ 21.30 & --    & Vega \\
Zeiss-1000 (SAO)   & $R$ & 21078  & 3000 & $>$ 19.50 & --    & Vega \\
Zeiss-1000 (SAO)   & $R$ & 115706 & 3600 & $>$ 21.00 & --    & Vega \\
Zeiss-1000 (SAO)   & $R$ & 189626 & 4200 & 21.88 & 0.07 &  Vega \\
Zeiss-1000 (SAO)   & $R$ & 274672 & 5400 & 22.15 & 0.07 &  Vega \\
\enddata
\tablecomments{
All magnitudes are reported without correction for Galactic or host-galaxy extinction.
\swift-UVOT limiting magnitudes correspond to 3$\sigma$ limits.
\swift-UVOT, OSN, and BOOTES-6 magnitudes are reported in the AB system.
BOOTES-6 $R$-band magnitudes were converted from Vega to AB magnitudes using a zero-point offset of 0.21 mag.
GRB IKI-FuN magnitudes are reported in the Vega system.
}
\end{deluxetable*}

\begin{figure}[ht]
    \centering
        \includegraphics[width = 0.36\textwidth, angle=-90]{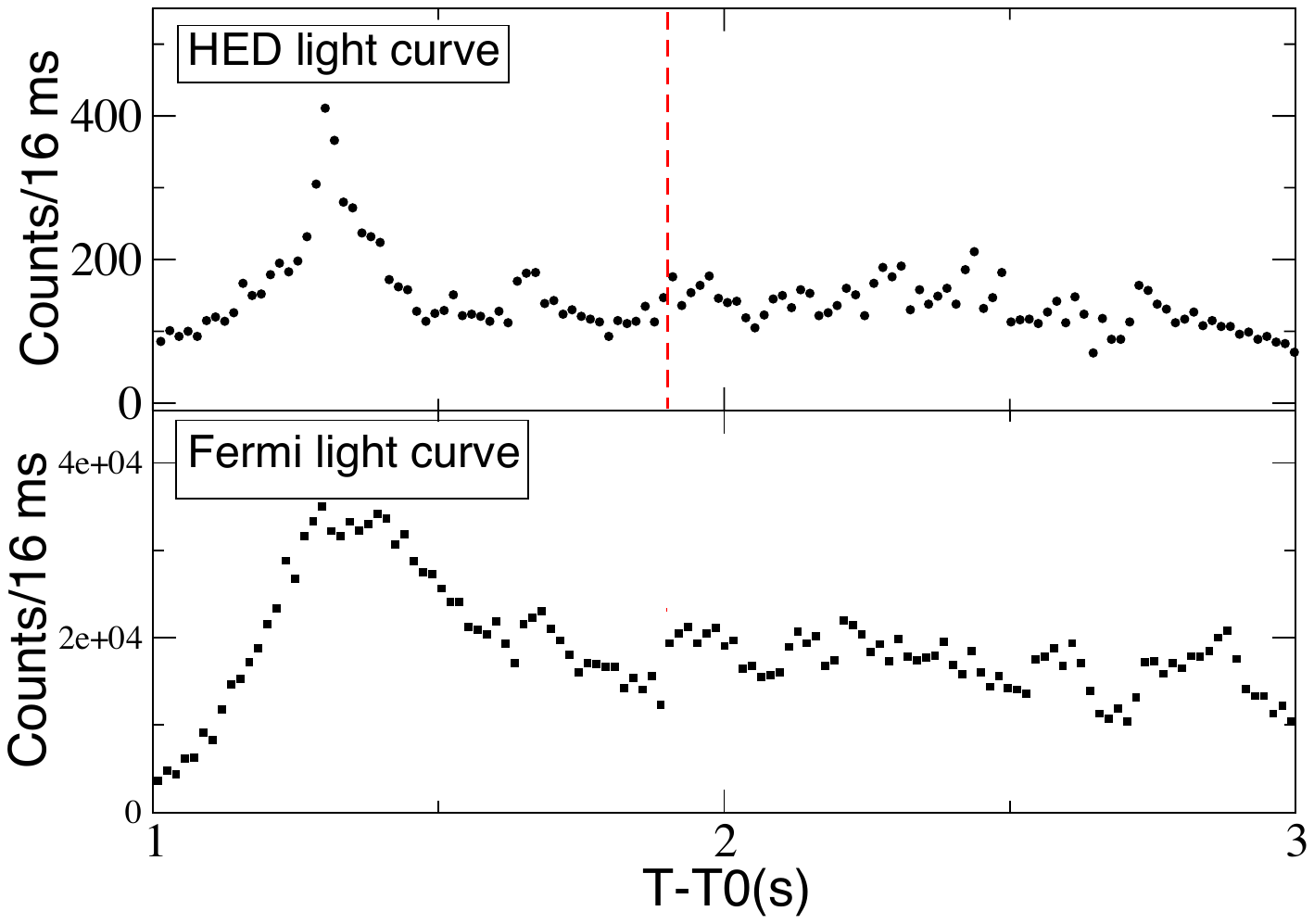}
                    \includegraphics[width = 0.36\textwidth, angle=-90]{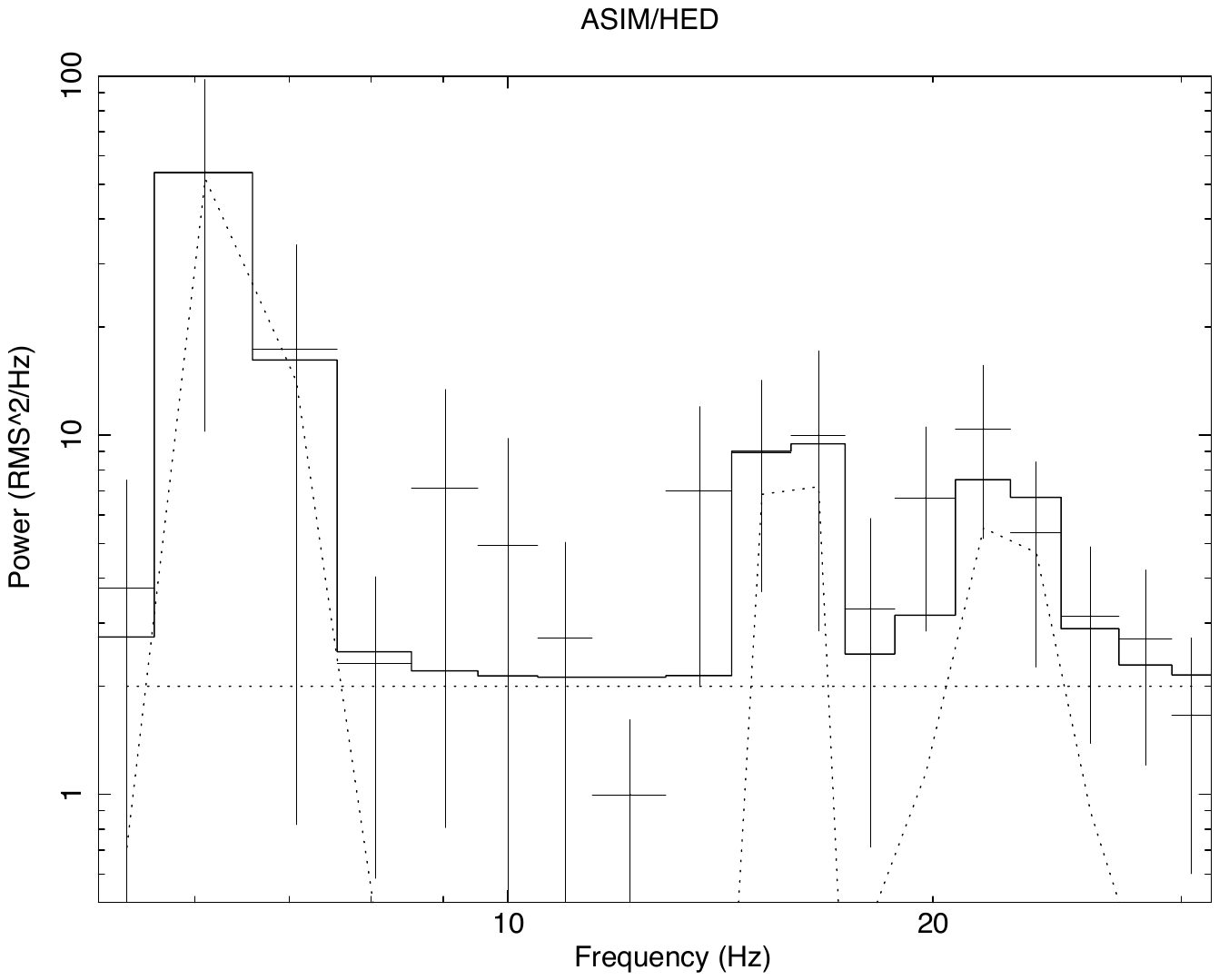}
            \includegraphics[width = 0.36\textwidth, angle=-90]{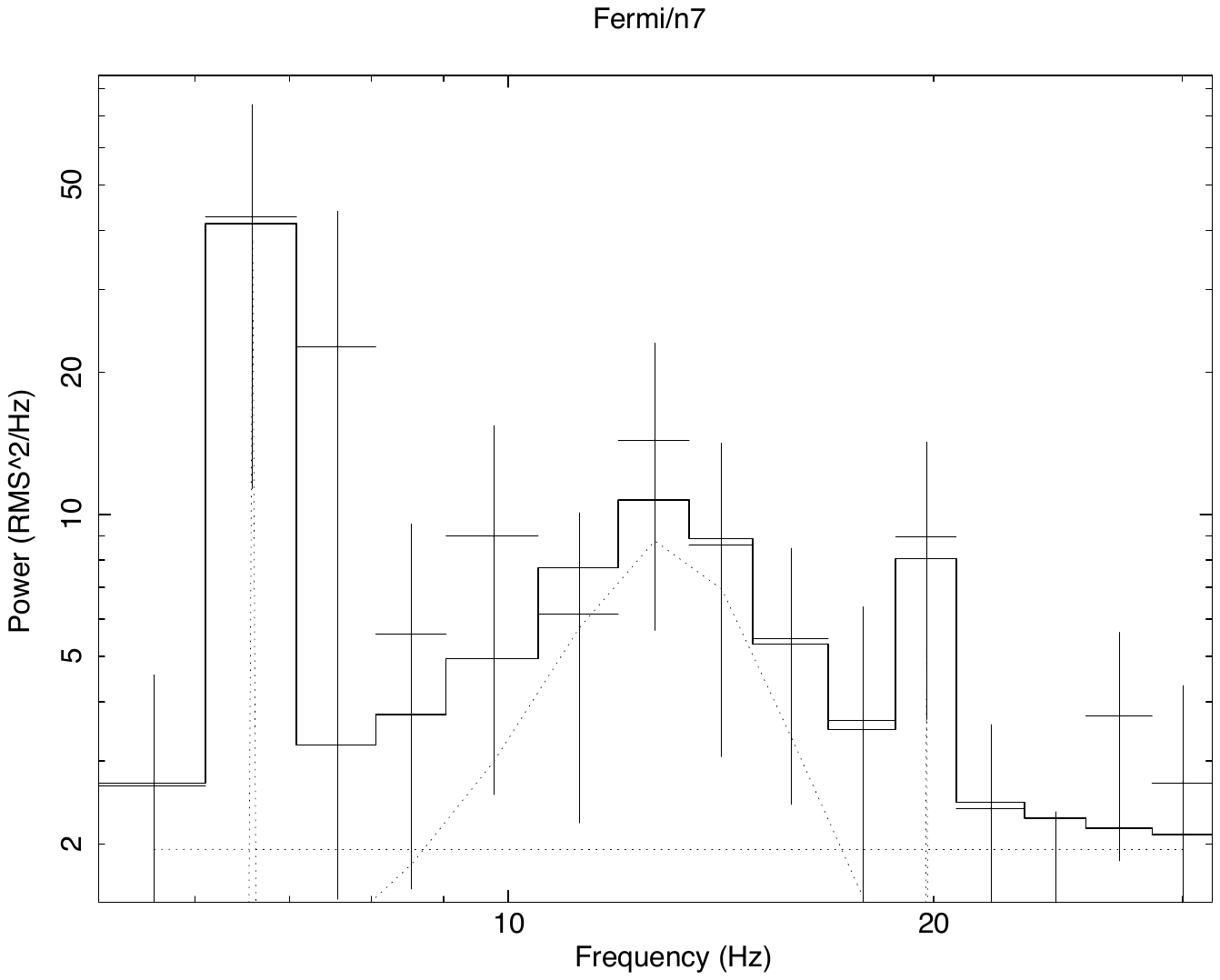}
\caption{Top: Barycentred ASIM (HED; 0.3-20\,MeV) and {\it Fermi} (GBM; 150\,keV-30\,MeV) light curves of \thisgrb showing the time-intervals (\fermiT+ 1.9 to \fermiT+ 3.0\,sec and \fermiT+ 1.0 to \fermiT+ 3.0\,sec) where the timing analysis was performed. Middle and Bottom: The PDS of the ASIM-HED (0.3-20\,MeV) and {\it Fermi}/NaI-7 (8-800\,keV) light curves of \thisgrb at the time intervals $1.9-3.0$\,sec and $1.0-3.0$\,sec where \fermiT is the {\it Fermi} trigger time.}
    \label{timing1}
\end{figure}

\begin{table*}
\caption{Power Density Spectra (PDS) timing analysis results of the light curve from \thisgrb using the \sw{lorentzian+lorentzian+lorentzian+powerlaw} function for {\it Fermi} (upper) and ASIM (lower) data. (T$_{\rm start}$ and T$_{\rm stop}$ have been referred with respect to the {\it Fermi} reference time (T$_{\rm 0, FERMI}$). The errors given are $1{\sigma}$.}
\label{Joint_timinganalysis_Table}
\begin{small}
\begin{center}
\begin{tabular}{|cc|cccccccc|c|} \hline
T$_{\rm start}$  & T$_{\rm stop}$   & \boldmath $\it \nu_{\rm QPO1}$ & \boldmath   $\rm FWHM_{\rm QPO1}$ & $\rm N_{\rm QPO1}$ & \boldmath $\it \nu_{\rm QPO2}$ & \boldmath $\rm FWHM_{\rm QPO2}$ & \boldmath $\rm N_{\rm QPO2}$  &  \boldmath $\it \Gamma_{\rm P}$  &  \boldmath $\rm N_{\rm P}$  & \bf $\chi^2$  (d.o.f.)   \\ 
(sec) &  (sec)  & ${\rm (Hz)}$ & ${\rm (Hz)}$ & & ${\rm (Hz)}$ & ${\rm (Hz)}$ &   & ${\rm (Hz)}$ & ${\rm (Hz)}$ &     \\ \hline
1.0 & 3.0 &  $6.2_{-0.06}^{+0.7}$   &   $(1{\pm}0.1){\rm E}-3$   & $37{\pm}30$  & \boldmath $19.5_{-1.0}^{+1.6}$  & $(1{\pm}0.1){\rm E}-3$  &  $10{\pm}8$  &  0\,(f) & $1.95{\pm}0.8$    &   3.5\,(5)        \\ 

1.9 & 3.0 &  $6.5_{-0.7}^{+0.10}$  &  $0.08_{-0.07}^{+0.4}$    & $61{\pm}50$      & \boldmath $21.7{\pm}1.0$  & $2.1_{-1.1}^{+2.1}$  & $33{\pm}20$ &  0\,(f)   &  $1.75{\pm}0.8$  &   5.5\,(8)        \\ \hline
\end{tabular}
\end{center}
\tablecomments{$\nu_\mathrm{QPO1}$, $\nu_\mathrm{QPO2}$ are the frequencies of the QPO features, while FWHM$_\mathrm{QPO1}$, FWHM$_\mathrm{QPO2}$ and N$_\mathrm{QPO1}$, N$_\mathrm{QPO2}$ are its full-width at half-maximum and normalization for both the QPOs, respectively. $\Gamma_{\rm P}$ and N$_{\rm P}$ represent the high-frequency power-law index and normalization, respectively. $\chi^2$  (d.o.f.): Chi-squared statistic value with degrees of freedom (d.o.f.) in parentheses, indicating the goodness of fit of the model to the data.
}
\end{small}
\end{table*}

\begin{figure*}
\centering
\includegraphics[width = .8\textwidth]{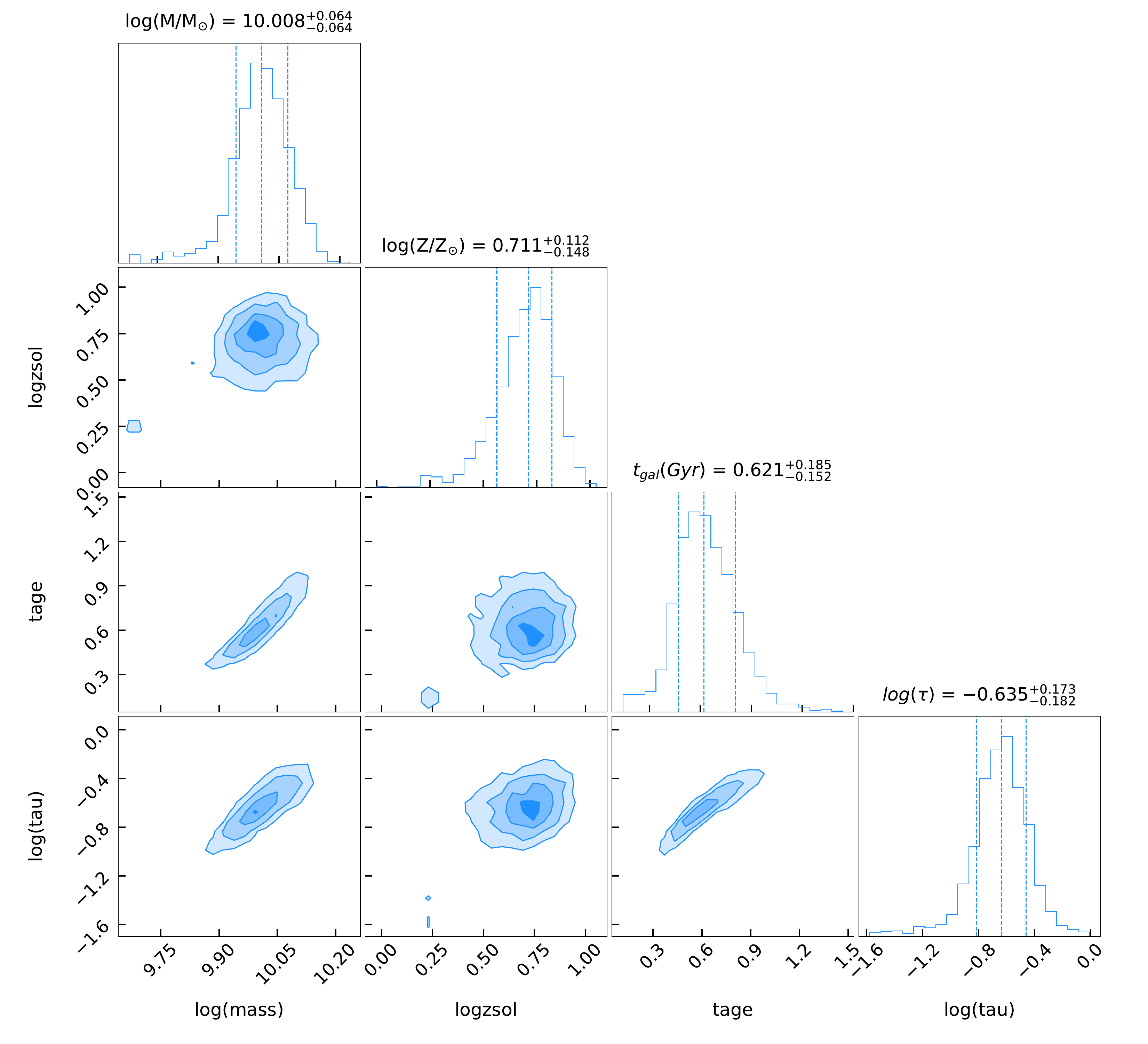}
\caption{Corner plot illustrating the posterior distributions of host SED model parameters obtained using \sw{Prospector}: stellar mass ($\log (\mathrm{M}/M_\odot) $) and star formation timescale ($\log (\tau/\mathrm{yr}) $), dust, from the MCMC fitting process. The contours represent 68\% and 95\% confidence intervals. The best-fit values are marked with blue lines.}
\label{host_sed_corner}
\end{figure*}

\begin{deluxetable*}{ccccc}[ht]
\tablecaption{Late-time \textit{Swift}/UVOT Observations for SN Search Associated with GRB~240825A. \label{tab:uvot_sn}}
\tablehead{
\colhead{Start Time (days, T$-$T$_0$)} & 
\colhead{Stop Time (days, T$-$T$_0$)} & 
\colhead{Exposure (sec)} & 
\colhead{Filter} & 
\colhead{Mag Limit (AB, 3$\sigma$)}
}
\startdata
24.73 & 33.14 & 6030  & U & 22.75 \\
3.94  & 33.14 & 7532  & U & 22.87 \\
4.73  & 33.14 & 11070 & U & 23.09 \\
2.62  & 33.14 & 14487 & U & 23.24 \\
1.57  & 33.14 & 18232 & U & 23.37 \\
0.47  & 33.14 & 18872 & U & 23.39 \\
\enddata
\tablecomments{All observations were performed in the UVOT $U$ band. Magnitudes represent 3$\sigma$ upper limits and have not been corrected for Galactic extinction.}
\end{deluxetable*}

\bibliography{GRB240825A}{}

@STRING{aap = "Astronomy \& Astrophysics"}

@STRING{apj = "The Astrophysical Journal"}

@STRING{mnras = "Monthly Notices of the Royal Astronomical Society"}

@ARTICLE{2024GCN.37314....1L,
       author = {{Laskar}, T. and {Alexander}, K.~D. and {Christy}, C. and {Pe{\~n}a}, C. and {Schroeder}, G. and {Berger}, E. and {Chornock}, R. and {Fong}, W. and {Margutti}, R. and {Schady}, P.},
        title = "{GRB 240825A: ALMA detection}",
      journal = {GRB Coordinates Network},
         year = 2024,
        month = aug,
       volume = {37314},
        pages = {1},
       adsurl = {https://ui.adsabs.harvard.edu/abs/2024GCN.37314....1L}
}

@ARTICLE{2024GCN.37322....1P,
       author = {{Pe{\~n}a}, C. and {Laskar}, T. and {Schroeder}, G. and {Alexander}, K.~D. and {Christy}, C. and {Berger}, E. and {Chornock}, R. and {Fong}, W. and {Margutti}, R. and {Schady}, P.},
        title = "{GRB 240825A: VLA detection}",
      journal = {GRB Coordinates Network},
         year = 2024,
        month = aug,
       volume = {37322},
        pages = {1},
       adsurl = {https://ui.adsabs.harvard.edu/abs/2024GCN.37322....1P}
}

@ARTICLE{2024GCN.37353....1P,
       author = {{Pe{\~n}a}, C. and {Schroeder}, G. and {Laskar}, T. and {Alexander}, K.~D. and {Christy}, C. and {Berger}, E. and {Chornock}, R. and {Fong}, W. and {Margutti}, R. and {Schady}, P. and {de Wet}, S. and {Groot}, P.},
        title = "{GRB 240825A: MeerKAT detection}",
      journal = {GRB Coordinates Network},
         year = 2024,
        month = aug,
       volume = {37353},
        pages = {1},
       adsurl = {https://ui.adsabs.harvard.edu/abs/2024GCN.37353....1P}
}

@ARTICLE{1993ApJ...413L.101K,
       author = {{Kouveliotou}, Chryssa and {Meegan}, Charles A. and {Fishman}, Gerald J. and {Bhat}, Narayana P. and {Briggs}, Michael S. and {Koshut}, Thomas M. and {Paciesas}, William S. and {Pendleton}, Geoffrey N.},
        title = "{Identification of Two Classes of Gamma-Ray Bursts}",
      journal = {\apjl},
         year = 1993,
        month = aug,
       volume = {413},
        pages = {L101},
          doi = {10.1086/186969},
       adsurl = {https://ui.adsabs.harvard.edu/abs/1993ApJ...413L.101K}
}

@ARTICLE{2002AJ....123.1111B,
       author = {{Bloom}, J.~S. and {Kulkarni}, S.~R. and {Djorgovski}, S.~G.},
        title = "{The Observed Offset Distribution of Gamma-Ray Bursts from Their Host Galaxies: A Robust Clue to the Nature of the Progenitors}",
      journal = {\aj},
         year = 2002,
        month = mar,
       volume = {123},
       number = {3},
        pages = {1111-1148},
          doi = {10.1086/338893},
archivePrefix = {arXiv},
       eprint = {astro-ph/0010176},
 primaryClass = {astro-ph},
       adsurl = {https://ui.adsabs.harvard.edu/abs/2002AJ....123.1111B}
}

@ARTICLE{2014ApJ...780...82U,
       author = {{Uhm}, Z. Lucas and {Zhang}, Bing},
        title = "{On the Non-existence of a Sharp Cooling Break in Gamma-Ray Burst Afterglow Spectra}",
      journal = {\apj},
         year = 2014,
        month = jan,
       volume = {780},
       number = {1},
          eid = {82},
        pages = {82},
          doi = {10.1088/0004-637X/780/1/82},
archivePrefix = {arXiv},
       eprint = {1301.0291},
 primaryClass = {astro-ph.HE},
       adsurl = {https://ui.adsabs.harvard.edu/abs/2014ApJ...780...82U}
}

@ARTICLE{2009MNRAS.394.2164V,
       author = {{van Eerten}, H.~J. and {Wijers}, R.~A.~M.~J.},
        title = "{Gamma-ray burst afterglow scaling coefficients for general density profiles}",
      journal = {\mnras},
         year = 2009,
        month = apr,
       volume = {394},
       number = {4},
        pages = {2164-2174},
          doi = {10.1111/j.1365-2966.2009.14482.x},
archivePrefix = {arXiv},
       eprint = {0810.2250},
 primaryClass = {astro-ph},
       adsurl = {https://ui.adsabs.harvard.edu/abs/2009MNRAS.394.2164V}
}

@ARTICLE{2022MNRAS.515.4890O,
       author = {{O'Connor}, B. and {Troja}, E. and {Dichiara}, S. and {Beniamini}, P. and {Cenko}, S.~B. and {Kouveliotou}, C. and {Gonz{\'a}lez}, J.~B. and {Durbak}, J. and {Gatkine}, P. and {Kutyrev}, A. and {Sakamoto}, T. and {S{\'a}nchez-Ram{\'\i}rez}, R. and {Veilleux}, S.},
        title = "{A deep survey of short GRB host galaxies over z 0-2: implications for offsets, redshifts, and environments}",
      journal = {\mnras},
         year = 2022,
        month = oct,
       volume = {515},
       number = {4},
        pages = {4890-4928},
          doi = {10.1093/mnras/stac1982},
archivePrefix = {arXiv},
       eprint = {2204.09059},
 primaryClass = {astro-ph.HE},
       adsurl = {https://ui.adsabs.harvard.edu/abs/2022MNRAS.515.4890O}
}

@ARTICLE{2006ARA&A..44..507W,
       author = {{Woosley}, S.~E. and {Bloom}, J.~S.},
        title = "{The Supernova Gamma-Ray Burst Connection}",
      journal = {\araa},
         year = 2006,
        month = sep,
       volume = {44},
       number = {1},
        pages = {507-556},
          doi = {10.1146/annurev.astro.43.072103.150558},
archivePrefix = {arXiv},
       eprint = {astro-ph/0609142},
 primaryClass = {astro-ph},
       adsurl = {https://ui.adsabs.harvard.edu/abs/2006ARA&A..44..507W}
}

@misc{2015ascl.soft04004B,
       author = {{Becker}, Andrew},
        title = "{HOTPANTS: High Order Transform of PSF ANd Template Subtraction}",
 howpublished = {Astrophysics Source Code Library, record ascl:1504.004},
         year = 2015,
        month = apr,
          eid = {ascl:1504.004},
       adsurl = {https://ui.adsabs.harvard.edu/abs/2015ascl.soft04004B}
}

@ARTICLE{1998Natur.395..670G,
       author = {{Galama}, T.~J. and {Vreeswijk}, P.~M. and {van Paradijs}, J. and {Kouveliotou}, C. and {Augusteijn}, T. and {B{\"o}hnhardt}, H. and {Brewer}, J.~P. and {Doublier}, V. and {Gonzalez}, J. -F. and {Leibundgut}, B. and {Lidman}, C. and {Hainaut}, O.~R. and {Patat}, F. and {Heise}, J. and {in't Zand}, J. and {Hurley}, K. and {Groot}, P.~J. and {Strom}, R.~G. and {Mazzali}, P.~A. and {Iwamoto}, K. and {Nomoto}, K. and {Umeda}, H. and {Nakamura}, T. and {Young}, T.~R. and {Suzuki}, T. and {Shigeyama}, T. and {Koshut}, T. and {Kippen}, M. and {Robinson}, C. and {de Wildt}, P. and {Wijers}, R.~A.~M.~J. and {Tanvir}, N. and {Greiner}, J. and {Pian}, E. and {Palazzi}, E. and {Frontera}, F. and {Masetti}, N. and {Nicastro}, L. and {Feroci}, M. and {Costa}, E. and {Piro}, L. and {Peterson}, B.~A. and {Tinney}, C. and {Boyle}, B. and {Cannon}, R. and {Stathakis}, R. and {Sadler}, E. and {Begam}, M.~C. and {Ianna}, P.},
        title = "{An unusual supernova in the error box of the {\ensuremath{\gamma}}-ray burst of 25 April 1998}",
      journal = {\nat},
         year = 1998,
        month = oct,
       volume = {395},
       number = {6703},
        pages = {670-672},
          doi = {10.1038/27150},
archivePrefix = {arXiv},
       eprint = {astro-ph/9806175},
 primaryClass = {astro-ph},
       adsurl = {https://ui.adsabs.harvard.edu/abs/1998Natur.395..670G}
}

@ARTICLE{2006Natur.442.1008C,
       author = {{Campana}, S. and {Mangano}, V. and {Blustin}, A.~J. and {Brown}, P. and {Burrows}, D.~N. and {Chincarini}, G. and {Cummings}, J.~R. and {Cusumano}, G. and {Della Valle}, M. and {Malesani}, D. and {M{\'e}sz{\'a}ros}, P. and {Nousek}, J.~A. and {Page}, M. and {Sakamoto}, T. and {Waxman}, E. and {Zhang}, B. and {Dai}, Z.~G. and {Gehrels}, N. and {Immler}, S. and {Marshall}, F.~E. and {Mason}, K.~O. and {Moretti}, A. and {O'Brien}, P.~T. and {Osborne}, J.~P. and {Page}, K.~L. and {Romano}, P. and {Roming}, P.~W.~A. and {Tagliaferri}, G. and {Cominsky}, L.~R. and {Giommi}, P. and {Godet}, O. and {Kennea}, J.~A. and {Krimm}, H. and {Angelini}, L. and {Barthelmy}, S.~D. and {Boyd}, P.~T. and {Palmer}, D.~M. and {Wells}, A.~A. and {White}, N.~E.},
        title = "{The association of GRB 060218 with a supernova and the evolution of the shock wave}",
      journal = {\nat},
         year = 2006,
        month = aug,
       volume = {442},
       number = {7106},
        pages = {1008-1010},
          doi = {10.1038/nature04892},
archivePrefix = {arXiv},
       eprint = {astro-ph/0603279},
 primaryClass = {astro-ph},
       adsurl = {https://ui.adsabs.harvard.edu/abs/2006Natur.442.1008C}
}

@ARTICLE{2019Natur.565..324I,
       author = {{Izzo}, L. and {de Ugarte Postigo}, A. and {Maeda}, K. and {Th{\"o}ne}, C.~C. and {Kann}, D.~A. and {Della Valle}, M. and {Sagues Carracedo}, A. and {Micha{\l}owski}, M.~J. and {Schady}, P. and {Schmidl}, S. and {Selsing}, J. and {Starling}, R.~L.~C. and {Suzuki}, A. and {Bensch}, K. and {Bolmer}, J. and {Campana}, S. and {Cano}, Z. and {Covino}, S. and {Fynbo}, J.~P.~U. and {Hartmann}, D.~H. and {Heintz}, K.~E. and {Hjorth}, J. and {Japelj}, J. and {Kami{\'n}ski}, K. and {Kaper}, L. and {Kouveliotou}, C. and {Kru{\.Z}y{\'n}ski}, M. and {Kwiatkowski}, T. and {Leloudas}, G. and {Levan}, A.~J. and {Malesani}, D.~B. and {Micha{\l}owski}, T. and {Piranomonte}, S. and {Pugliese}, G. and {Rossi}, A. and {S{\'a}nchez-Ram{\'\i}rez}, R. and {Schulze}, S. and {Steeghs}, D. and {Tanvir}, N.~R. and {Ulaczyk}, K. and {Vergani}, S.~D. and {Wiersema}, K.},
        title = "{Signatures of a jet cocoon in early spectra of a supernova associated with a {\ensuremath{\gamma}}-ray burst}",
      journal = {\nat},
         year = 2019,
        month = jan,
       volume = {565},
       number = {7739},
        pages = {324-327},
          doi = {10.1038/s41586-018-0826-3},
archivePrefix = {arXiv},
       eprint = {1901.05500},
 primaryClass = {astro-ph.HE},
       adsurl = {https://ui.adsabs.harvard.edu/abs/2019Natur.565..324I}
}

@ARTICLE{2021A&A...646A..50H,
       author = {{Hu}, Y. -D. and {Castro-Tirado}, A.~J. and {Kumar}, A. and {Gupta}, R. and {Valeev}, A.~F. and {Pandey}, S.~B. and {Kann}, D.~A. and {Castell{\'o}n}, A. and {Agudo}, I. and {Aryan}, A. and {Caballero-Garc{\'\i}a}, M.~D. and {Guziy}, S. and {Martin-Carrillo}, A. and {Oates}, S.~R. and {Pian}, E. and {S{\'a}nchez-Ram{\'\i}rez}, R. and {Sokolov}, V.~V. and {Zhang}, B. -B.},
        title = "{10.4 m GTC observations of the nearby VHE-detected GRB 190829A/SN 2019oyw}",
      journal = {\aap},
         year = 2021,
        month = feb,
       volume = {646},
          eid = {A50},
        pages = {A50},
          doi = {10.1051/0004-6361/202039349},
archivePrefix = {arXiv},
       eprint = {2009.04021},
 primaryClass = {astro-ph.HE},
       adsurl = {https://ui.adsabs.harvard.edu/abs/2021A&A...646A..50H}
}

@ARTICLE{2024A&A...692A...3S,
       author = {{S{\'a}nchez-Ram{\'\i}rez}, R. and {Lang}, R.~G. and {Pozanenko}, A. and {Mart{\'\i}nez-Huerta}, H. and {Hu}, Y. -D. and {Pandey}, S.~B. and {Gupta}, R. and {Ror}, A.~K. and {Zhang}, B. -B. and {Caballero-Garc{\'\i}a}, M.~D. and {Oates}, S.~R. and {P{\'e}rez-Garc{\'\i}a}, I. and {Guziy}, S. and {Fern{\'a}ndez-Garc{\'\i}a}, E.~J. and {Wu}, S. -Y. and {Almeida}, L. and {Aryan}, A. and {Belkin}, S. and {Bom}, C.~R. and {Butner}, M. and {Burkhonov}, O. and {Carrasco-Garc{\'\i}a}, I. and {Castell{\'o}n}, A. and {Castro Tirado}, M.~A. and {Chelovekov}, I. and {Egamberdiyev}, Sh. A. and {Garc{\'\i}a-Benito}, R. and {Garc{\'\i}a Gonz{\'a}lez}, S.~A. and {Grebenev}, S. and {Kilpatrick}, C.~D. and {Klunko}, E. and {Makler}, M. and {Minaev}, P. and {Mkrtchyan}, A. and {Moskvitin}, A. and {Navarete}, F. and {Novichonok}, A. and {Pankov}, N. and {Passas-Varo}, M. and {P{\'e}rez del Pulgar}, C. and {Reina Terol}, A. and {Smith}, J.~A. and {Tinyanont}, S. and {Tucker}, D.~L. and {Uklein}, R. and {Volnova}, A. and {Wiesner}, M.~P. and {Gritsevich}, M. and {Castro-Tirado}, A.~J.},
        title = "{Early photometric and spectroscopic observations of the extraordinarily bright INTEGRAL-detected GRB 221009A}",
      journal = {\aap},
         year = 2024,
        month = dec,
       volume = {692},
          eid = {A3},
        pages = {A3},
          doi = {10.1051/0004-6361/202449783},
       adsurl = {https://ui.adsabs.harvard.edu/abs/2024A&A...692A...3S}
}

@ARTICLE{2022MNRAS.511.1694G,
       author = {{Gupta}, Rahul and {Gupta}, S. and {Chattopadhyay}, T. and {Lipunov}, V. and {Castro-Tirado}, A.~J. and {Bhattacharya}, D. and {Pandey}, S.~B. and {Oates}, S.~R. and {Kumar}, Amit and {Hu}, Y. -D. and {Valeev}, A.~F. and {Minaev}, P. Yu and {Kumar}, H. and {Vinko}, J. and {Dimple} and {Sharma}, V. and {Aryan}, A. and {Castell{\'o}n}, A. and {Gabovich}, A. and {Moskvitin}, A. and {Ordasi}, A. and {P{\'a}l}, A. and {Pozanenko}, A. and {Zhang}, B. -B. and {Kumar}, B. and {Svinkin}, D. and {Saraogi}, D. and {Vlasenko}, D. and {Fern{\'a}ndez-Garc{\'\i}a}, E. and {Gorbovskoy}, E. and {Anupama}, G.~C. and {Misra}, K. and {S{\'a}rneczky}, K. and {Kriskovics}, L. and {Castro-Tirado}, M. {\'A}. and {Caballero-Garc{\'\i}a}, M.~D. and {Tiurina}, N. and {Balanutsa}, P. and {Lopez}, R.~R. and {S{\'a}nchez-Ram{\'\i}rez}, R. and {Szak{\'a}ts}, R. and {Belkin}, S. and {Guziy}, S. and {Iyyani}, S. and {Tiwari}, S.~N. and {Vadawale}, Santosh V. and {Sun}, T. and {Bhalerao}, V. and {Kornilov}, V. and {Sokolov}, V.~V.},
        title = "{Probing into emission mechanisms of GRB 190530A using time-resolved spectra and polarization studies: synchrotron origin?}",
      journal = {\mnras},
         year = 2022,
        month = apr,
       volume = {511},
       number = {2},
        pages = {1694-1713},
          doi = {10.1093/mnras/stac015},
archivePrefix = {arXiv},
       eprint = {2201.01167},
 primaryClass = {astro-ph.HE},
       adsurl = {https://ui.adsabs.harvard.edu/abs/2022MNRAS.511.1694G}
}

@ARTICLE{2005SSRv..120...95R,
       author = {{Roming}, Peter W.~A. and {Kennedy}, Thomas E. and {Mason}, Keith O. and {Nousek}, John A. and {Ahr}, Lindy and {Bingham}, Richard E. and {Broos}, Patrick S. and {Carter}, Mary J. and {Hancock}, Barry K. and {Huckle}, Howard E. and {Hunsberger}, S.~D. and {Kawakami}, Hajime and {Killough}, Ronnie and {Koch}, T. Scott and {McLelland}, Michael K. and {Smith}, Kelly and {Smith}, Philip J. and {Soto}, Juan Carlos and {Boyd}, Patricia T. and {Breeveld}, Alice A. and {Holland}, Stephen T. and {Ivanushkina}, Mariya and {Pryzby}, Michael S. and {Still}, Martin D. and {Stock}, Joseph},
        title = "{The Swift Ultra-Violet/Optical Telescope}",
      journal = {\ssr},
         year = 2005,
        month = oct,
       volume = {120},
       number = {3-4},
        pages = {95-142},
          doi = {10.1007/s11214-005-5095-4},
archivePrefix = {arXiv},
       eprint = {astro-ph/0507413},
 primaryClass = {astro-ph},
       adsurl = {https://ui.adsabs.harvard.edu/abs/2005SSRv..120...95R}
}

@ARTICLE{2024GCN.37293....1M,
       author = {{Martin-Carrillo}, A. and {Schneider}, B. and {Pugliese}, G. and {Izzo}, L. and {Malesani}, D.~B. and {Saccardi}, A. and {Laskar}, T. and {Agui Fernandez}, J.~F. and {Vergani}, S.~D. and {Stargate Collaboration}},
        title = "{GRB 240825A: VLT/X-shooter redshift}",
      journal = {GRB Coordinates Network},
         year = 2024,
        month = aug,
       volume = {37293},
        pages = {1},
       adsurl = {https://ui.adsabs.harvard.edu/abs/2024GCN.37293....1M}
}

@ARTICLE{2024GCN.37287....1I,
       author = {{Izzo}, L. and {Malesani}, D.~B.},
        title = "{GRB 240825A: LCO optical observations}",
      journal = {GRB Coordinates Network},
         year = 2024,
        month = aug,
       volume = {37287},
        pages = {1},
       adsurl = {https://ui.adsabs.harvard.edu/abs/2024GCN.37287....1I}
}

@ARTICLE{2024GCN.37306....1W,
       author = {{Wang}, B. -T. and {Song}, F. -F. and {Li}, R. -Z. and {Mao}, J. and {Xin}, Y. -X. and {Bai}, J. -M.},
        title = "{GRB 240825A: GMG Continued Optical Observation on the Second Night}",
      journal = {GRB Coordinates Network},
         year = 2024,
        month = aug,
       volume = {37306},
        pages = {1},
       adsurl = {https://ui.adsabs.harvard.edu/abs/2024GCN.37306....1W}
}

@ARTICLE{2024GCN.37335....1G,
       author = {{Romanov}, Filipp Dmitrievich},
        title = "{GRB 240825A: iTelescope optical observations}",
      journal = {GRB Coordinates Network},
         year = 2024,
        month = aug,
       volume = {37335},
        pages = {1},
       adsurl = {https://ui.adsabs.harvard.edu/abs/2024GCN.37335....1G}
}

@ARTICLE{2024GCN.37300....1L,
       author = {{Le Floc'h}, E. and {Adami}, C. and {Schneider}, B. and {Saccardi}, A. and {Basa}, S. and {Dennefeld}, M. and {Sch{\"u}ssler}, F. and {Mistral Grb Collaboration}},
        title = "{GRB 240825A : MISTRAL/T193 OHP optical follow-up of the afterglow}",
      journal = {GRB Coordinates Network},
         year = 2024,
        month = aug,
       volume = {37300},
        pages = {1},
       adsurl = {https://ui.adsabs.harvard.edu/abs/2024GCN.37300....1L}
}

@ARTICLE{2024GCN.37295....1B,
       author = {{Brivio}, R. and {Ferro}, M. and {D'Avanzo}, P. and {Covino}, S. and {Fugazza}, D. and {REM Team}},
        title = "{GRB 240825A: REM detection of the optical/NIR afterglow}",
      journal = {GRB Coordinates Network},
         year = 2024,
        month = aug,
       volume = {37295},
        pages = {1},
       adsurl = {https://ui.adsabs.harvard.edu/abs/2024GCN.37295....1B}
}

@ARTICLE{2024GCN.37373....1S,
       author = {{SVOM/C-GFT Team} and {Chao Wu} and {Kang}, Zhe and {Xin}, Liping and {Han}, Xuhui and {Zhang}, Pinpin and {Lu}, Xiaomeng and {Li}, Zhenwei and {Lv}, You and {Zhang}, Ruosong and {Xiao}, Yujie and {SVOM JSWG} and {Wei}, Jian-Yan and {Cordier}, Bertrand and {Zhang}, Shuang-Nan and {Basa}, St{\'e}phane and {Att{\'e}ia}, Jean-Luc and {Claret}, Arnaud and {Dai}, Zi-Gao and {Daigne}, Fr{\'e}d{\'e}ric and {Deng}, Jin-Song and {Goldwurm}, Andrea and {G{\"o}tz}, Diego and {Han}, Xu-Hui and {Lachaud}, Cyril and {Liang}, En-Wei and {Qiu}, Yu-Lei and {Vergani}, Susanna and {Wang}, Jing and {Wu}, Chao and {Xin}, Li-Ping and {Zhang}, Bing and {SVOM team}},
        title = "{GRB 240825A: Magnitude Correction in SVOM/C-GFT Optical Observations}",
      journal = {GRB Coordinates Network},
         year = 2024,
        month = sep,
       volume = {37373},
        pages = {1},
       adsurl = {https://ui.adsabs.harvard.edu/abs/2024GCN.37373....1S}
}

@ARTICLE{2025ApJ...979...38C,
       author = {{Cheng}, Yehao and {Pan}, Yu and {Yang}, Yuan-Pei and {Zhang}, Jinghua and {Du}, Guowang and {Fang}, Yuan and {Kumar}, Brajesh and {Guo}, Helong and {Er}, Xinzhong and {Chen}, Xinlei and {Liu}, Chenxu and {Wang}, Tao and {Qin}, Zhenfei and {Jin}, Yicheng and {Zou}, Xingzhu and {Han}, Xuhui and {Zhang}, Pinpin and {Xin}, Liping and {Wu}, Chao and {Lian}, Jianhui and {Liu}, Xiangkun and {Liu}, Xiaowei},
        title = "{Simultaneous Multiband Photometry of the Early Optical Afterglow of GRB 240825A with Mephisto}",
      journal = {\apj},
         year = 2025,
        month = jan,
       volume = {979},
       number = {1},
          eid = {38},
        pages = {38},
          doi = {10.3847/1538-4357/ad9ea1},
archivePrefix = {arXiv},
       eprint = {2409.14716},
 primaryClass = {astro-ph.HE},
       adsurl = {https://ui.adsabs.harvard.edu/abs/2025ApJ...979...38C}
}

@ARTICLE{2019arXiv190612178N,
       author = {{Neubert}, Torsten and {{\O}stgaard}, Nikolai and {Reglero}, Victor and {Blanc}, Elisabeth and {Chanrion}, Olivier and {Oxborrow}, Carol Anne and {Orr}, Astrid and {Tacconi}, Matteo and {Hartnack}, Ole and {Bhanderi}, Dan D.~V.},
        title = "{The ASIM Mission on the International Space Station}",
      journal = {arXiv e-prints},
         year = 2019,
        month = jun,
          eid = {arXiv:1906.12178},
        pages = {arXiv:1906.12178},
          doi = {10.48550/arXiv.1906.12178},
archivePrefix = {arXiv},
       eprint = {1906.12178},
 primaryClass = {astro-ph.IM},
       adsurl = {https://ui.adsabs.harvard.edu/abs/2019arXiv190612178N}
}

@ARTICLE{2024ApJ...972..166G,
       author = {{Gupta}, Rahul and {Pandey}, S.~B. and {Gupta}, S. and {Chattopadhayay}, T. and {Bhattacharya}, D. and {Bhalerao}, V. and {Castro-Tirado}, A.~J. and {Valeev}, A. and {Ror}, A.~K. and {Sharma}, V. and {Racusin}, J. and {Aryan}, A. and {Iyyani}, S. and {Vadawale}, S.},
        title = "{A Detailed Time-resolved and Energy-resolved Spectro-polarimetric Study of Bright Gamma-Ray Bursts Detected by AstroSat CZTI in Its First Year of Operation}",
      journal = {\apj},
         year = 2024,
        month = sep,
       volume = {972},
       number = {2},
          eid = {166},
        pages = {166},
          doi = {10.3847/1538-4357/ad5a92},
archivePrefix = {arXiv},
       eprint = {2406.13755},
 primaryClass = {astro-ph.HE},
       adsurl = {https://ui.adsabs.harvard.edu/abs/2024ApJ...972..166G}
}

@ARTICLE{2010Natur.463..513S,
       author = {{Soderberg}, A.~M. and {Chakraborti}, S. and {Pignata}, G. and {Chevalier}, R.~A. and {Chandra}, P. and {Ray}, A. and {Wieringa}, M.~H. and {Copete}, A. and {Chaplin}, V. and {Connaughton}, V. and {Barthelmy}, S.~D. and {Bietenholz}, M.~F. and {Chugai}, N. and {Stritzinger}, M.~D. and {Hamuy}, M. and {Fransson}, C. and {Fox}, O. and {Levesque}, E.~M. and {Grindlay}, J.~E. and {Challis}, P. and {Foley}, R.~J. and {Kirshner}, R.~P. and {Milne}, P.~A. and {Torres}, M.~A.~P.},
        title = "{A relativistic type Ibc supernova without a detected {\ensuremath{\gamma}}-ray burst}",
      journal = {\nat},
         year = 2010,
        month = jan,
       volume = {463},
       number = {7280},
        pages = {513-515},
          doi = {10.1038/nature08714},
archivePrefix = {arXiv},
       eprint = {0908.2817},
 primaryClass = {astro-ph.HE},
       adsurl = {https://ui.adsabs.harvard.edu/abs/2010Natur.463..513S}
}

@ARTICLE{2017A&A...605A.107C,
       author = {{Cano}, Z. and {Izzo}, L. and {de Ugarte Postigo}, A. and {Th{\"o}ne}, C.~C. and {Kr{\"u}hler}, T. and {Heintz}, K.~E. and {Malesani}, D. and {Geier}, S. and {Fuentes}, C. and {Chen}, T. -W. and {Covino}, S. and {D'Elia}, V. and {Fynbo}, J.~P.~U. and {Goldoni}, P. and {Gomboc}, A. and {Hjorth}, J. and {Jakobsson}, P. and {Kann}, D.~A. and {Milvang-Jensen}, B. and {Pugliese}, G. and {S{\'a}nchez-Ram{\'\i}rez}, R. and {Schulze}, S. and {Sollerman}, J. and {Tanvir}, N.~R. and {Wiersema}, K.},
        title = "{GRB 161219B/SN 2016jca: A low-redshift gamma-ray burst supernova powered by radioactive heating}",
      journal = {\aap},
         year = 2017,
        month = sep,
       volume = {605},
          eid = {A107},
        pages = {A107},
          doi = {10.1051/0004-6361/201731005},
archivePrefix = {arXiv},
       eprint = {1704.05401},
 primaryClass = {astro-ph.HE},
       adsurl = {https://ui.adsabs.harvard.edu/abs/2017A&A...605A.107C}
}

@ARTICLE{schady07,
   author = {{Schady}, P. and {Mason}, K.~O. and {Page}, M.~J. and {De Pasquale}, M. and 
	{Morris}, D.~C. and {Romano}, P. and {Roming}, P.~W.~A. and 
	{Immler}, S. and {vanden Berk}, D.~E.},
    title = "{Dust and gas in the local environments of gamma-ray bursts}",
  journal = mnras,
   eprint = {arXiv:astro-ph/0702122},
     year = 2007,
    month = may,
   volume = 377,
    pages = {273-284},
      doi = {10.1111/j.1365-2966.2007.11592.x},
   adsurl = {http://adsabs.harvard.edu/abs/2007MNRAS.377..273S}
}

@ARTICLE{sch10,
   author = {{Schady}, P. and {Page}, M.~J. and {Oates}, S.~R. and {Still}, M. and 
	{de Pasquale}, M. and {Dwelly}, T. and {Kuin}, N.~P.~M. and 
	{Holland}, S.~T. and {Marshall}, F.~E. and {Roming}, P.~W.~A.
	},
    title = "{Dust and metal column densities in gamma-ray burst host galaxies}",
  journal = {\mnras},
archivePrefix = "arXiv",
   eprint = {0910.2590},
 primaryClass = "astro-ph.CO",
     year = 2010,
    month = feb,
   volume = 401,
    pages = {2773-2792},
      doi = {10.1111/j.1365-2966.2009.15861.x},
   adsurl = {http://adsabs.harvard.edu/abs/2010MNRAS.401.2773S}
}

@ARTICLE{kalberla,
   author = {{Kalberla}, P.~M.~W. and {Burton}, W.~B. and {Hartmann}, D. and 
	{Arnal}, E.~M. and {Bajaja}, E. and {Morras}, R. and {P{\"o}ppel}, W.~G.~L.
	},
    title = "{The Leiden/Argentine/Bonn (LAB) Survey of Galactic HI. Final data release of the combined LDS and IAR surveys with improved stray-radiation corrections}",
  journal = aap,
   eprint = {arXiv:astro-ph/0504140},
     year = 2005,
    month = sep,
   volume = 440,
    pages = {775-782},
      doi = {10.1051/0004-6361:20041864},
   adsurl = {http://adsabs.harvard.edu/abs/2005A%26A...440..775K}
}

@ARTICLE{2026arXiv260628267G,
       author = {{Gupta}, Rahul and {Racusin}, Judith and {Oganesyan}, Gor and {Gill}, Ramandeep and {Karpov}, Sergey and {Oates}, Samantha R. and {Gritsevich}, Maria and {Castro-Tirado}, Alberto J. and {Lipunov}, Vladimir and {Gompertz}, Benjamin P. and {Dainotti}, Maria Giovanna and {Cenko}, S. Bradley and {Zhang}, Bing and {Komesh}, Toktarkhan and {Jel{\'\i}nek}, Martin and {Basa}, St{\'e}phane and {Sakamoto}, Takanori and {Akerlof}, Carl W. and {Carrillo}, Antonio Martin and {Shilling}, Sam and {Seeb}, Ryan and {Pandey}, Shashi B. and {Bhalerao}, Varun and {Caballero-Garcia}, M.~D. and {Panchenko}, Ivan and {Balanutsa}, Pavel and {Kuznetsov}, Artem and {Zhirkov}, Kirill and {Tyurina}, Natalia},
        title = "{Early Optical Follow-up of Gamma-Ray Bursts: The Critical Role of Robotic Telescopes}",
      journal = {arXiv e-prints},
         year = 2026,
        month = jun,
          eid = {arXiv:2606.28267},
        pages = {arXiv:2606.28267},
          doi = {10.48550/arXiv.2606.28267},
archivePrefix = {arXiv},
       eprint = {2606.28267},
 primaryClass = {astro-ph.HE},
       adsurl = {https://ui.adsabs.harvard.edu/abs/2026arXiv260628267G}
}

@ARTICLE{2024Natur.626..742Y,
       author = {{Yang}, Yu-Han and {Troja}, Eleonora and {O'Connor}, Brendan and {Fryer}, Chris L. and {Im}, Myungshin and {Durbak}, Joe and {Paek}, Gregory S.~H. and {Ricci}, Roberto and {Bom}, Cl{\'e}cio R. and {Gillanders}, James H. and {Castro-Tirado}, Alberto J. and {Peng}, Zong-Kai and {Dichiara}, Simone and {Ryan}, Geoffrey and {van Eerten}, Hendrik and {Dai}, Zi-Gao and {Chang}, Seo-Won and {Choi}, Hyeonho and {De}, Kishalay and {Hu}, Youdong and {Kilpatrick}, Charles D. and {Kutyrev}, Alexander and {Jeong}, Mankeun and {Lee}, Chung-Uk and {Makler}, Martin and {Navarete}, Felipe and {P{\'e}rez-Garc{\'\i}a}, Ignacio},
        title = "{A lanthanide-rich kilonova in the aftermath of a long gamma-ray burst}",
      journal = {\nat},
         year = 2024,
        month = feb,
       volume = {626},
       number = {8000},
        pages = {742-745},
          doi = {10.1038/s41586-023-06979-5},
archivePrefix = {arXiv},
       eprint = {2308.00638},
 primaryClass = {astro-ph.HE},
       adsurl = {https://ui.adsabs.harvard.edu/abs/2024Natur.626..742Y}
}

@ARTICLE{2024arXiv240904871G,
       author = {{Gupta}, Rahul and {Ror}, A.~K. and {Pandey}, S.~B. and {Racusin}, J. and {Moss}, M. and {Aryan}, A. and {Klingler}, N. and {Castro-Tirado}, A.~J.},
        title = "{An Intermediate Luminosity GRB 210210A: The early onset of the external forward shock in the X-ray?}",
      journal = {arXiv e-prints},
         year = 2024,
        month = sep,
          eid = {arXiv:2409.04871},
        pages = {arXiv:2409.04871},
          doi = {10.48550/arXiv.2409.04871},
archivePrefix = {arXiv},
       eprint = {2409.04871},
 primaryClass = {astro-ph.HE},
       adsurl = {https://ui.adsabs.harvard.edu/abs/2024arXiv240904871G}
}

@ARTICLE{2021MNRAS.505.4086G,
       author = {{Gupta}, Rahul and {Oates}, S.~R. and {Pandey}, S.~B. and {Castro-Tirado}, A.~J. and {Joshi}, Jagdish C. and {Hu}, Y. -D. and {Valeev}, A.~F. and {Zhang}, B.~B. and {Zhang}, Z. and {Kumar}, Amit and {Aryan}, A. and {Lien}, A. and {Kumar}, B. and {Cui}, Ch and {Wang}, Ch and {Dimple} and {Bhattacharya}, D. and {Sonbas}, E. and {Bai}, J. and {Tello}, J.~C. and {Gorosabel}, J. and {Castro Cer{\'o}n}, J.~M. and {Porto}, J.~R.~F. and {Misra}, K. and {De Pasquale}, M. and {Caballero-Garc{\'\i}a}, M.~D. and {Jel{\'\i}nek}, M. and {Kub{\'a}nek}, P. and {Minaev}, P. Yu and {Cunniffe}, R. and {S{\'a}nchez-Ram{\'\i}rez}, R. and {Guziy}, S. and {Jeong}, S. and {Tiwari}, S.~N. and {Razzaque}, S. and {Bhalerao}, V. and {Pintado}, V.~C. and {Sokolov}, V.~V. and {Zhao}, X. and {Fan}, Y. and {Xin}, Y.},
        title = "{GRB 140102A: insight into prompt spectral evolution and early optical afterglow emission}",
      journal = {\mnras},
         year = 2021,
        month = aug,
       volume = {505},
       number = {3},
        pages = {4086-4105},
          doi = {10.1093/mnras/stab1573},
archivePrefix = {arXiv},
       eprint = {2105.13145},
 primaryClass = {astro-ph.HE},
       adsurl = {https://ui.adsabs.harvard.edu/abs/2021MNRAS.505.4086G}
}

@ARTICLE{2005ApJ...628..315Z,
       author = {{Zhang}, Bing and {Kobayashi}, Shiho},
        title = "{Gamma-Ray Burst Early Afterglows: Reverse Shock Emission from an Arbitrarily Magnetized Ejecta}",
      journal = {\apj},
         year = 2005,
        month = jul,
       volume = {628},
       number = {1},
        pages = {315-334},
          doi = {10.1086/429787},
archivePrefix = {arXiv},
       eprint = {astro-ph/0404140},
 primaryClass = {astro-ph},
       adsurl = {https://ui.adsabs.harvard.edu/abs/2005ApJ...628..315Z}
}

@ARTICLE{1998ApJ...497L..17S,
       author = {{Sari}, Re'em and {Piran}, Tsvi and {Narayan}, Ramesh},
        title = "{Spectra and Light Curves of Gamma-Ray Burst Afterglows}",
      journal = {\apjl},
         year = 1998,
        month = apr,
       volume = {497},
       number = {1},
        pages = {L17-L20},
          doi = {10.1086/311269},
archivePrefix = {arXiv},
       eprint = {astro-ph/9712005},
 primaryClass = {astro-ph},
       adsurl = {https://ui.adsabs.harvard.edu/abs/1998ApJ...497L..17S}
}

@ARTICLE{schlegel,
   author = {{Schlegel}, D.~J. and {Finkbeiner}, D.~P. and {Davis}, M.},
    title = "{Maps of Dust Infrared Emission for Use in Estimation of Reddening and Cosmic Microwave Background Radiation Foregrounds}",
  journal = apj,
   eprint = {arXiv:astro-ph/9710327},
     year = 1998,
    month = jun,
   volume = 500,
    pages = {525-+},
      doi = {10.1086/305772},
   adsurl = {http://adsabs.harvard.edu/abs/1998ApJ...500..525S}
}

@ARTICLE{1993ApJ...405..273W,
       author = {{Woosley}, S.~E.},
        title = "{Gamma-Ray Bursts from Stellar Mass Accretion Disks around Black Holes}",
      journal = {\apj},
         year = 1993,
        month = mar,
       volume = {405},
        pages = {273},
          doi = {10.1086/172359},
       adsurl = {https://ui.adsabs.harvard.edu/abs/1993ApJ...405..273W}
}

@ARTICLE{1999ApJ...524..262M,
       author = {{MacFadyen}, A.~I. and {Woosley}, S.~E.},
        title = "{Collapsars: Gamma-Ray Bursts and Explosions in ``Failed Supernovae''}",
      journal = {\apj},
         year = 1999,
        month = oct,
       volume = {524},
       number = {1},
        pages = {262-289},
          doi = {10.1086/307790},
archivePrefix = {arXiv},
       eprint = {astro-ph/9810274},
 primaryClass = {astro-ph},
       adsurl = {https://ui.adsabs.harvard.edu/abs/1999ApJ...524..262M}
}

@ARTICLE{2003Natur.423..847H,
       author = {{Hjorth}, Jens and {Sollerman}, Jesper and {M{\o}ller}, Palle and {Fynbo}, Johan P.~U. and {Woosley}, Stan E. and {Kouveliotou}, Chryssa and {Tanvir}, Nial R. and {Greiner}, Jochen and {Andersen}, Michael I. and {Castro-Tirado}, Alberto J. and {Castro Cer{\'o}n}, Jos{\'e} Mar{\'\i}a and {Fruchter}, Andrew S. and {Gorosabel}, Javier and {Jakobsson}, P{\'a}ll and {Kaper}, Lex and {Klose}, Sylvio and {Masetti}, Nicola and {Pedersen}, Holger and {Pedersen}, Kristian and {Pian}, Elena and {Palazzi}, Eliana and {Rhoads}, James E. and {Rol}, Evert and {van den Heuvel}, Edward P.~J. and {Vreeswijk}, Paul M. and {Watson}, Darach and {Wijers}, Ralph A.~M.~J.},
        title = "{A very energetic supernova associated with the {\ensuremath{\gamma}}-ray burst of 29 March 2003}",
      journal = {\nat},
         year = 2003,
        month = jun,
       volume = {423},
       number = {6942},
        pages = {847-850},
          doi = {10.1038/nature01750},
archivePrefix = {arXiv},
       eprint = {astro-ph/0306347},
 primaryClass = {astro-ph},
       adsurl = {https://ui.adsabs.harvard.edu/abs/2003Natur.423..847H}
}

@ARTICLE{2014ARA&A..52...43B,
       author = {{Berger}, Edo},
        title = "{Short-Duration Gamma-Ray Bursts}",
      journal = {\araa},
         year = 2014,
        month = aug,
       volume = {52},
        pages = {43-105},
          doi = {10.1146/annurev-astro-081913-035926},
archivePrefix = {arXiv},
       eprint = {1311.2603},
 primaryClass = {astro-ph.HE},
       adsurl = {https://ui.adsabs.harvard.edu/abs/2014ARA&A..52...43B}
}

@ARTICLE{2022ApJ...932....1R,
       author = {{Rossi}, A. and {Rothberg}, B. and {Palazzi}, E. and {Kann}, D.~A. and {D'Avanzo}, P. and {Amati}, L. and {Klose}, S. and {Perego}, A. and {Pian}, E. and {Guidorzi}, C. and {Pozanenko}, A.~S. and {Savaglio}, S. and {Stratta}, G. and {Agapito}, G. and {Covino}, S. and {Cusano}, F. and {D'Elia}, V. and {De Pasquale}, M. and {Della Valle}, M. and {Kuhn}, O. and {Izzo}, L. and {Loffredo}, E. and {Masetti}, N. and {Melandri}, A. and {Minaev}, P.~Y. and {Guelbenzu}, A. Nicuesa and {Paris}, D. and {Paiano}, S. and {Plantet}, C. and {Rossi}, F. and {Salvaterra}, R. and {Schulze}, S. and {Veillet}, C. and {Volnova}, A.~A.},
        title = "{The Peculiar Short-duration GRB 200826A and Its Supernova}",
      journal = {\apj},
         year = 2022,
        month = jun,
       volume = {932},
       number = {1},
          eid = {1},
        pages = {1},
          doi = {10.3847/1538-4357/ac60a2},
archivePrefix = {arXiv},
       eprint = {2105.03829},
 primaryClass = {astro-ph.HE},
       adsurl = {https://ui.adsabs.harvard.edu/abs/2022ApJ...932....1R}
}

@INPROCEEDINGS{2006AIPC..836..540S,
       author = {{Savaglio}, S. and {Glazebrook}, K. and {Le Borgne}, D.},
        title = "{GRB Host Studies (GHostS)}",
    booktitle = {Gamma-Ray Bursts in the Swift Era},
         year = 2006,
       editor = {{Holt}, S.~S. and {Gehrels}, N. and {Nousek}, J.~A.},
       series = {American Institute of Physics Conference Series},
       volume = {836},
        month = may,
    publisher = {AIP},
        pages = {540-545},
          doi = {10.1063/1.2207951},
archivePrefix = {arXiv},
       eprint = {astro-ph/0601528},
 primaryClass = {astro-ph},
       adsurl = {https://ui.adsabs.harvard.edu/abs/2006AIPC..836..540S}
}

@ARTICLE{2022ApJ...940...57N,
       author = {{Nugent}, Anya E. and {Fong}, Wen-Fai and {Dong}, Yuxin and {Leja}, Joel and {Berger}, Edo and {Zevin}, Michael and {Chornock}, Ryan and {Cobb}, Bethany E. and {Kelley}, Luke Zoltan and {Kilpatrick}, Charles D. and {Levan}, Andrew and {Margutti}, Raffaella and {Paterson}, Kerry and {Perley}, Daniel and {Escorial}, Alicia Rouco and {Smith}, Nathan and {Tanvir}, Nial},
        title = "{Short GRB Host Galaxies. II. A Legacy Sample of Redshifts, Stellar Population Properties, and Implications for Their Neutron Star Merger Origins}",
      journal = {\apj},
         year = 2022,
        month = nov,
       volume = {940},
       number = {1},
          eid = {57},
        pages = {57},
          doi = {10.3847/1538-4357/ac91d1},
archivePrefix = {arXiv},
       eprint = {2206.01764},
 primaryClass = {astro-ph.GA},
       adsurl = {https://ui.adsabs.harvard.edu/abs/2022ApJ...940...57N}
}

@ARTICLE{2021MNRAS.503.3931T,
       author = {{Taggart}, K. and {Perley}, D.~A.},
        title = "{Core-collapse, superluminous, and gamma-ray burst supernova host galaxy populations at low redshift: the importance of dwarf and starbursting galaxies}",
      journal = {\mnras},
         year = 2021,
        month = may,
       volume = {503},
       number = {3},
        pages = {3931-3952},
          doi = {10.1093/mnras/stab174},
archivePrefix = {arXiv},
       eprint = {1911.09112},
 primaryClass = {astro-ph.HE},
       adsurl = {https://ui.adsabs.harvard.edu/abs/2021MNRAS.503.3931T}
}

@ARTICLE{2025arXiv250702806W,
       author = {{Wu}, Chao and {Wang}, Yun and {Li}, Hua-Li and {Xin}, Li-Ping and {Xu}, Dong and {Schneider}, Benjamin and {de Ugarte Postigo}, Antonio and {Lamb}, Gavin and {Reguitti}, Andrea and {Saccardi}, Andrea and {Gao}, Xing and {Li}, Xing-Ling and {Wang}, Qiu-Li and {Zhang}, Bing and {Wei}, Jian-Yan and {Zhang}, Shuang-Nan and {Daigne}, Fr{\'e}d{\'e}ric and {Atteia}, Jean-Luc and {Bernardini}, Maria-Grazia and {Cai}, Hong-bo and {Claret}, Arnaud and {Cordier}, Bertrand and {Deng}, Jin-Song and {Godet}, Olivier and {G{\"o}tz}, Diego and {Han}, Xu-Hui and {Kang}, Zhe and {Li}, Guang-Wei and {Li}, Zhen-Wei and {Liu}, Cheng-Zhi and {Lu}, Xiao-Meng and {Lv}, You and {Osborne}, Julian P and {Palmerio}, Jesse and {Qiu}, Yu-Lei and {Schanne}, St{\'e}phane and {Turpin}, Damien and {Vergani}, Susanna Diana and {Wang}, Jing and {Xiao}, Yu-Jie and {Xie}, Wen-Jin and {Xu}, Yang and {Yao}, Zhu-Heng and {Zhang}, Pin-Pin and {Zhang}, Ruo-Son and {Zhu}, Cheng-Wei and {Brivio}, Riccardo and {Covino}, Stefano and {D'Avanzo}, Paolo and {Ferro}, Matteo and {Melandri}, Andrea and {Rossi}, Andrea and {Ag{\"u}{\'\i} Fern{\'a}ndez}, Jos{\'e} Feliciano and {Th{\"o}ne}, Christina C. and {Bai}, Chun-Hai and {Esamdin}, Ali and {Iskandar}, Abdusamatjan and {Yaqup}, Shahidin and {Zhang}, Yu and {Zhong}, Tu-Hong and {Fu}, Shao-Yu and {Jiang}, Shuai-Qing and {Liu}, Xing and {An}, Jie and {Zhu}, Zi-Pei and {Cao}, Jia-Xin and {Liang}, En-Wei and {Lin}, Da-Bin and {Wang}, Xiang-Gao and {Du}, Guo-Wang and {Er}, Xin-Zhong and {Fang}, Yuan and {Liu}, Xiao-Wei and {Adami}, Christophe and {Dennefeld}, Michel and {Le Floc'h}, Emeric and {Uldall Fynbo}, Johan Peter and {Jakobsson}, P{\'a}ll and {Bj{\o}rn Malesani}, Daniele and {Jin}, Zhi-Ping and {Ren}, Jia and {Wang}, Hao and {Wei}, Da-Ming and {Zhou}, Hao and {Campana}, Sergio and {Kobayashi}, Shiho and {De Pasquale}, Massimiliano},
        title = "{GRB 240825A: Early Reverse Shock and Its Physical Implications}",
      journal = {arXiv e-prints},
         year = 2025,
        month = jul,
          eid = {arXiv:2507.02806},
        pages = {arXiv:2507.02806},
          doi = {10.48550/arXiv.2507.02806},
archivePrefix = {arXiv},
       eprint = {2507.02806},
 primaryClass = {astro-ph.HE},
       adsurl = {https://ui.adsabs.harvard.edu/abs/2025arXiv250702806W}
}

@ARTICLE{2025MNRAS.538L.100C,
       author = {{Caballero-Garc{\'\i}a}, M.~D. and {G{\"o}{\v{g}}{\"u}{\c{s}}}, E. and {Navarro-Gonz{\'a}lez}, J. and {Atapin}, K.~E. and {Sonbas}, E. and {Uzuner}, M. and {Castro-Tirado}, A.~J. and {Pandey}, S.~B. and {Gupta}, Rahul and {Ror}, A.~K. and {Hu}, Y. -D. and {Wu}, S. -Y. and {S{\'a}nchez-Ramirez}, R. and {Guziy}, S. and {Christiansen}, F. and {Connell}, P.~H. and {Neubert}, T. and {{\O}stgaard}, N. and {Adsuara}, J.~E. and {Gordillo-V{\'a}zquez}, F.~J. and {Fern{\'a}ndez-Garcia}, E. and {P{\'e}rez-Garcia}, I. and {Reglero}, V.},
        title = "{X-ray and gamma-ray timing of GRB 180720B, GRB 181222B, GRB 211211A, and GRB 220910A observed with Fermi and ASIM}",
      journal = {\mnras},
         year = 2025,
        month = mar,
       volume = {538},
       number = {1},
        pages = {L100-L108},
          doi = {10.1093/mnrasl/slaf016},
archivePrefix = {arXiv},
       eprint = {2411.18243},
 primaryClass = {astro-ph.HE},
       adsurl = {https://ui.adsabs.harvard.edu/abs/2025MNRAS.538L.100C}
}

@ARTICLE{2023ApJ...959...44L,
       author = {{Luo}, Jia-Wei and {Wang}, Fei-Fei and {Zhu-Ge}, Jia-Ming and {Li}, Ye and {Zou}, Yuan-Chuan and {Zhang}, Bing},
        title = "{Identifying the Physical Origin of Gamma-Ray Bursts with Supervised Machine Learning}",
      journal = {\apj},
         year = 2023,
        month = dec,
       volume = {959},
       number = {1},
          eid = {44},
        pages = {44},
          doi = {10.3847/1538-4357/ad03ec},
archivePrefix = {arXiv},
       eprint = {2211.16451},
 primaryClass = {astro-ph.HE},
       adsurl = {https://ui.adsabs.harvard.edu/abs/2023ApJ...959...44L}
}

@ARTICLE{2009ApJ...703.1696Z,
       author = {{Zhang}, Bing and {Zhang}, Bin-Bin and {Virgili}, Francisco J. and {Liang}, En-Wei and {Kann}, D. Alexander and {Wu}, Xue-Feng and {Proga}, Daniel and {Lv}, Hou-Jun and {Toma}, Kenji and {M{\'e}sz{\'a}ros}, Peter and {Burrows}, David N. and {Roming}, Peter W.~A. and {Gehrels}, Neil},
        title = "{Discerning the Physical Origins of Cosmological Gamma-ray Bursts Based on Multiple Observational Criteria: The Cases of z = 6.7 GRB 080913, z = 8.2 GRB 090423, and Some Short/Hard GRBs}",
      journal = {\apj},
         year = 2009,
        month = oct,
       volume = {703},
       number = {2},
        pages = {1696-1724},
          doi = {10.1088/0004-637X/703/2/1696},
archivePrefix = {arXiv},
       eprint = {0902.2419},
 primaryClass = {astro-ph.HE},
       adsurl = {https://ui.adsabs.harvard.edu/abs/2009ApJ...703.1696Z}
}

@ARTICLE{2013MNRAS.432..857M,
       author = {{MacLachlan}, G.~A. and {Shenoy}, A. and {Sonbas}, E. and {Dhuga}, K.~S. and {Cobb}, B.~E. and {Ukwatta}, T.~N. and {Morris}, D.~C. and {Eskandarian}, A. and {Maximon}, L.~C. and {Parke}, W.~C.},
        title = "{Minimum variability time-scales of long and short GRBs}",
      journal = {\mnras},
         year = 2013,
        month = jun,
       volume = {432},
       number = {2},
        pages = {857-865},
          doi = {10.1093/mnras/stt241},
archivePrefix = {arXiv},
       eprint = {1201.4431},
 primaryClass = {astro-ph.HE},
       adsurl = {https://ui.adsabs.harvard.edu/abs/2013MNRAS.432..857M}
}

@ARTICLE{2006Natur.441..463F,
       author = {{Fruchter}, A.~S. and {Levan}, A.~J. and {Strolger}, L. and {Vreeswijk}, P.~M. and {Thorsett}, S.~E. and {Bersier}, D. and {Burud}, I. and {Castro Cer{\'o}n}, J.~M. and {Castro-Tirado}, A.~J. and {Conselice}, C. and {Dahlen}, T. and {Ferguson}, H.~C. and {Fynbo}, J.~P.~U. and {Garnavich}, P.~M. and {Gibbons}, R.~A. and {Gorosabel}, J. and {Gull}, T.~R. and {Hjorth}, J. and {Holland}, S.~T. and {Kouveliotou}, C. and {Levay}, Z. and {Livio}, M. and {Metzger}, M.~R. and {Nugent}, P.~E. and {Petro}, L. and {Pian}, E. and {Rhoads}, J.~E. and {Riess}, A.~G. and {Sahu}, K.~C. and {Smette}, A. and {Tanvir}, N.~R. and {Wijers}, R.~A.~M.~J. and {Woosley}, S.~E.},
        title = "{Long {\ensuremath{\gamma}}-ray bursts and core-collapse supernovae have different environments}",
      journal = {\nat},
         year = 2006,
        month = may,
       volume = {441},
       number = {7092},
        pages = {463-468},
          doi = {10.1038/nature04787},
archivePrefix = {arXiv},
       eprint = {astro-ph/0603537},
 primaryClass = {astro-ph},
       adsurl = {https://ui.adsabs.harvard.edu/abs/2006Natur.441..463F}
}

@ARTICLE{2022ApJ...940...56F,
       author = {{Fong}, Wen-fai and {Nugent}, Anya E. and {Dong}, Yuxin and {Berger}, Edo and {Paterson}, Kerry and {Chornock}, Ryan and {Levan}, Andrew and {Blanchard}, Peter and {Alexander}, Kate D. and {Andrews}, Jennifer and {Cobb}, Bethany E. and {Cucchiara}, Antonino and {Fox}, Derek and {Fryer}, Chris L. and {Gordon}, Alexa C. and {Kilpatrick}, Charles D. and {Lunnan}, Ragnhild and {Margutti}, Raffaella and {Miller}, Adam and {Milne}, Peter and {Nicholl}, Matt and {Perley}, Daniel and {Rastinejad}, Jillian and {Escorial}, Alicia Rouco and {Schroeder}, Genevieve and {Smith}, Nathan and {Tanvir}, Nial and {Terreran}, Giacomo},
        title = "{Short GRB Host Galaxies. I. Photometric and Spectroscopic Catalogs, Host Associations, and Galactocentric Offsets}",
      journal = {\apj},
         year = 2022,
        month = nov,
       volume = {940},
       number = {1},
          eid = {56},
        pages = {56},
          doi = {10.3847/1538-4357/ac91d0},
archivePrefix = {arXiv},
       eprint = {2206.01763},
 primaryClass = {astro-ph.GA},
       adsurl = {https://ui.adsabs.harvard.edu/abs/2022ApJ...940...56F}
}

@ARTICLE{2016ApJ...829....7L,
       author = {{Lien}, Amy and {Sakamoto}, Takanori and {Barthelmy}, Scott D. and {Baumgartner}, Wayne H. and {Cannizzo}, John K. and {Chen}, Kevin and {Collins}, Nicholas R. and {Cummings}, Jay R. and {Gehrels}, Neil and {Krimm}, Hans A. and {Markwardt}, Craig. B. and {Palmer}, David M. and {Stamatikos}, Michael and {Troja}, Eleonora and {Ukwatta}, T.~N.},
        title = "{The Third Swift Burst Alert Telescope Gamma-Ray Burst Catalog}",
      journal = {\apj},
         year = 2016,
        month = sep,
       volume = {829},
       number = {1},
          eid = {7},
        pages = {7},
          doi = {10.3847/0004-637X/829/1/7},
archivePrefix = {arXiv},
       eprint = {1606.01956},
 primaryClass = {astro-ph.HE},
       adsurl = {https://ui.adsabs.harvard.edu/abs/2016ApJ...829....7L}
}

@ARTICLE{2023NatAs...7.1136C,
       author = {{Castro-Tirado}, Alberto J.},
        title = "{Tracking transients night and day}",
      journal = {Nature Astronomy},
         year = 2023,
        month = sep,
       volume = {7},
        pages = {1136-1136},
          doi = {10.1038/s41550-023-02075-w},
       adsurl = {https://ui.adsabs.harvard.edu/abs/2023NatAs...7.1136C}
}

@ARTICLE{1982ApJ...253..785A,
       author = {{Arnett}, W.~D.},
        title = "{Type I supernovae. I - Analytic solutions for the early part of the light curve}",
      journal = {\apj},
         year = 1982,
        month = feb,
       volume = {253},
        pages = {785-797},
          doi = {10.1086/159681},
       adsurl = {https://ui.adsabs.harvard.edu/abs/1982ApJ...253..785A}
}

@ARTICLE{2008MNRAS.383.1485V,
       author = {{Valenti}, S. and {Benetti}, S. and {Cappellaro}, E. and {Patat}, F. and {Mazzali}, P. and {Turatto}, M. and {Hurley}, K. and {Maeda}, K. and {Gal-Yam}, A. and {Foley}, R.~J. and {Filippenko}, A.~V. and {Pastorello}, A. and {Challis}, P. and {Frontera}, F. and {Harutyunyan}, A. and {Iye}, M. and {Kawabata}, K. and {Kirshner}, R.~P. and {Li}, W. and {Lipkin}, Y.~M. and {Matheson}, T. and {Nomoto}, K. and {Ofek}, E.~O. and {Ohyama}, Y. and {Pian}, E. and {Poznanski}, D. and {Salvo}, M. and {Sauer}, D.~N. and {Schmidt}, B.~P. and {Soderberg}, A. and {Zampieri}, L.},
        title = "{The broad-lined Type Ic supernova 2003jd}",
      journal = {\mnras},
         year = 2008,
        month = feb,
       volume = {383},
       number = {4},
        pages = {1485-1500},
          doi = {10.1111/j.1365-2966.2007.12647.x},
archivePrefix = {arXiv},
       eprint = {0710.5173},
 primaryClass = {astro-ph},
       adsurl = {https://ui.adsabs.harvard.edu/abs/2008MNRAS.383.1485V}
}

@ARTICLE{1998Natur.395..672I,
       author = {{Iwamoto}, K. and {Mazzali}, P.~A. and {Nomoto}, K. and {Umeda}, H. and {Nakamura}, T. and {Patat}, F. and {Danziger}, I.~J. and {Young}, T.~R. and {Suzuki}, T. and {Shigeyama}, T. and {Augusteijn}, T. and {Doublier}, V. and {Gonzalez}, J. -F. and {Boehnhardt}, H. and {Brewer}, J. and {Hainaut}, O.~R. and {Lidman}, C. and {Leibundgut}, B. and {Cappellaro}, E. and {Turatto}, M. and {Galama}, T.~J. and {Vreeswijk}, P.~M. and {Kouveliotou}, C. and {van Paradijs}, J. and {Pian}, E. and {Palazzi}, E. and {Frontera}, F.},
        title = "{A hypernova model for the supernova associated with the {\ensuremath{\gamma}}-ray burst of 25 April 1998}",
      journal = {\nat},
         year = 1998,
        month = oct,
       volume = {395},
       number = {6703},
        pages = {672-674},
          doi = {10.1038/27155},
archivePrefix = {arXiv},
       eprint = {astro-ph/9806382},
 primaryClass = {astro-ph},
       adsurl = {https://ui.adsabs.harvard.edu/abs/1998Natur.395..672I}
}

@ARTICLE{Volnova21,
       author = {{Volnova}, A. and {Pozanenko}, A. and {Mazaeva}, E. and {Belkin}, S. and {Molotov}, I. and {Elenin}, L. and {Tungalag}, N. and {Buckley}, D.},
        title = "{IKI GRB-FuN: observations of GRBs with small-aperture telescopes}",
      journal = {Anais da Academia Brasileira de Ciencias},
         year = 2021,
        month = may,
       volume = {93},
       number = {1},
        pages = {},
          doi = {/10.1590/0001-3765202120200883},
archivePrefix = {arXiv},
       eprint = {},
 primaryClass = {},
       adsurl = {}
}

@ARTICLE{2024AstL...50..701B,
       author = {{Belkin}, S.~O. and {Pozanenko}, A.~S.},
        title = "{Search for Correlations and Study of Selection Effects When Analyzing a Sample of Supernovae Associated with Gamma-Ray Bursts}",
      journal = {Astronomy Letters},
         year = 2024,
        month = dec,
       volume = {50},
       number = {12},
        pages = {701-731},
          doi = {10.1134/S1063773725700069},
       adsurl = {https://ui.adsabs.harvard.edu/abs/2024AstL...50..701B}
}

@ARTICLE{2003ApJ...599L..95M,
       author = {{Mazzali}, Paolo A. and {Deng}, Jinsong and {Tominaga}, Nozomu and {Maeda}, Keiichi and {Nomoto}, Ken'ichi and {Matheson}, Thomas and {Kawabata}, Koji S. and {Stanek}, Krzysztof Z. and {Garnavich}, Peter M.},
        title = "{The Type Ic Hypernova SN 2003dh/GRB 030329}",
      journal = {\apjl},
         year = 2003,
        month = dec,
       volume = {599},
       number = {2},
        pages = {L95-L98},
          doi = {10.1086/381259},
archivePrefix = {arXiv},
       eprint = {astro-ph/0309555},
 primaryClass = {astro-ph},
       adsurl = {https://ui.adsabs.harvard.edu/abs/2003ApJ...599L..95M}
}

@BOOK{2017suex.book.....B,
       author = {{Branch}, David and {Wheeler}, J. Craig},
        title = "{Supernova Explosions}",
         year = 2017,
          doi = {10.1007/978-3-662-55054-0},
       adsurl = {https://ui.adsabs.harvard.edu/abs/2017suex.book.....B}
}

@ARTICLE{2022PASP..134e4201B,
       author = {{Bora}, Zs{\'o}fia and {Vink{\'o}}, J{\'o}zsef and {K{\"o}nyves-T{\'o}th}, R{\'e}ka},
        title = "{Initial Ni-56 Masses in Type Ia Supernovae}",
      journal = {\pasp},
         year = 2022,
        month = may,
       volume = {134},
       number = {1035},
          eid = {054201},
        pages = {054201},
          doi = {10.1088/1538-3873/ac63e7},
archivePrefix = {arXiv},
       eprint = {2306.05820},
 primaryClass = {astro-ph.SR},
       adsurl = {https://ui.adsabs.harvard.edu/abs/2022PASP..134e4201B}
}

@ARTICLE{2016MNRAS.458.2973P,
       author = {{Prentice}, S.~J. and {Mazzali}, P.~A. and {Pian}, E. and {Gal-Yam}, A. and {Kulkarni}, S.~R. and {Rubin}, A. and {Corsi}, A. and {Fremling}, C. and {Sollerman}, J. and {Yaron}, O. and {Arcavi}, I. and {Zheng}, W. and {Kasliwal}, M.~M. and {Filippenko}, A.~V. and {Cenko}, S.~B. and {Cao}, Y. and {Nugent}, P.~E.},
        title = "{The bolometric light curves and physical parameters of stripped-envelope supernovae}",
      journal = {\mnras},
         year = 2016,
        month = may,
       volume = {458},
       number = {3},
        pages = {2973-3002},
          doi = {10.1093/mnras/stw299},
archivePrefix = {arXiv},
       eprint = {1602.01736},
 primaryClass = {astro-ph.HE},
       adsurl = {https://ui.adsabs.harvard.edu/abs/2016MNRAS.458.2973P}
}

@ARTICLE{2023FrASS..10.2887H,
       author = {{Hu}, Y. -D. and {Fern{\'a}ndez-Garc{\'\i}a}, E. and {Caballero-Garc{\'\i}a}, M.~D. and {P{\'e}rez-Garc{\'\i}a}, I. and {Carrasco-Garc{\'\i}a}, I.~M. and {Castell{\'o}n}, A. and {P{\'e}rez del Pulgar}, C. and {Reina Terol}, A.~J. and {Castro-Tirado}, A.~J.},
        title = "{The burst observer and optical transient exploring system in the multi-messenger astronomy era}",
      journal = {Frontiers in Astronomy and Space Sciences},
         year = 2023,
        month = may,
       volume = {10},
          eid = {952887},
        pages = {952887},
          doi = {10.3389/fspas.2023.952887},
archivePrefix = {arXiv},
       eprint = {2302.06565},
 primaryClass = {astro-ph.HE},
       adsurl = {https://ui.adsabs.harvard.edu/abs/2023FrASS..10.2887H}
}

@ARTICLE{2025ApJ...979..159S,
       author = {{Stratta}, G. and {Nicuesa Guelbenzu}, A.~M. and {Klose}, S. and {Rossi}, A. and {Singh}, P. and {Palazzi}, E. and {Guidorzi}, C. and {Camisasca}, A. and {Bernuzzi}, S. and {Rau}, A. and {Bulla}, M. and {Ragosta}, F. and {Maiorano}, E. and {Paris}, D.},
        title = "{The Puzzling Long GRB 191019A: Evidence for Kilonova Light}",
      journal = {\apj},
         year = 2025,
        month = feb,
       volume = {979},
       number = {2},
          eid = {159},
        pages = {159},
          doi = {10.3847/1538-4357/ad9b7b},
archivePrefix = {arXiv},
       eprint = {2412.04059},
 primaryClass = {astro-ph.HE},
       adsurl = {https://ui.adsabs.harvard.edu/abs/2025ApJ...979..159S}
}

@ARTICLE{2022ApJ...931L..23L,
       author = {{L{\"u}}, Hou-Jun and {Yuan}, Hao-Yu and {Yi}, Ting-Feng and {Wang}, Xiang-Gao and {Hu}, You-Dong and {Yuan}, Yong and {Rice}, Jared and {Wang}, Jian-Guo and {Cao}, Jia-Xin and {Kong}, De-Feng and {Fernandez-Garc{\'\i}a}, Emilio and {Castro-Tirado}, Alberto J. and {Lian}, Ji-Shun and {Gan}, Wen-Pei and {Wang}, Shan-Qin and {Xin}, Li-Ping and {Caballero-Garc{\'\i}a}, M.~D. and {Fan}, Yu-Feng and {Liang}, En-Wei},
        title = "{GRB 211227A as a Peculiar Long Gamma-Ray Burst from a Compact Star Merger}",
      journal = {\apjl},
         year = 2022,
        month = jun,
       volume = {931},
       number = {2},
          eid = {L23},
        pages = {L23},
          doi = {10.3847/2041-8213/ac6e3a},
archivePrefix = {arXiv},
       eprint = {2201.06395},
 primaryClass = {astro-ph.HE},
       adsurl = {https://ui.adsabs.harvard.edu/abs/2022ApJ...931L..23L}
}

@ARTICLE{2013ApJ...763...15Q,
       author = {{Qin}, Ying and {Liang}, En-Wei and {Liang}, Yun-Feng and {Yi}, Shuang-Xi and {Lin}, Lin and {Zhang}, Bin-Bin and {Zhang}, Jin and {L{\"u}}, Hou-Jun and {Lu}, Rui-Jing and {L{\"u}}, Lian-Zhong and {Zhang}, Bing},
        title = "{A Comprehensive Analysis of Fermi Gamma-Ray Burst Data. III. Energy-dependent T $_{90}$ Distributions of GBM GRBs and Instrumental Selection Effect on Duration Classification}",
      journal = {\apj},
         year = 2013,
        month = jan,
       volume = {763},
       number = {1},
          eid = {15},
        pages = {15},
          doi = {10.1088/0004-637X/763/1/15},
archivePrefix = {arXiv},
       eprint = {1205.1188},
 primaryClass = {astro-ph.HE},
       adsurl = {https://ui.adsabs.harvard.edu/abs/2013ApJ...763...15Q}
}

@ARTICLE{2000ApJ...541L..51K,
       author = {{Kumar}, Pawan and {Panaitescu}, Alin},
        title = "{Afterglow Emission from Naked Gamma-Ray Bursts}",
      journal = {\apjl},
         year = 2000,
        month = oct,
       volume = {541},
       number = {2},
        pages = {L51-L54},
          doi = {10.1086/312905},
archivePrefix = {arXiv},
       eprint = {astro-ph/0006317},
 primaryClass = {astro-ph},
       adsurl = {https://ui.adsabs.harvard.edu/abs/2000ApJ...541L..51K}
}

@ARTICLE{1999ApJ...520..641S,
       author = {{Sari}, Re'em and {Piran}, Tsvi},
        title = "{Predictions for the Very Early Afterglow and the Optical Flash}",
      journal = {\apj},
         year = 1999,
        month = aug,
       volume = {520},
       number = {2},
        pages = {641-649},
          doi = {10.1086/307508},
archivePrefix = {arXiv},
       eprint = {astro-ph/9901338},
 primaryClass = {astro-ph},
       adsurl = {https://ui.adsabs.harvard.edu/abs/1999ApJ...520..641S}
}

@ARTICLE{2011MNRAS.411.2792S,
       author = {{Starling}, R.~L.~C. and {Wiersema}, K. and {Levan}, A.~J. and {Sakamoto}, T. and {Bersier}, D. and {Goldoni}, P. and {Oates}, S.~R. and {Rowlinson}, A. and {Campana}, S. and {Sollerman}, J. and {Tanvir}, N.~R. and {Malesani}, D. and {Fynbo}, J.~P.~U. and {Covino}, S. and {D'Avanzo}, P. and {O'Brien}, P.~T. and {Page}, K.~L. and {Osborne}, J.~P. and {Vergani}, S.~D. and {Barthelmy}, S. and {Burrows}, D.~N. and {Cano}, Z. and {Curran}, P.~A. and {de Pasquale}, M. and {D'Elia}, V. and {Evans}, P.~A. and {Flores}, H. and {Fruchter}, A.~S. and {Garnavich}, P. and {Gehrels}, N. and {Gorosabel}, J. and {Hjorth}, J. and {Holland}, S.~T. and {van der Horst}, A.~J. and {Hurkett}, C.~P. and {Jakobsson}, P. and {Kamble}, A.~P. and {Kouveliotou}, C. and {Kuin}, N.~P.~M. and {Kaper}, L. and {Mazzali}, P.~A. and {Nugent}, P.~E. and {Pian}, E. and {Stamatikos}, M. and {Th{\"o}ne}, C.~C. and {Woosley}, S.~E.},
        title = "{Discovery of the nearby long, soft GRB 100316D with an associated supernova}",
      journal = {\mnras},
         year = 2011,
        month = mar,
       volume = {411},
       number = {4},
        pages = {2792-2803},
          doi = {10.1111/j.1365-2966.2010.17879.x},
archivePrefix = {arXiv},
       eprint = {1004.2919},
 primaryClass = {astro-ph.CO},
       adsurl = {https://ui.adsabs.harvard.edu/abs/2011MNRAS.411.2792S}
}

@ARTICLE{2017AdAst2017E...5C,
       author = {{Cano}, Zach and {Wang}, Shan-Qin and {Dai}, Zi-Gao and {Wu}, Xue-Feng},
        title = "{The Observer's Guide to the Gamma-Ray Burst Supernova Connection}",
      journal = {Advances in Astronomy},
         year = 2017,
        month = jan,
       volume = {2017},
          eid = {8929054},
        pages = {8929054},
          doi = {10.1155/2017/8929054},
archivePrefix = {arXiv},
       eprint = {1604.03549},
 primaryClass = {astro-ph.HE},
       adsurl = {https://ui.adsabs.harvard.edu/abs/2017AdAst2017E...5C}
}

@ARTICLE{2007PhR...442..166N,
       author = {{Nakar}, Ehud},
        title = "{Short-hard gamma-ray bursts}",
      journal = {\physrep},
         year = 2007,
        month = apr,
       volume = {442},
       number = {1-6},
        pages = {166-236},
          doi = {10.1016/j.physrep.2007.02.005},
archivePrefix = {arXiv},
       eprint = {astro-ph/0701748},
 primaryClass = {astro-ph},
       adsurl = {https://ui.adsabs.harvard.edu/abs/2007PhR...442..166N}
}

@ARTICLE{2002ApJ...568..820G,
       author = {{Granot}, Jonathan and {Sari}, Re'em},
        title = "{The Shape of Spectral Breaks in Gamma-Ray Burst Afterglows}",
      journal = {\apj},
         year = 2002,
        month = apr,
       volume = {568},
       number = {2},
        pages = {820-829},
          doi = {10.1086/338966},
archivePrefix = {arXiv},
       eprint = {astro-ph/0108027},
 primaryClass = {astro-ph},
       adsurl = {https://ui.adsabs.harvard.edu/abs/2002ApJ...568..820G}
}

@ARTICLE{2006Natur.444.1044G,
       author = {{Gehrels}, N. and {Norris}, J.~P. and {Barthelmy}, S.~D. and {Granot}, J. and {Kaneko}, Y. and {Kouveliotou}, C. and {Markwardt}, C.~B. and {M{\'e}sz{\'a}ros}, P. and {Nakar}, E. and {Nousek}, J.~A. and {O'Brien}, P.~T. and {Page}, M. and {Palmer}, D.~M. and {Parsons}, A.~M. and {Roming}, P.~W.~A. and {Sakamoto}, T. and {Sarazin}, C.~L. and {Schady}, P. and {Stamatikos}, M. and {Woosley}, S.~E.},
        title = "{A new {\ensuremath{\gamma}}-ray burst classification scheme from GRB060614}",
      journal = {\nat},
         year = 2006,
        month = dec,
       volume = {444},
       number = {7122},
        pages = {1044-1046},
          doi = {10.1038/nature05376},
archivePrefix = {arXiv},
       eprint = {astro-ph/0610635},
 primaryClass = {astro-ph},
       adsurl = {https://ui.adsabs.harvard.edu/abs/2006Natur.444.1044G}
}

@ARTICLE{2006Natur.444.1050D,
       author = {{Della Valle}, M. and {Chincarini}, G. and {Panagia}, N. and {Tagliaferri}, G. and {Malesani}, D. and {Testa}, V. and {Fugazza}, D. and {Campana}, S. and {Covino}, S. and {Mangano}, V. and {Antonelli}, L.~A. and {D'Avanzo}, P. and {Hurley}, K. and {Mirabel}, I.~F. and {Pellizza}, L.~J. and {Piranomonte}, S. and {Stella}, L.},
        title = "{An enigmatic long-lasting {\ensuremath{\gamma}}-ray burst not accompanied by a bright supernova}",
      journal = {\nat},
         year = 2006,
        month = dec,
       volume = {444},
       number = {7122},
        pages = {1050-1052},
          doi = {10.1038/nature05374},
archivePrefix = {arXiv},
       eprint = {astro-ph/0608322},
 primaryClass = {astro-ph},
       adsurl = {https://ui.adsabs.harvard.edu/abs/2006Natur.444.1050D}
}

@ARTICLE{2022Natur.612..223R,
       author = {{Rastinejad}, Jillian C. and {Gompertz}, Benjamin P. and {Levan}, Andrew J. and {Fong}, Wen-fai and {Nicholl}, Matt and {Lamb}, Gavin P. and {Malesani}, Daniele B. and {Nugent}, Anya E. and {Oates}, Samantha R. and {Tanvir}, Nial R. and {de Ugarte Postigo}, Antonio and {Kilpatrick}, Charles D. and {Moore}, Christopher J. and {Metzger}, Brian D. and {Ravasio}, Maria Edvige and {Rossi}, Andrea and {Schroeder}, Genevieve and {Jencson}, Jacob and {Sand}, David J. and {Smith}, Nathan and {Ag{\"u}{\'\i} Fern{\'a}ndez}, Jos{\'e} Feliciano and {Berger}, Edo and {Blanchard}, Peter K. and {Chornock}, Ryan and {Cobb}, Bethany E. and {De Pasquale}, Massimiliano and {Fynbo}, Johan P.~U. and {Izzo}, Luca and {Kann}, D. Alexander and {Laskar}, Tanmoy and {Marini}, Ester and {Paterson}, Kerry and {Escorial}, Alicia Rouco and {Sears}, Huei M. and {Th{\"o}ne}, Christina C.},
        title = "{A kilonova following a long-duration gamma-ray burst at 350 Mpc}",
      journal = {\nat},
         year = 2022,
        month = dec,
       volume = {612},
       number = {7939},
        pages = {223-227},
          doi = {10.1038/s41586-022-05390-w},
archivePrefix = {arXiv},
       eprint = {2204.10864},
 primaryClass = {astro-ph.HE},
       adsurl = {https://ui.adsabs.harvard.edu/abs/2022Natur.612..223R}
}

@ARTICLE{2022Natur.612..228T,
       author = {{Troja}, E. and {Fryer}, C.~L. and {O'Connor}, B. and {Ryan}, G. and {Dichiara}, S. and {Kumar}, A. and {Ito}, N. and {Gupta}, R. and {Wollaeger}, R.~T. and {Norris}, J.~P. and {Kawai}, N. and {Butler}, N.~R. and {Aryan}, A. and {Misra}, K. and {Hosokawa}, R. and {Murata}, K.~L. and {Niwano}, M. and {Pandey}, S.~B. and {Kutyrev}, A. and {van Eerten}, H.~J. and {Chase}, E.~A. and {Hu}, Y. -D. and {Caballero-Garcia}, M.~D. and {Castro-Tirado}, A.~J.},
        title = "{A nearby long gamma-ray burst from a merger of compact objects}",
      journal = {\nat},
         year = 2022,
        month = dec,
       volume = {612},
       number = {7939},
        pages = {228-231},
          doi = {10.1038/s41586-022-05327-3},
archivePrefix = {arXiv},
       eprint = {2209.03363},
 primaryClass = {astro-ph.HE},
       adsurl = {https://ui.adsabs.harvard.edu/abs/2022Natur.612..228T}
}

@ARTICLE{2024Natur.626..737L,
       author = {{Levan}, Andrew J. and {Gompertz}, Benjamin P. and {Salafia}, Om Sharan and {Bulla}, Mattia and {Burns}, Eric and {Hotokezaka}, Kenta and {Izzo}, Luca and {Lamb}, Gavin P. and {Malesani}, Daniele B. and {Oates}, Samantha R. and {Ravasio}, Maria Edvige and {Rouco Escorial}, Alicia and {Schneider}, Benjamin and {Sarin}, Nikhil and {Schulze}, Steve and {Tanvir}, Nial R. and {Ackley}, Kendall and {Anderson}, Gemma and {Brammer}, Gabriel B. and {Christensen}, Lise and {Dhillon}, Vikram S. and {Evans}, Phil A. and {Fausnaugh}, Michael and {Fong}, Wen-fai and {Fruchter}, Andrew S. and {Fryer}, Chris and {Fynbo}, Johan P.~U. and {Gaspari}, Nicola and {Heintz}, Kasper E. and {Hjorth}, Jens and {Kennea}, Jamie A. and {Kennedy}, Mark R. and {Laskar}, Tanmoy and {Leloudas}, Giorgos and {Mandel}, Ilya and {Martin-Carrillo}, Antonio and {Metzger}, Brian D. and {Nicholl}, Matt and {Nugent}, Anya and {Palmerio}, Jesse T. and {Pugliese}, Giovanna and {Rastinejad}, Jillian and {Rhodes}, Lauren and {Rossi}, Andrea and {Saccardi}, Andrea and {Smartt}, Stephen J. and {Stevance}, Heloise F. and {Tohuvavohu}, Aaron and {van der Horst}, Alexander and {Vergani}, Susanna D. and {Watson}, Darach and {Barclay}, Thomas and {Bhirombhakdi}, Kornpob and {Breedt}, Elm{\'e} and {Breeveld}, Alice A. and {Brown}, Alexander J. and {Campana}, Sergio and {Chrimes}, Ashley A. and {D'Avanzo}, Paolo and {D'Elia}, Valerio and {De Pasquale}, Massimiliano and {Dyer}, Martin J. and {Galloway}, Duncan K. and {Garbutt}, James A. and {Green}, Matthew J. and {Hartmann}, Dieter H. and {Jakobsson}, P{\'a}ll and {Kerry}, Paul and {Kouveliotou}, Chryssa and {Langeroodi}, Danial and {Le Floc'h}, Emeric and {Leung}, James K. and {Littlefair}, Stuart P. and {Munday}, James and {O'Brien}, Paul and {Parsons}, Steven G. and {Pelisoli}, Ingrid and {Sahman}, David I. and {Salvaterra}, Ruben and {Sbarufatti}, Boris and {Steeghs}, Danny and {Tagliaferri}, Gianpiero and {Th{\"o}ne}, Christina C. and {de Ugarte Postigo}, Antonio and {Kann}, David Alexander},
        title = "{Heavy-element production in a compact object merger observed by JWST}",
      journal = {\nat},
         year = 2024,
        month = feb,
       volume = {626},
       number = {8000},
        pages = {737-741},
          doi = {10.1038/s41586-023-06759-1},
archivePrefix = {arXiv},
       eprint = {2307.02098},
 primaryClass = {astro-ph.HE},
       adsurl = {https://ui.adsabs.harvard.edu/abs/2024Natur.626..737L}
}

@ARTICLE{2002A&A...390...81A,
       author = {{Amati}, L. and {Frontera}, F. and {Tavani}, M. and {in't Zand}, J.~J.~M. and {Antonelli}, A. and {Costa}, E. and {Feroci}, M. and {Guidorzi}, C. and {Heise}, J. and {Masetti}, N. and {Montanari}, E. and {Nicastro}, L. and {Palazzi}, E. and {Pian}, E. and {Piro}, L. and {Soffitta}, P.},
        title = "{Intrinsic spectra and energetics of BeppoSAX Gamma-Ray Bursts with known redshifts}",
      journal = {\aap},
         year = 2002,
        month = jul,
       volume = {390},
        pages = {81-89},
          doi = {10.1051/0004-6361:20020722},
archivePrefix = {arXiv},
       eprint = {astro-ph/0205230},
 primaryClass = {astro-ph},
       adsurl = {https://ui.adsabs.harvard.edu/abs/2002A&A...390...81A}
}

@ARTICLE{2000ApJ...534..248N,
       author = {{Norris}, J.~P. and {Marani}, G.~F. and {Bonnell}, J.~T.},
        title = "{Connection between Energy-dependent Lags and Peak Luminosity in Gamma-Ray Bursts}",
      journal = {\apj},
         year = 2000,
        month = may,
       volume = {534},
       number = {1},
        pages = {248-257},
          doi = {10.1086/308725},
archivePrefix = {arXiv},
       eprint = {astro-ph/9903233},
 primaryClass = {astro-ph},
       adsurl = {https://ui.adsabs.harvard.edu/abs/2000ApJ...534..248N}
}

@ARTICLE{Fruchter2006,
       author = {{Fruchter}, A.~S. and {Levan}, A.~J. and {Strolger}, L. and {Vreeswijk}, P.~M. and {Thorsett}, S.~E. and {Bersier}, D. and {Burud}, I. and {Castro Cer{\'o}n}, J.~M. and {Castro-Tirado}, A.~J. and {Conselice}, C. and {Dahlen}, T. and {Ferguson}, H.~C. and {Fynbo}, J.~P.~U. and {Garnavich}, P.~M. and {Gibbons}, R.~A. and {Gorosabel}, J. and {Gull}, T.~R. and {Hjorth}, J. and {Holland}, S.~T. and {Kouveliotou}, C. and {Levay}, Z. and {Livio}, M. and {Metzger}, M.~R. and {Nugent}, P.~E. and {Petro}, L. and {Pian}, E. and {Rhoads}, J.~E. and {Riess}, A.~G. and {Sahu}, K.~C. and {Smette}, A. and {Tanvir}, N.~R. and {Wijers}, R.~A.~M.~J. and {Woosley}, S.~E.},
        title = "{Long {\ensuremath{\gamma}}-ray bursts and core-collapse supernovae have different environments}",
      journal = {\nat},
         year = 2006,
        month = may,
       volume = {441},
       number = {7092},
        pages = {463-468},
          doi = {10.1038/nature04787},
archivePrefix = {arXiv},
       eprint = {astro-ph/0603537},
 primaryClass = {astro-ph},
       adsurl = {https://ui.adsabs.harvard.edu/abs/2006Natur.441..463F}
}

@ARTICLE{Savaglio2009,
       author = {{Savaglio}, S. and {Glazebrook}, K. and {Le Borgne}, D.},
        title = "{The Galaxy Population Hosting Gamma-Ray Bursts}",
      journal = {\apj},
         year = 2009,
        month = jan,
       volume = {691},
       number = {1},
        pages = {182-211},
          doi = {10.1088/0004-637X/691/1/182},
archivePrefix = {arXiv},
       eprint = {0803.2718},
 primaryClass = {astro-ph},
       adsurl = {https://ui.adsabs.harvard.edu/abs/2009ApJ...691..182S}
}

@ARTICLE{2022JApA...43...82G,
       author = {{Gupta}, Rahul and {Pandey}, Shashi Bhushan and {Kumar}, Amit and {Aryan}, Amar and {Ror}, Amit Kumar and {Sharma}, Saurabh and {Misra}, Kuntal and {Castro-Tirado}, A.~J. and {Tiwari}, Sugriva Nath},
        title = "{Photometric studies on the host galaxies of gamma-ray bursts using 3.6m Devasthal optical telescope}",
      journal = {Journal of Astrophysics and Astronomy},
         year = 2022,
        month = dec,
       volume = {43},
       number = {2},
          eid = {82},
        pages = {82},
          doi = {10.1007/s12036-022-09865-0},
archivePrefix = {arXiv},
       eprint = {2205.13940},
 primaryClass = {astro-ph.HE},
       adsurl = {https://ui.adsabs.harvard.edu/abs/2022JApA...43...82G}
}

@ARTICLE{2024GCN.37302....1F,
       author = {{Frederiks}, D. and {Lysenko}, A. and {Ridnaia}, A. and {Svinkin}, D. and {Tsvetkova}, A. and {Ulanov}, M. and {Cline}, T. and {Konus-Wind Team}},
        title = "{Konus-Wind detection of GRB 240825A (bright/long)}",
      journal = {GRB Coordinates Network},
         year = 2024,
        month = aug,
       volume = {37302},
        pages = {1},
       adsurl = {https://ui.adsabs.harvard.edu/abs/2024GCN.37302....1F}
}

@ARTICLE{2024GCN.37298....1J,
       author = {{Joshi}, J. and {Waratkar}, G. and {Vibhute}, A. and {Bhalerao}, V. and {Bhattacharya}, D. and {Rao}, A.~R. and {Vadawale}, S. and {AstroSat CZTI Collaboration}},
        title = "{GRB 240825A: AstroSat CZTI detection of the long bright burst}",
      journal = {GRB Coordinates Network},
         year = 2024,
        month = aug,
       volume = {37298},
        pages = {1},
       adsurl = {https://ui.adsabs.harvard.edu/abs/2024GCN.37298....1J}
}

@ARTICLE{2016ApJ...817..144B,
       author = {{Blanchard}, Peter K. and {Berger}, Edo and {Fong}, Wen-fai},
        title = "{The Offset and Host Light Distributions of Long Gamma-Ray Bursts: A New View From HST Observations of Swift Bursts}",
      journal = {\apj},
         year = 2016,
        month = feb,
       volume = {817},
       number = {2},
          eid = {144},
        pages = {144},
          doi = {10.3847/0004-637X/817/2/144},
archivePrefix = {arXiv},
       eprint = {1509.07866},
 primaryClass = {astro-ph.HE},
       adsurl = {https://ui.adsabs.harvard.edu/abs/2016ApJ...817..144B}
}

@ARTICLE{2015MNRAS.446.1129B,
       author = {{Bernardini}, M.~G. and {Ghirlanda}, G. and {Campana}, S. and {Covino}, S. and {Salvaterra}, R. and {Atteia}, J. -L. and {Burlon}, D. and {Calderone}, G. and {D'Avanzo}, P. and {D'Elia}, V. and {Ghisellini}, G. and {Heussaff}, V. and {Lazzati}, D. and {Melandri}, A. and {Nava}, L. and {Vergani}, S.~D. and {Tagliaferri}, G.},
        title = "{Comparing the spectral lag of short and long gamma-ray bursts and its relation with the luminosity}",
      journal = {\mnras},
         year = 2015,
        month = jan,
       volume = {446},
       number = {2},
        pages = {1129-1138},
          doi = {10.1093/mnras/stu2153},
archivePrefix = {arXiv},
       eprint = {1410.5216},
 primaryClass = {astro-ph.HE},
       adsurl = {https://ui.adsabs.harvard.edu/abs/2015MNRAS.446.1129B}
}

@ARTICLE{Fong2013,
       author = {{Fong}, W. and {Berger}, E. and {Chornock}, R. and {Margutti}, R. and {Levan}, A.~J. and {Tanvir}, N.~R. and {Tunnicliffe}, R.~L. and {Czekala}, I. and {Fox}, D.~B. and {Perley}, D.~A. and {Cenko}, S.~B. and {Zauderer}, B.~A. and {Laskar}, T. and {Persson}, S.~E. and {Monson}, A.~J. and {Kelson}, D.~D. and {Birk}, C. and {Murphy}, D. and {Servillat}, M. and {Anglada}, G.},
        title = "{Demographics of the Galaxies Hosting Short-duration Gamma-Ray Bursts}",
      journal = {\apj},
         year = 2013,
        month = may,
       volume = {769},
       number = {1},
          eid = {56},
        pages = {56},
          doi = {10.1088/0004-637X/769/1/56},
archivePrefix = {arXiv},
       eprint = {1302.3221},
 primaryClass = {astro-ph.HE},
       adsurl = {https://ui.adsabs.harvard.edu/abs/2013ApJ...769...56F}
}

@ARTICLE{Berger2014,
       author = {{Berger}, Edo},
        title = "{Short-Duration Gamma-Ray Bursts}",
      journal = {\araa},
         year = 2014,
        month = aug,
       volume = {52},
        pages = {43-105},
          doi = {10.1146/annurev-astro-081913-035926},
archivePrefix = {arXiv},
       eprint = {1311.2603},
 primaryClass = {astro-ph.HE},
       adsurl = {https://ui.adsabs.harvard.edu/abs/2014ARA&A..52...43B}
}

@ARTICLE{2013A&A...558A..33A,
       author = {{Astropy Collaboration} and {Robitaille}, Thomas P. and {Tollerud}, Erik J. and {Greenfield}, Perry and {Droettboom}, Michael and {Bray}, Erik and {Aldcroft}, Tom and {Davis}, Matt and {Ginsburg}, Adam and {Price-Whelan}, Adrian M. and {Kerzendorf}, Wolfgang E. and {Conley}, Alexander and {Crighton}, Neil and {Barbary}, Kyle and {Muna}, Demitri and {Ferguson}, Henry and {Grollier}, Fr{\'e}d{\'e}ric and {Parikh}, Madhura M. and {Nair}, Prasanth H. and {Unther}, Hans M. and {Deil}, Christoph and {Woillez}, Julien and {Conseil}, Simon and {Kramer}, Roban and {Turner}, James E.~H. and {Singer}, Leo and {Fox}, Ryan and {Weaver}, Benjamin A. and {Zabalza}, Victor and {Edwards}, Zachary I. and {Azalee Bostroem}, K. and {Burke}, D.~J. and {Casey}, Andrew R. and {Crawford}, Steven M. and {Dencheva}, Nadia and {Ely}, Justin and {Jenness}, Tim and {Labrie}, Kathleen and {Lim}, Pey Lian and {Pierfederici}, Francesco and {Pontzen}, Andrew and {Ptak}, Andy and {Refsdal}, Brian and {Servillat}, Mathieu and {Streicher}, Ole},
        title = "{Astropy: A community Python package for astronomy}",
      journal = {\aap},
         year = 2013,
        month = oct,
       volume = {558},
          eid = {A33},
        pages = {A33},
          doi = {10.1051/0004-6361/201322068},
archivePrefix = {arXiv},
       eprint = {1307.6212},
 primaryClass = {astro-ph.IM},
       adsurl = {https://ui.adsabs.harvard.edu/abs/2013A&A...558A..33A}
}

@ARTICLE{2018AJ....156..123A,
       author = {{Astropy Collaboration} and {Price-Whelan}, A.~M. and {Sip{\H{o}}cz}, B.~M. and {G{\"u}nther}, H.~M. and {Lim}, P.~L. and {Crawford}, S.~M. and {Conseil}, S. and {Shupe}, D.~L. and {Craig}, M.~W. and {Dencheva}, N. and {Ginsburg}, A. and {VanderPlas}, J.~T. and {Bradley}, L.~D. and {P{\'e}rez-Su{\'a}rez}, D. and {de Val-Borro}, M. and {Aldcroft}, T.~L. and {Cruz}, K.~L. and {Robitaille}, T.~P. and {Tollerud}, E.~J. and {Ardelean}, C. and {Babej}, T. and {Bach}, Y.~P. and {Bachetti}, M. and {Bakanov}, A.~V. and {Bamford}, S.~P. and {Barentsen}, G. and {Barmby}, P. and {Baumbach}, A. and {Berry}, K.~L. and {Biscani}, F. and {Boquien}, M. and {Bostroem}, K.~A. and {Bouma}, L.~G. and {Brammer}, G.~B. and {Bray}, E.~M. and {Breytenbach}, H. and {Buddelmeijer}, H. and {Burke}, D.~J. and {Calderone}, G. and {Cano Rodr{\'\i}guez}, J.~L. and {Cara}, M. and {Cardoso}, J.~V.~M. and {Cheedella}, S. and {Copin}, Y. and {Corrales}, L. and {Crichton}, D. and {D'Avella}, D. and {Deil}, C. and {Depagne}, {\'E}. and {Dietrich}, J.~P. and {Donath}, A. and {Droettboom}, M. and {Earl}, N. and {Erben}, T. and {Fabbro}, S. and {Ferreira}, L.~A. and {Finethy}, T. and {Fox}, R.~T. and {Garrison}, L.~H. and {Gibbons}, S.~L.~J. and {Goldstein}, D.~A. and {Gommers}, R. and {Greco}, J.~P. and {Greenfield}, P. and {Groener}, A.~M. and {Grollier}, F. and {Hagen}, A. and {Hirst}, P. and {Homeier}, D. and {Horton}, A.~J. and {Hosseinzadeh}, G. and {Hu}, L. and {Hunkeler}, J.~S. and {Ivezi{\'c}}, {\v{Z}}. and {Jain}, A. and {Jenness}, T. and {Kanarek}, G. and {Kendrew}, S. and {Kern}, N.~S. and {Kerzendorf}, W.~E. and {Khvalko}, A. and {King}, J. and {Kirkby}, D. and {Kulkarni}, A.~M. and {Kumar}, A. and {Lee}, A. and {Lenz}, D. and {Littlefair}, S.~P. and {Ma}, Z. and {Macleod}, D.~M. and {Mastropietro}, M. and {McCully}, C. and {Montagnac}, S. and {Morris}, B.~M. and {Mueller}, M. and {Mumford}, S.~J. and {Muna}, D. and {Murphy}, N.~A. and {Nelson}, S. and {Nguyen}, G.~H. and {Ninan}, J.~P. and {N{\"o}the}, M. and {Ogaz}, S. and {Oh}, S. and {Parejko}, J.~K. and {Parley}, N. and {Pascual}, S. and {Patil}, R. and {Patil}, A.~A. and {Plunkett}, A.~L. and {Prochaska}, J.~X. and {Rastogi}, T. and {Reddy Janga}, V. and {Sabater}, J. and {Sakurikar}, P. and {Seifert}, M. and {Sherbert}, L.~E. and {Sherwood-Taylor}, H. and {Shih}, A.~Y. and {Sick}, J. and {Silbiger}, M.~T. and {Singanamalla}, S. and {Singer}, L.~P. and {Sladen}, P.~H. and {Sooley}, K.~A. and {Sornarajah}, S. and {Streicher}, O. and {Teuben}, P. and {Thomas}, S.~W. and {Tremblay}, G.~R. and {Turner}, J.~E.~H. and {Terr{\'o}n}, V. and {van Kerkwijk}, M.~H. and {de la Vega}, A. and {Watkins}, L.~L. and {Weaver}, B.~A. and {Whitmore}, J.~B. and {Woillez}, J. and {Zabalza}, V. and {Astropy Contributors}},
        title = "{The Astropy Project: Building an Open-science Project and Status of the v2.0 Core Package}",
      journal = {\aj},
         year = 2018,
        month = sep,
       volume = {156},
       number = {3},
          eid = {123},
        pages = {123},
          doi = {10.3847/1538-3881/aabc4f},
archivePrefix = {arXiv},
       eprint = {1801.02634},
 primaryClass = {astro-ph.IM},
       adsurl = {https://ui.adsabs.harvard.edu/abs/2018AJ....156..123A}
}

@ARTICLE{2006ApJ...641..993K,
       author = {{Kann}, D.~A. and {Klose}, S. and {Zeh}, A.},
        title = "{Signatures of Extragalactic Dust in Pre-Swift GRB Afterglows}",
      journal = {\apj},
         year = 2006,
        month = apr,
       volume = {641},
       number = {2},
        pages = {993-1009},
          doi = {10.1086/500652},
archivePrefix = {arXiv},
       eprint = {astro-ph/0512575},
 primaryClass = {astro-ph},
       adsurl = {https://ui.adsabs.harvard.edu/abs/2006ApJ...641..993K}
}

@ARTICLE{2010ApJ...720.1513K,
       author = {{Kann}, D.~A. and {Klose}, S. and {Zhang}, B. and {Malesani}, D. and {Nakar}, E. and {Pozanenko}, A. and {Wilson}, A.~C. and {Butler}, N.~R. and {Jakobsson}, P. and {Schulze}, S. and {Andreev}, M. and {Antonelli}, L.~A. and {Bikmaev}, I.~F. and {Biryukov}, V. and {B{\"o}ttcher}, M. and {Burenin}, R.~A. and {Castro Cer{\'o}n}, J.~M. and {Castro-Tirado}, A.~J. and {Chincarini}, G. and {Cobb}, B.~E. and {Covino}, S. and {D'Avanzo}, P. and {D'Elia}, V. and {Della Valle}, M. and {de Ugarte Postigo}, A. and {Efimov}, Yu. and {Ferrero}, P. and {Fugazza}, D. and {Fynbo}, J.~P.~U. and {G{\r{a}}lfalk}, M. and {Grundahl}, F. and {Gorosabel}, J. and {Gupta}, S. and {Guziy}, S. and {Hafizov}, B. and {Hjorth}, J. and {Holhjem}, K. and {Ibrahimov}, M. and {Im}, M. and {Israel}, G.~L. and {Je{\'l}inek}, M. and {Jensen}, B.~L. and {Karimov}, R. and {Khamitov}, I.~M. and {Kizilo{\v{g}}lu}, {\"U}. and {Klunko}, E. and {Kub{\'a}nek}, P. and {Kutyrev}, A.~S. and {Laursen}, P. and {Levan}, A.~J. and {Mannucci}, F. and {Martin}, C.~M. and {Mescheryakov}, A. and {Mirabal}, N. and {Norris}, J.~P. and {Ovaldsen}, J. -E. and {Paraficz}, D. and {Pavlenko}, E. and {Piranomonte}, S. and {Rossi}, A. and {Rumyantsev}, V. and {Salinas}, R. and {Sergeev}, A. and {Sharapov}, D. and {Sollerman}, J. and {Stecklum}, B. and {Stella}, L. and {Tagliaferri}, G. and {Tanvir}, N.~R. and {Telting}, J. and {Testa}, V. and {Updike}, A.~C. and {Volnova}, A. and {Watson}, D. and {Wiersema}, K. and {Xu}, D.},
        title = "{The Afterglows of Swift-era Gamma-ray Bursts. I. Comparing pre-Swift and Swift-era Long/Soft (Type II) GRB Optical Afterglows}",
      journal = {\apj},
         year = 2010,
        month = sep,
       volume = {720},
       number = {2},
        pages = {1513-1558},
          doi = {10.1088/0004-637X/720/2/1513},
archivePrefix = {arXiv},
       eprint = {0712.2186},
 primaryClass = {astro-ph},
       adsurl = {https://ui.adsabs.harvard.edu/abs/2010ApJ...720.1513K}
}

@ARTICLE{2013MNRAS.430..163Q,
       author = {{Qin}, Yi-Ping and {Chen}, Zhi-Fu},
        title = "{Statistical classification of gamma-ray bursts based on the Amati relation}",
      journal = {\mnras},
         year = 2013,
        month = mar,
       volume = {430},
       number = {1},
        pages = {163-173},
          doi = {10.1093/mnras/sts547},
archivePrefix = {arXiv},
       eprint = {1311.4713},
 primaryClass = {astro-ph.HE},
       adsurl = {https://ui.adsabs.harvard.edu/abs/2013MNRAS.430..163Q}
}

@ARTICLE{2000ApJS..126...19P,
       author = {{Preece}, R.~D. and {Briggs}, M.~S. and {Mallozzi}, R.~S. and {Pendleton}, G.~N. and {Paciesas}, W.~S. and {Band}, D.~L.},
        title = "{The BATSE Gamma-Ray Burst Spectral Catalog. I. High Time Resolution Spectroscopy of Bright Bursts Using High Energy Resolution Data}",
      journal = {\apjs},
         year = 2000,
        month = jan,
       volume = {126},
       number = {1},
        pages = {19-36},
          doi = {10.1086/313289},
archivePrefix = {arXiv},
       eprint = {astro-ph/9908119},
 primaryClass = {astro-ph},
       adsurl = {https://ui.adsabs.harvard.edu/abs/2000ApJS..126...19P}
}

@ARTICLE{2021ApJ...913...60P,
       author = {{Poolakkil}, S. and {Preece}, R. and {Fletcher}, C. and {Goldstein}, A. and {Bhat}, P.~N. and {Bissaldi}, E. and {Briggs}, M.~S. and {Burns}, E. and {Cleveland}, W.~H. and {Giles}, M.~M. and {Hui}, C.~M. and {Kocevski}, D. and {Lesage}, S. and {Mailyan}, B. and {Malacaria}, C. and {Paciesas}, W.~S. and {Roberts}, O.~J. and {Veres}, P. and {von Kienlin}, A. and {Wilson-Hodge}, C.~A.},
        title = "{The Fermi-GBM Gamma-Ray Burst Spectral Catalog: 10 yr of Data}",
      journal = {\apj},
         year = 2021,
        month = may,
       volume = {913},
       number = {1},
          eid = {60},
        pages = {60},
          doi = {10.3847/1538-4357/abf24d},
archivePrefix = {arXiv},
       eprint = {2103.13528},
 primaryClass = {astro-ph.HE},
       adsurl = {https://ui.adsabs.harvard.edu/abs/2021ApJ...913...60P}
}

@ARTICLE{2011A&A...534A.108K,
       author = {{Kr{\"u}hler}, T. and {Greiner}, J. and {Schady}, P. and {Savaglio}, S. and {Afonso}, P.~M.~J. and {Clemens}, C. and {Elliott}, J. and {Filgas}, R. and {Gruber}, D. and {Kann}, D.~A. and {Klose}, S. and {K{\"u}pc{\"u}-Yolda{\c{s}}}, A. and {McBreen}, S. and {Olivares}, F. and {Pierini}, D. and {Rau}, A. and {Rossi}, A. and {Nardini}, M. and {Nicuesa Guelbenzu}, A. and {Sudilovsky}, V. and {Updike}, A.~C.},
        title = "{The SEDs and host galaxies of the dustiest GRB afterglows}",
      journal = {\aap},
         year = 2011,
        month = oct,
       volume = {534},
          eid = {A108},
        pages = {A108},
          doi = {10.1051/0004-6361/201117428},
archivePrefix = {arXiv},
       eprint = {1108.0674},
 primaryClass = {astro-ph.CO},
       adsurl = {https://ui.adsabs.harvard.edu/abs/2011A&A...534A.108K}
}

@ARTICLE{2012A&A...537A..15S,
       author = {{Schady}, P. and {Dwelly}, T. and {Page}, M.~J. and {Kr{\"u}hler}, T. and {Greiner}, J. and {Oates}, S.~R. and {de Pasquale}, M. and {Nardini}, M. and {Roming}, P.~W.~A. and {Rossi}, A. and {Still}, M.},
        title = "{The dust extinction curves of gamma-ray burst host galaxies}",
      journal = {\aap},
         year = 2012,
        month = jan,
       volume = {537},
          eid = {A15},
        pages = {A15},
          doi = {10.1051/0004-6361/201117414},
archivePrefix = {arXiv},
       eprint = {1110.3218},
 primaryClass = {astro-ph.CO},
       adsurl = {https://ui.adsabs.harvard.edu/abs/2012A&A...537A..15S}
}

@ARTICLE{2023ApJ...949L..39S,
       author = {{Srinivasaragavan}, Gokul P. and {O'Connor}, Brendan and {Cenko}, S. Bradley and {Dittmann}, Alexander J. and {Yang}, Sheng and {Sollerman}, Jesper and {Anupama}, G.~C. and {Barway}, Sudhanshu and {Bhalerao}, Varun and {Kumar}, Harsh and {Swain}, Vishwajeet and {Hammerstein}, Erica and {Holt}, Isiah and {Anand}, Shreya and {Andreoni}, Igor and {Coughlin}, Michael W. and {Dichiara}, Simone and {Gal-Yam}, Avishay and {Miller}, M. Coleman and {Soon}, Jaime and {Soria}, Roberto and {Durbak}, Joseph and {Gillanders}, James H. and {Laha}, Sibasish and {Moore}, Anna M. and {Ragosta}, Fabio and {Troja}, Eleonora},
        title = "{A Sensitive Search for Supernova Emission Associated with the Extremely Energetic and Nearby GRB 221009A}",
      journal = {\apjl},
         year = 2023,
        month = jun,
       volume = {949},
       number = {2},
          eid = {L39},
        pages = {L39},
          doi = {10.3847/2041-8213/accf97},
archivePrefix = {arXiv},
       eprint = {2303.12849},
 primaryClass = {astro-ph.HE},
       adsurl = {https://ui.adsabs.harvard.edu/abs/2023ApJ...949L..39S}
}

@ARTICLE{2026arXiv260518973O,
       author = {{O'Connor}, Brendan and {Busmann}, Malte and {Hall}, Xander J. and {Taguchi}, Kenta and {Tanaka}, Masaomi and {Gruen}, Daniel and {Toshikage}, Seiji and {Amsellem}, Ariel J. and {Zhu}, Ziyuan and {Palmese}, Antonella and {Green}, Dylan and {Banovetz}, John and {Yang}, Yu-Han and {Troja}, Eleonora and {van Eerten}, Hendrik and {Gassert}, Julius and {Maleki}, Mitra and {Bailey}, Steven and {BenZvi}, Segev and {Cabrera}, Tomas and {Kunnumkai}, Keerthi and {Myers}, Adam and {Ries}, Christoph and {Schlegel}, David and {Schmidt}, Michael and {Wilke}, Silona and {Yadav}, Muskan},
        title = "{GRB 260310A/SN 2026fgk: Photometric and Spectroscopic Evolution of a Nearby GRB-Supernova and an Exceptionally Bright Afterglow at z=0.153}",
      journal = {arXiv e-prints},
         year = 2026,
        month = may,
          eid = {arXiv:2605.18973},
        pages = {arXiv:2605.18973},
          doi = {10.48550/arXiv.2605.18973},
archivePrefix = {arXiv},
       eprint = {2605.18973},
 primaryClass = {astro-ph.HE},
       adsurl = {https://ui.adsabs.harvard.edu/abs/2026arXiv260518973O}
}

@ARTICLE{2025A&C....5200954F,
       author = {{Finneran}, Gabriel and {Cotter}, Laura and {Martin-Carrillo}, Antonio},
        title = "{The GRBSN webtool: An open-source repository for gamma-ray burst-supernova associations}",
      journal = {Astronomy and Computing},
         year = 2025,
        month = jul,
       volume = {52},
          eid = {100954},
        pages = {100954},
          doi = {10.1016/j.ascom.2025.100954},
archivePrefix = {arXiv},
       eprint = {2411.08866},
 primaryClass = {astro-ph.HE},
       adsurl = {https://ui.adsabs.harvard.edu/abs/2025A&C....5200954F}
}

@ARTICLE{2022A&A...667A..62B,
       author = {{Brennan}, S.~J. and {Fraser}, M.},
        title = "{The Automated Photometry of Transients pipeline (AUTOPHOT)}",
      journal = {\aap},
         year = 2022,
        month = nov,
       volume = {667},
          eid = {A62},
        pages = {A62},
          doi = {10.1051/0004-6361/202243067},
archivePrefix = {arXiv},
       eprint = {2201.02635},
 primaryClass = {astro-ph.IM},
       adsurl = {https://ui.adsabs.harvard.edu/abs/2022A&A...667A..62B}
}

@ARTICLE{2024ApJ...966..219B,
       author = {{Bianco}, C.~L. and {Mirtorabi}, M.~T. and {Moradi}, R. and {Rastegarnia}, F. and {Rueda}, J.~A. and {Ruffini}, R. and {Wang}, Y. and {Della Valle}, M. and {Li}, Liang and {Zhang}, S.~R.},
        title = "{Probing Electromagnetic Gravitational-wave Emission Coincidence in a Type I Binary-driven Hypernova Family of Long Gamma-Ray Bursts at Very High Redshift}",
      journal = {\apj},
         year = 2024,
        month = may,
       volume = {966},
       number = {2},
          eid = {219},
        pages = {219},
          doi = {10.3847/1538-4357/ad2fa9},
archivePrefix = {arXiv},
       eprint = {2306.05855},
 primaryClass = {astro-ph.HE},
       adsurl = {https://ui.adsabs.harvard.edu/abs/2024ApJ...966..219B}
}

@ARTICLE{2024GCN.37536....1R,
       author = {{Ruffini}, R. and {Bianco}, C.~L. and {Della Valle}, M. and {Li}, Liang and {Mirtorabi}, M.~T. and {Moradi}, R. and {Rastegar Nia}, F. and {Rueda}, J.~A. and {Wang}, Y. and {Icranet Team}},
        title = "{GRB 240825A: The nature of the afterglow motivates the search of the associated supernova}",
      journal = {GRB Coordinates Network},
         year = 2024,
        month = sep,
       volume = {37536},
        pages = {1},
       adsurl = {https://ui.adsabs.harvard.edu/abs/2024GCN.37536....1R}
}

@ARTICLE{1996A&AS..117..393B,
       author = {{Bertin}, E. and {Arnouts}, S.},
        title = "{SExtractor: Software for source extraction.}",
      journal = {\aaps},
         year = 1996,
        month = jun,
       volume = {117},
        pages = {393-404},
          doi = {10.1051/aas:1996164},
       adsurl = {https://ui.adsabs.harvard.edu/abs/1996A&AS..117..393B}
}

@INPROCEEDINGS{1996ASPC..101...17A,
       author = {{Arnaud}, K.~A.},
        title = "{XSPEC: The First Ten Years}",
    booktitle = {Astronomical Data Analysis Software and Systems V},
         year = 1996,
       editor = {{Jacoby}, George H. and {Barnes}, Jeannette},
       series = {Astronomical Society of the Pacific Conference Series},
       volume = {101},
        month = jan,
        pages = {17},
       adsurl = {https://ui.adsabs.harvard.edu/abs/1996ASPC..101...17A}
}

@ARTICLE{2022ApJ...935..167A,
       author = {{Astropy Collaboration} and {Price-Whelan}, Adrian M. and {Lim}, Pey Lian and {Earl}, Nicholas and {Starkman}, Nathaniel and {Bradley}, Larry and {Shupe}, David L. and {Patil}, Aarya A. and {Corrales}, Lia and {Brasseur}, C.~E. and {N{\"o}the}, Maximilian and {Donath}, Axel and {Tollerud}, Erik and {Morris}, Brett M. and {Ginsburg}, Adam and {Vaher}, Eero and {Weaver}, Benjamin A. and {Tocknell}, James and {Jamieson}, William and {van Kerkwijk}, Marten H. and {Robitaille}, Thomas P. and {Merry}, Bruce and {Bachetti}, Matteo and {G{\"u}nther}, H. Moritz and {Aldcroft}, Thomas L. and {Alvarado-Montes}, Jaime A. and {Archibald}, Anne M. and {B{\'o}di}, Attila and {Bapat}, Shreyas and {Barentsen}, Geert and {Baz{\'a}n}, Juanjo and {Biswas}, Manish and {Boquien}, M{\'e}d{\'e}ric and {Burke}, D.~J. and {Cara}, Daria and {Cara}, Mihai and {Conroy}, Kyle E. and {Conseil}, Simon and {Craig}, Matthew W. and {Cross}, Robert M. and {Cruz}, Kelle L. and {D'Eugenio}, Francesco and {Dencheva}, Nadia and {Devillepoix}, Hadrien A.~R. and {Dietrich}, J{\"o}rg P. and {Eigenbrot}, Arthur Davis and {Erben}, Thomas and {Ferreira}, Leonardo and {Foreman-Mackey}, Daniel and {Fox}, Ryan and {Freij}, Nabil and {Garg}, Suyog and {Geda}, Robel and {Glattly}, Lauren and {Gondhalekar}, Yash and {Gordon}, Karl D. and {Grant}, David and {Greenfield}, Perry and {Groener}, Austen M. and {Guest}, Steve and {Gurovich}, Sebastian and {Handberg}, Rasmus and {Hart}, Akeem and {Hatfield-Dodds}, Zac and {Homeier}, Derek and {Hosseinzadeh}, Griffin and {Jenness}, Tim and {Jones}, Craig K. and {Joseph}, Prajwel and {Kalmbach}, J. Bryce and {Karamehmetoglu}, Emir and {Ka{\l}uszy{\'n}ski}, Miko{\l}aj and {Kelley}, Michael S.~P. and {Kern}, Nicholas and {Kerzendorf}, Wolfgang E. and {Koch}, Eric W. and {Kulumani}, Shankar and {Lee}, Antony and {Ly}, Chun and {Ma}, Zhiyuan and {MacBride}, Conor and {Maljaars}, Jakob M. and {Muna}, Demitri and {Murphy}, N.~A. and {Norman}, Henrik and {O'Steen}, Richard and {Oman}, Kyle A. and {Pacifici}, Camilla and {Pascual}, Sergio and {Pascual-Granado}, J. and {Patil}, Rohit R. and {Perren}, Gabriel I. and {Pickering}, Timothy E. and {Rastogi}, Tanuj and {Roulston}, Benjamin R. and {Ryan}, Daniel F. and {Rykoff}, Eli S. and {Sabater}, Jose and {Sakurikar}, Parikshit and {Salgado}, Jes{\'u}s and {Sanghi}, Aniket and {Saunders}, Nicholas and {Savchenko}, Volodymyr and {Schwardt}, Ludwig and {Seifert-Eckert}, Michael and {Shih}, Albert Y. and {Jain}, Anany Shrey and {Shukla}, Gyanendra and {Sick}, Jonathan and {Simpson}, Chris and {Singanamalla}, Sudheesh and {Singer}, Leo P. and {Singhal}, Jaladh and {Sinha}, Manodeep and {Sip{\H{o}}cz}, Brigitta M. and {Spitler}, Lee R. and {Stansby}, David and {Streicher}, Ole and {{\v{S}}umak}, Jani and {Swinbank}, John D. and {Taranu}, Dan S. and {Tewary}, Nikita and {Tremblay}, Grant R. and {de Val-Borro}, Miguel and {Van Kooten}, Samuel J. and {Vasovi{\'c}}, Zlatan and {Verma}, Shresth and {de Miranda Cardoso}, Jos{\'e} Vin{\'\i}cius and {Williams}, Peter K.~G. and {Wilson}, Tom J. and {Winkel}, Benjamin and {Wood-Vasey}, W.~M. and {Xue}, Rui and {Yoachim}, Peter and {Zhang}, Chen and {Zonca}, Andrea and {Astropy Project Contributors}},
        title = "{The Astropy Project: Sustaining and Growing a Community-oriented Open-source Project and the Latest Major Release (v5.0) of the Core Package}",
      journal = {\apj},
         year = 2022,
        month = aug,
       volume = {935},
       number = {2},
          eid = {167},
        pages = {167},
          doi = {10.3847/1538-4357/ac7c74},
archivePrefix = {arXiv},
       eprint = {2206.14220},
 primaryClass = {astro-ph.IM},
       adsurl = {https://ui.adsabs.harvard.edu/abs/2022ApJ...935..167A}
}

@ARTICLE{2011ApJ...737..103S,
       author = {{Schlafly}, Edward F. and {Finkbeiner}, Douglas P.},
        title = "{Measuring Reddening with Sloan Digital Sky Survey Stellar Spectra and Recalibrating SFD}",
      journal = {\apj},
         year = 2011,
        month = aug,
       volume = {737},
       number = {2},
          eid = {103},
        pages = {103},
          doi = {10.1088/0004-637X/737/2/103},
archivePrefix = {arXiv},
       eprint = {1012.4804},
 primaryClass = {astro-ph.GA},
       adsurl = {https://ui.adsabs.harvard.edu/abs/2011ApJ...737..103S}
}

@ARTICLE{Fong2010,
       author = {{Fong}, W. and {Berger}, E. and {Fox}, D.~B.},
        title = "{Hubble Space Telescope Observations of Short Gamma-Ray Burst Host Galaxies: Morphologies, Offsets, and Local Environments}",
      journal = {\apj},
         year = 2010,
        month = jan,
       volume = {708},
       number = {1},
        pages = {9-25},
          doi = {10.1088/0004-637X/708/1/9},
archivePrefix = {arXiv},
       eprint = {0909.1804},
 primaryClass = {astro-ph.HE},
       adsurl = {https://ui.adsabs.harvard.edu/abs/2010ApJ...708....9F}
}

@ARTICLE{2023arXiv231216265G,
       author = {{Gupta}, Rahul},
        title = "{Multiwavelength Observations of Gamma Ray Bursts}",
      journal = {arXiv e-prints},
         year = 2023,
        month = dec,
          eid = {arXiv:2312.16265},
        pages = {arXiv:2312.16265},
          doi = {10.48550/arXiv.2312.16265},
archivePrefix = {arXiv},
       eprint = {2312.16265},
 primaryClass = {astro-ph.HE},
       adsurl = {https://ui.adsabs.harvard.edu/abs/2023arXiv231216265G}
}

@ARTICLE{2024A&A...683A..55C,
       author = {{Castro-Tirado}, A.~J. and {Gupta}, R. and {Pandey}, S.~B. and {Nicuesa Guelbenzu}, A. and {Eikenberry}, S. and {Ackley}, K. and {Gerarts}, A. and {Valeev}, A.~F. and {Jeong}, S. and {Park}, I.~H. and {Oates}, S.~R. and {Zhang}, B. -B. and {S{\'a}nchez-Ram{\'\i}rez}, R. and {Mart{\'\i}n-Carrillo}, A. and {Tello}, J.~C. and {Jel{\'\i}nek}, M. and {Hu}, Y. -D. and {Cunniffe}, R. and {Sokolov}, V.~V. and {Guziy}, S. and {Ferrero}, P. and {Caballero-Garc{\'\i}a}, M.~D. and {Ror}, A.~K. and {Aryan}, A. and {Castro Tirado}, M.~A. and {Fern{\'a}ndez-Garc{\'\i}a}, E. and {Gritsevich}, M. and {Olivares}, I. and {P{\'e}rez-Garc{\'\i}a}, I. and {Castro Cer{\'o}n}, J.~M. and {Cepa}, J.},
        title = "{Revealing the characteristics of the dark GRB 150309A: Dust extinguished or high-z?}",
      journal = {\aap},
         year = 2024,
        month = mar,
       volume = {683},
          eid = {A55},
        pages = {A55},
          doi = {10.1051/0004-6361/202346042},
archivePrefix = {arXiv},
       eprint = {2310.08503},
 primaryClass = {astro-ph.HE},
       adsurl = {https://ui.adsabs.harvard.edu/abs/2024A&A...683A..55C}
}

@ARTICLE{2002ApJ...579..386N,
       author = {{Norris}, J.~P.},
        title = "{Implications of the Lag-Luminosity Relationship for Unified Gamma-Ray Burst Paradigms}",
      journal = {\apj},
         year = 2002,
        month = nov,
       volume = {579},
       number = {1},
        pages = {386-403},
          doi = {10.1086/342747},
archivePrefix = {arXiv},
       eprint = {astro-ph/0201503},
 primaryClass = {astro-ph},
       adsurl = {https://ui.adsabs.harvard.edu/abs/2002ApJ...579..386N}
}

@ARTICLE{2010ApJ...725.1965L,
       author = {{L{\"u}}, Hou-Jun and {Liang}, En-Wei and {Zhang}, Bin-Bin and {Zhang}, Bing},
        title = "{A New Classification Method for Gamma-ray Bursts}",
      journal = {\apj},
         year = 2010,
        month = dec,
       volume = {725},
       number = {2},
        pages = {1965-1970},
          doi = {10.1088/0004-637X/725/2/1965},
archivePrefix = {arXiv},
       eprint = {1001.0598},
 primaryClass = {astro-ph.HE},
       adsurl = {https://ui.adsabs.harvard.edu/abs/2010ApJ...725.1965L}
}

@ARTICLE{2008A&A...482..349G,
       author = {{Giallongo}, E. and {Ragazzoni}, R. and {Grazian}, A. and {Baruffolo}, A. and {Beccari}, G. and {de Santis}, C. and {Diolaiti}, E. and {di Paola}, A. and {Farinato}, J. and {Fontana}, A. and {Gallozzi}, S. and {Gasparo}, F. and {Gentile}, G. and {Green}, R. and {Hill}, J. and {Kuhn}, O. and {Pasian}, F. and {Pedichini}, F. and {Radovich}, M. and {Salinari}, P. and {Smareglia}, R. and {Speziali}, R. and {Testa}, V. and {Thompson}, D. and {Vernet}, E. and {Wagner}, R.~M.},
        title = "{The performance of the blue prime focus large binocular camera at the large binocular telescope}",
      journal = {\aap},
         year = 2008,
        month = apr,
       volume = {482},
       number = {1},
        pages = {349-357},
          doi = {10.1051/0004-6361:20078402},
archivePrefix = {arXiv},
       eprint = {0801.1474},
 primaryClass = {astro-ph},
       adsurl = {https://ui.adsabs.harvard.edu/abs/2008A&A...482..349G}
}

@ARTICLE{2024MNRAS.532.1434Z,
       author = {{Zhu}, Si-Yuan and {Sun}, Wan-Peng and {Ma}, Da-Ling and {Zhang}, Fu-Wen},
        title = "{Classification of Fermi gamma-ray bursts based on machine learning}",
      journal = {\mnras},
         year = 2024,
        month = aug,
       volume = {532},
       number = {2},
        pages = {1434-1443},
          doi = {10.1093/mnras/stae1594},
archivePrefix = {arXiv},
       eprint = {2406.05357},
 primaryClass = {astro-ph.HE},
       adsurl = {https://ui.adsabs.harvard.edu/abs/2024MNRAS.532.1434Z}
}

@ARTICLE{2025ApJ...981...14N,
       author = {{Negro}, Michela and {Cibrario}, Nicol{\'o} and {Burns}, Eric and {Wood}, Joshua and {Goldstein}, Adam and {Dal Canton}, Tito},
        title = "{Prompt Gamma-Ray Burst Recognition through Waterfalls and Deep Learning}",
      journal = {\apj},
         year = 2025,
        month = mar,
       volume = {981},
       number = {1},
          eid = {14},
        pages = {14},
          doi = {10.3847/1538-4357/ada8a9},
archivePrefix = {arXiv},
       eprint = {2406.03643},
 primaryClass = {astro-ph.HE},
       adsurl = {https://ui.adsabs.harvard.edu/abs/2025ApJ...981...14N}
}

@ARTICLE{2024ApJ...974..120N,
       author = {{Nuessle}, P. and {Racusin}, J.~L. and {White}, N.~E.},
        title = "{GRB Progenitor Classification from Gamma-Ray Burst Prompt and Afterglow Observations}",
      journal = {\apj},
         year = 2024,
        month = oct,
       volume = {974},
       number = {1},
          eid = {120},
        pages = {120},
          doi = {10.3847/1538-4357/ad6a56},
archivePrefix = {arXiv},
       eprint = {2407.08857},
 primaryClass = {astro-ph.HE},
       adsurl = {https://ui.adsabs.harvard.edu/abs/2024ApJ...974..120N}
}

@ARTICLE{2010ApJ...711.1073U,
       author = {{Ukwatta}, T.~N. and {Stamatikos}, M. and {Dhuga}, K.~S. and {Sakamoto}, T. and {Barthelmy}, S.~D. and {Eskandarian}, A. and {Gehrels}, N. and {Maximon}, L.~C. and {Norris}, J.~P. and {Parke}, W.~C.},
        title = "{Spectral Lags and the Lag-Luminosity Relation: An Investigation with Swift BAT Gamma-ray Bursts}",
      journal = {\apj},
         year = 2010,
        month = mar,
       volume = {711},
       number = {2},
        pages = {1073-1086},
          doi = {10.1088/0004-637X/711/2/1073},
archivePrefix = {arXiv},
       eprint = {0908.2370},
 primaryClass = {astro-ph.HE},
       adsurl = {https://ui.adsabs.harvard.edu/abs/2010ApJ...711.1073U}
}

@ARTICLE{2018A&A...615A.136T,
       author = {{Tanga}, M. and {Kr{\"u}hler}, T. and {Schady}, P. and {Klose}, S. and {Graham}, J.~F. and {Greiner}, J. and {Kann}, D.~A. and {Nardini}, M.},
        title = "{The environment of the SN-less GRB 111005A at z = 0.0133}",
      journal = {\aap},
         year = 2018,
        month = jul,
       volume = {615},
          eid = {A136},
        pages = {A136},
          doi = {10.1051/0004-6361/201731799},
archivePrefix = {arXiv},
       eprint = {1708.06270},
 primaryClass = {astro-ph.GA},
       adsurl = {https://ui.adsabs.harvard.edu/abs/2018A&A...615A.136T}
}

@ARTICLE{2025JHEAp..4700384L,
       author = {{Lenart}, Aleksander {\L}. and {Dainotti}, Maria G. and {Khatiya}, Nikita and {Bal}, Dhruv and {Hartmann}, Dieter H. and {Fraija}, Nissim and {Zhang}, Bing},
        title = "{The multiwavelength correlations quest for central engines of GRB plateaus: Magnetar vs black hole spin-down}",
      journal = {Journal of High Energy Astrophysics},
         year = 2025,
        month = jul,
       volume = {47},
          eid = {100384},
        pages = {100384},
          doi = {10.1016/j.jheap.2025.100384},
archivePrefix = {arXiv},
       eprint = {2502.16204},
 primaryClass = {astro-ph.HE},
       adsurl = {https://ui.adsabs.harvard.edu/abs/2025JHEAp..4700384L}
}

@ARTICLE{2020ApJ...904...97D,
       author = {{Dainotti}, M.~G. and {Lenart}, A. {\L}. and {Sarracino}, G. and {Nagataki}, S. and {Capozziello}, S. and {Fraija}, N.},
        title = "{The X-Ray Fundamental Plane of the Platinum Sample, the Kilonovae, and the SNe Ib/c Associated with GRBs}",
      journal = {\apj},
         year = 2020,
        month = dec,
       volume = {904},
       number = {2},
          eid = {97},
        pages = {97},
          doi = {10.3847/1538-4357/abbe8a},
archivePrefix = {arXiv},
       eprint = {2010.02092},
 primaryClass = {astro-ph.HE},
       adsurl = {https://ui.adsabs.harvard.edu/abs/2020ApJ...904...97D}
}

@ARTICLE{2019MNRAS.485.5294P,
       author = {{Pandey}, S.~B. and {Hu}, Y. and {Castro-Tirado}, Ao J. and {Pozanenko}, A.~S. and {S{\'a}nchez-Ram{\'\i}rez}, R. and {Gorosabel}, J. and {Guziy}, S. and {Jelinek}, M. and {Tello}, J.~C. and {Jeong}, S. and {Oates}, S.~R. and {Zhang}, B.-B. and {Mazaeva}, E.~D. and {Volnova}, A.~A. and {Minaev}, P. Yu and {van Eerten}, H.~J. and {Caballero-Garc{\'\i}a}, M.~D. and {P{\'e}rez-Ram{\'\i}rez}, D. and {Bremer}, M. and {Winters}, J.-M. and {Park}, I.~H. and {Guelbenzu}, A. Nicuesa and {Klose}, S. and {Moskvitin}, A. and {Sokolov}, V.~V. and {Sonbas}, E. and {Ayala}, A. and {Cepa}, J. and {Butler}, N. and {Troja}, E. and {Chernenko}, A.~M. and {Molkov}, S.~V. and {Volvach}, A.~E. and {Inasaridze}, R. Ya and {Egamberdiyev}, Sh A. and {Burkhonov}, O. and {Reva}, I.~V. and {Polyakov}, K.~A. and {Matkin}, A.~A. and {Ivanov}, A.~L. and {Molotov}, I. and {Guver}, T. and {Watson}, A.~M. and {Kutyrev}, A. and {Lee}, W.~H. and {Fox}, O. and {Littlejohns}, O. and {Cucchiara}, A. and {Gonzalez}, J. and {Richer}, M.~G. and {Rom{\'a}n-Z{\'u}{\~n}iga}, C.~G. and {Tanvir}, N.~R. and {Bloom}, J.~S. and {Prochaska}, J.~X. and {Gehrels}, N. and {Moseley}, H. and {de Diego}, J.~A. and {Ram{\'\i}rez-Ruiz}, E. and {Klunko}, E.~V. and {Fan}, Y. and {Zhao}, X. and {Bai}, J. and {Wang}, Ch and {Xin}, Y. and {Cui}, Ch and {Tungalag}, N. and {Peng}, Z.-K. and {Kumar}, Amit and {Gupta}, Rahul and {Aryan}, Amar and {Kumar}, Brajesh and {Volvach}, L.~N. and {Lamb}, G.~P. and {Valeev}, A.~F.},
        title = "{A multiwavelength analysis of a collection of short-duration GRBs observed between 2012 and 2015}",
      journal = {\mnras},
         year = 2019,
        month = jun,
       volume = {485},
       number = {4},
        pages = {5294-5318},
          doi = {10.1093/mnras/stz530},
archivePrefix = {arXiv},
       eprint = {1902.07900},
 primaryClass = {astro-ph.HE},
       adsurl = {https://ui.adsabs.harvard.edu/abs/2019MNRAS.485.5294P}
}

@ARTICLE{2018A&A...616A.169M,
       author = {{Micha{\l}owskI}, Micha{\l} J. and {Xu}, Dong and {Stevens}, Jamie and {Levan}, Andrew and {Yang}, Jun and {Paragi}, Zsolt and {Kamble}, Atish and {Tsai}, An-Li and {Dannerbauer}, Helmut and {van der Horst}, Alexander J. and {Shao}, Lang and {Crosby}, David and {Gentile}, Gianfranco and {Stanway}, Elizabeth and {Wiersema}, Klaas and {Fynbo}, Johan P.~U. and {Tanvir}, Nial R. and {Kamphuis}, Peter and {Garrett}, Michael and {Bartczak}, Przemys{\l}aw},
        title = "{The second-closest gamma-ray burst: sub-luminous GRB 111005A with no supernova in a super-solar metallicity environment}",
      journal = {\aap},
         year = 2018,
        month = sep,
       volume = {616},
          eid = {A169},
        pages = {A169},
          doi = {10.1051/0004-6361/201629942},
archivePrefix = {arXiv},
       eprint = {1610.06928},
 primaryClass = {astro-ph.HE},
       adsurl = {https://ui.adsabs.harvard.edu/abs/2018A&A...616A.169M}
}

@ARTICLE{2020MNRAS.492.1919M,
       author = {{Minaev}, P.~Y. and {Pozanenko}, A.~S.},
        title = "{The E$_{p,I}$-E$_{iso}$ correlation: type I gamma-ray bursts and the new classification method}",
      journal = {\mnras},
         year = 2020,
        month = feb,
       volume = {492},
       number = {2},
        pages = {1919-1936},
          doi = {10.1093/mnras/stz3611},
archivePrefix = {arXiv},
       eprint = {1912.09810},
 primaryClass = {astro-ph.HE},
       adsurl = {https://ui.adsabs.harvard.edu/abs/2020MNRAS.492.1919M}
}

@ARTICLE{2024GCN.37290....1E,
       author = {{Evans}, P.~A. and {Goad}, M.~R. and {Osborne}, J.~P. and {Beardmore}, A.~P. and {Swift-XRT Team.}},
        title = "{GRB 240825A: Enhanced Swift-XRT position}",
      journal = {GRB Coordinates Network},
         year = 2024,
        month = aug,
       volume = {37290},
        pages = {1},
       adsurl = {https://ui.adsabs.harvard.edu/abs/2024GCN.37290....1E}
}

@ARTICLE{2023A&A...671A.112C,
       author = {{Camisasca}, A.~E. and {Guidorzi}, C. and {Amati}, L. and {Frontera}, F. and {Song}, X.~Y. and {Xiao}, S. and {Xiong}, S.~L. and {Zhang}, S.~N. and {Margutti}, R. and {Kobayashi}, S. and {Mundell}, C.~G. and {Ge}, M.~Y. and {Gomboc}, A. and {Jia}, S.~M. and {Jordana-Mitjans}, N. and {Li}, C.~K. and {Li}, X.~B. and {Maccary}, R. and {Shrestha}, M. and {Xue}, W.~C. and {Zhang}, S.},
        title = "{GRB minimum variability timescale with Insight-HXMT and Swift. Implications for progenitor models, dissipation physics, and GRB classifications}",
      journal = {\aap},
         year = 2023,
        month = mar,
       volume = {671},
          eid = {A112},
        pages = {A112},
          doi = {10.1051/0004-6361/202245657},
archivePrefix = {arXiv},
       eprint = {2301.01176},
 primaryClass = {astro-ph.HE},
       adsurl = {https://ui.adsabs.harvard.edu/abs/2023A&A...671A.112C}
}

@ARTICLE{2022Natur.612..232Y,
       author = {{Yang}, Jun and {Ai}, Shunke and {Zhang}, Bin-Bin and {Zhang}, Bing and {Liu}, Zi-Ke and {Wang}, Xiangyu Ivy and {Yang}, Yu-Han and {Yin}, Yi-Han and {Li}, Ye and {L{\"u}}, Hou-Jun},
        title = "{A long-duration gamma-ray burst with a peculiar origin}",
      journal = {\nat},
         year = 2022,
        month = dec,
       volume = {612},
       number = {7939},
        pages = {232-235},
          doi = {10.1038/s41586-022-05403-8},
archivePrefix = {arXiv},
       eprint = {2204.12771},
 primaryClass = {astro-ph.HE},
       adsurl = {https://ui.adsabs.harvard.edu/abs/2022Natur.612..232Y}
}

@ARTICLE{2025NSRev..12E.401S,
       author = {{Sun}, H. and {Wang}, C. -W. and {Yang}, J. and {Zhang}, B. -B. and {Xiong}, S. -L. and {Yin}, Y. -H.~I. and {Liu}, Y. and {Li}, Y. and {Xue}, W. -C. and {Yan}, Z. and {Zhang}, C. and {Tan}, W. -J. and {Pan}, H. -W. and {Liu}, J. -C. and {Cheng}, H. -Q. and {Zhang}, Y. -Q. and {Hu}, J. -W. and {Zheng}, C. and {An}, Z. -H. and {Cai}, C. and {Cai}, Z. -M. and {Hu}, L. and {Jin}, C. and {Li}, D. -Y. and {Li}, X. -Q. and {Liu}, H. -Y. and {Liu}, M. and {Peng}, W. -X. and {Song}, L. -M. and {Sun}, S. -L. and {Sun}, X. -J. and {Wang}, X. -L. and {Wen}, X. -Y. and {Xiao}, S. and {Yi}, S. -X. and {Zhang}, F. and {Zhang}, W. -D. and {Zhang}, X. -F. and {Zhang}, Y. -H. and {Zhao}, D. -H. and {Zheng}, S. -J. and {Ling}, Z. -X. and {Zhang}, S. -N. and {Yuan}, W. and {Zhang}, B.},
        title = "{Magnetar emergence in a peculiar gamma-ray burst from a compact star merger}",
      journal = {National Science Review},
         year = 2025,
        month = mar,
       volume = {12},
       number = {3},
          eid = {nwae401},
        pages = {nwae401},
          doi = {10.1093/nsr/nwae401},
archivePrefix = {arXiv},
       eprint = {2307.05689},
 primaryClass = {astro-ph.HE},
       adsurl = {https://ui.adsabs.harvard.edu/abs/2025NSRev..12E.401S}
}

@ARTICLE{2022ApJ...931L...2W,
       author = {{Wang}, Xiangyu Ivy and {Zhang}, Bin-Bin and {Lei}, Wei-Hua},
        title = "{GRB 200826A: A Precursor of a Long Gamma-Ray Burst with Missing Main Emission}",
      journal = {\apjl},
         year = 2022,
        month = may,
       volume = {931},
       number = {1},
          eid = {L2},
        pages = {L2},
          doi = {10.3847/2041-8213/ac6c7e},
archivePrefix = {arXiv},
       eprint = {2202.12460},
 primaryClass = {astro-ph.HE},
       adsurl = {https://ui.adsabs.harvard.edu/abs/2022ApJ...931L...2W}
}

@ARTICLE{2007A&A...469..379E,
       author = {{Evans}, P.~A. and {Beardmore}, A.~P. and {Page}, K.~L. and {Tyler}, L.~G. and {Osborne}, J.~P. and {Goad}, M.~R. and {O'Brien}, P.~T. and {Vetere}, L. and {Racusin}, J. and {Morris}, D. and {Burrows}, D.~N. and {Capalbi}, M. and {Perri}, M. and {Gehrels}, N. and {Romano}, P.},
        title = "{An online repository of Swift/XRT light curves of {\ensuremath{\gamma}}-ray bursts}",
      journal = {\aap},
         year = 2007,
        month = jul,
       volume = {469},
       number = {1},
        pages = {379-385},
          doi = {10.1051/0004-6361:20077530},
archivePrefix = {arXiv},
       eprint = {0704.0128},
 primaryClass = {astro-ph},
       adsurl = {https://ui.adsabs.harvard.edu/abs/2007A&A...469..379E}
}

@ARTICLE{2013NewAR..57..141G,
       author = {{Gao}, He and {Lei}, Wei-Hua and {Zou}, Yuan-Chuan and {Wu}, Xue-Feng and {Zhang}, Bing},
        title = "{A complete reference of the analytical synchrotron external shock models of gamma-ray bursts}",
      journal = {\nar},
         year = 2013,
        month = dec,
       volume = {57},
       number = {6},
        pages = {141-190},
          doi = {10.1016/j.newar.2013.10.001},
archivePrefix = {arXiv},
       eprint = {1310.2181},
 primaryClass = {astro-ph.HE},
       adsurl = {https://ui.adsabs.harvard.edu/abs/2013NewAR..57..141G}
}

@ARTICLE{2016MNRAS.462.3243Z,
       author = {{Zhang}, Zhi-Bin and {Yang}, En-Bo and {Choi}, Chul-Sung and {Chang}, Heon-Young},
        title = "{Classifying gamma-ray bursts with Gaussian Mixture Model}",
      journal = {\mnras},
         year = 2016,
        month = nov,
       volume = {462},
       number = {3},
        pages = {3243-3254},
          doi = {10.1093/mnras/stw1835},
       adsurl = {https://ui.adsabs.harvard.edu/abs/2016MNRAS.462.3243Z}
}

@ARTICLE{2022ApJ...927..157M,
       author = {{Moss}, Michael and {Lien}, Amy and {Guiriec}, Sylvain and {Cenko}, S. Bradley and {Sakamoto}, Takanori},
        title = "{Instrumental Tip-of-the-iceberg Effects on the Prompt Emission of Swift/BAT Gamma-ray Bursts}",
      journal = {\apj},
         year = 2022,
        month = mar,
       volume = {927},
       number = {2},
          eid = {157},
        pages = {157},
          doi = {10.3847/1538-4357/ac4d94},
archivePrefix = {arXiv},
       eprint = {2111.13392},
 primaryClass = {astro-ph.HE},
       adsurl = {https://ui.adsabs.harvard.edu/abs/2022ApJ...927..157M}
}

@ARTICLE{2009MNRAS.397.1177E,
       author = {{Evans}, P.~A. and {Beardmore}, A.~P. and {Page}, K.~L. and {Osborne}, J.~P. and {O'Brien}, P.~T. and {Willingale}, R. and {Starling}, R.~L.~C. and {Burrows}, D.~N. and {Godet}, O. and {Vetere}, L. and {Racusin}, J. and {Goad}, M.~R. and {Wiersema}, K. and {Angelini}, L. and {Capalbi}, M. and {Chincarini}, G. and {Gehrels}, N. and {Kennea}, J.~A. and {Margutti}, R. and {Morris}, D.~C. and {Mountford}, C.~J. and {Pagani}, C. and {Perri}, M. and {Romano}, P. and {Tanvir}, N.},
        title = "{Methods and results of an automatic analysis of a complete sample of Swift-XRT observations of GRBs}",
      journal = {\mnras},
         year = 2009,
        month = aug,
       volume = {397},
       number = {3},
        pages = {1177-1201},
          doi = {10.1111/j.1365-2966.2009.14913.x},
archivePrefix = {arXiv},
       eprint = {0812.3662},
 primaryClass = {astro-ph},
       adsurl = {https://ui.adsabs.harvard.edu/abs/2009MNRAS.397.1177E}
}

@ARTICLE{2024GCN.37296....1K,
       author = {{Kuin}, N.~P.~M. and {Gupta}, R. and {Swift/UVOT Team}},
        title = "{GRB 240825A: Swift/UVOT Detection}",
      journal = {GRB Coordinates Network},
         year = 2024,
        month = aug,
       volume = {37296},
        pages = {1},
       adsurl = {https://ui.adsabs.harvard.edu/abs/2024GCN.37296....1K}
}

@ARTICLE{2023ApJ...954L...5V,
       author = {{Veres}, P. and {Bhat}, P.~N. and {Burns}, E. and {Hamburg}, R. and {Fraija}, N. and {Kocevski}, D. and {Preece}, R. and {Poolakkil}, S. and {Christensen}, N. and {Bizouard}, M.~A. and {Dal Canton}, T. and {Bala}, S. and {Bissaldi}, E. and {Briggs}, M.~S. and {Cleveland}, W. and {Goldstein}, A. and {Hristov}, B.~A. and {Hui}, C.~M. and {Lesage}, S. and {Mailyan}, B. and {Roberts}, O.~J. and {Wilson-Hodge}, C.~A.},
        title = "{Extreme Variability in a Long-duration Gamma-Ray Burst Associated with a Kilonova}",
      journal = {\apjl},
         year = 2023,
        month = sep,
       volume = {954},
       number = {1},
          eid = {L5},
        pages = {L5},
          doi = {10.3847/2041-8213/ace82d},
archivePrefix = {arXiv},
       eprint = {2305.12262},
 primaryClass = {astro-ph.HE},
       adsurl = {https://ui.adsabs.harvard.edu/abs/2023ApJ...954L...5V}
}

@ARTICLE{2001PASP..113.1420V,
       author = {{van Dokkum}, Pieter G.},
        title = "{Cosmic-Ray Rejection by Laplacian Edge Detection}",
      journal = {\pasp},
         year = 2001,
        month = nov,
       volume = {113},
       number = {789},
        pages = {1420-1427},
          doi = {10.1086/323894},
archivePrefix = {arXiv},
       eprint = {astro-ph/0108003},
 primaryClass = {astro-ph},
       adsurl = {https://ui.adsabs.harvard.edu/abs/2001PASP..113.1420V}
}

@ARTICLE{2023MNRAS.519.3201C,
       author = {{Caballero-Garc{\'\i}a}, M.~D. and {Gupta}, Rahul and {Pandey}, S.~B. and {Oates}, S.~R. and {Marisaldi}, M. and {Ramsli}, A. and {Hu}, Y. -D. and {Castro-Tirado}, A.~J. and {S{\'a}nchez-Ram{\'\i}rez}, R. and {Connell}, P.~H. and {Christiansen}, F. and {Ror}, A. Kumar and {Aryan}, A. and {Bai}, J. -M. and {Castro-Tirado}, M.~A. and {Fan}, Y. -F. and {Fern{\'a}ndez-Garc{\'\i}a}, E. and {Kumar}, A. and {Lindanger}, A. and {Mezentsev}, A. and {Navarro-Gonz{\'a}lez}, J. and {Neubert}, T. and {{\O}stgaard}, N. and {P{\'e}rez-Garc{\'\i}a}, I. and {Reglero}, V. and {Sarria}, D. and {Sun}, T.~R. and {Xiong}, D. -R. and {Yang}, J. and {Yang}, Y. -H. and {Zhang}, B. -B.},
        title = "{Multiwavelength study of the luminous GRB 210619B observed with Fermi and ASIM}",
      journal = {\mnras},
         year = 2023,
        month = mar,
       volume = {519},
       number = {3},
        pages = {3201-3226},
          doi = {10.1093/mnras/stac3629},
archivePrefix = {arXiv},
       eprint = {2205.07790},
 primaryClass = {astro-ph.HE},
       adsurl = {https://ui.adsabs.harvard.edu/abs/2023MNRAS.519.3201C}
}
\bibliographystyle{aasjournalv7}

\end{document}